\documentclass[a4paper,11pt]{article}
\pdfoutput=1 
\usepackage{jinstpub} 
\usepackage{floatrow}
\usepackage{graphicx}
\usepackage{pdfpages}
\usepackage{placeins}
\usepackage{subcaption}
\usepackage{tabularx}
\usepackage{textcomp}

\title{\boldmath ARIADNE - A Novel Optical LArTPC: Technical Design Report and Initial Characterisation using a Secondary Beam from the CERN PS and Cosmic Muons}

\author{D. Hollywood,}
\author{K. Majumdar,}
\author[1]{K. Mavrokoridis,\note{Corresponding author}}
\author{K. J. McCormick,}
\author{B. Philippou,}
\author{S. Powell,}
\author{A. Roberts,}
\author{N. A. Smith,}
\author{G. Stavrakis,}
\author{C. Touramanis,}
\author{and J. Vann}

\affiliation{University of Liverpool, Department of Physics, Oliver Lodge Bld, Oxford Street, \\ Liverpool, L69 7ZE, UK}

\emailAdd{k.mavrokoridis@liverpool.ac.uk}

\abstract{\\ARIADNE is a 1-ton (330~kg fiducial mass) dual-phase liquid argon (LAr) time projection chamber (TPC) featuring a novel optical readout. Four electron-multiplying charge-coupled device (EMCCD) cameras are mounted externally, and these capture the secondary scintillation light produced in the holes of a thick electron gas multiplier (THGEM).  Track reconstruction using this novel readout approach is demonstrated. Optical readout has the potential to be a cost effective alternative to charge readout in future LArTPCs.

In this paper, the technical design of the detector is detailed. Results of mixed particle detection using a secondary beam from the CERN PS (representing the first ever optical images of argon interactions in a dual-phase LArTPC at a beamline) and cosmic muon detection at the University of Liverpool are also presented.}

\keywords{Time projection Chambers (TPC), Noble liquid detectors, Micropattern gaseous detectors, Photon detectors for UV, visible and IR photons (solid-state).}

\arxivnumber{1910.03406} 

\collaboration{\hspace{37mm} \includegraphics[height=35mm]{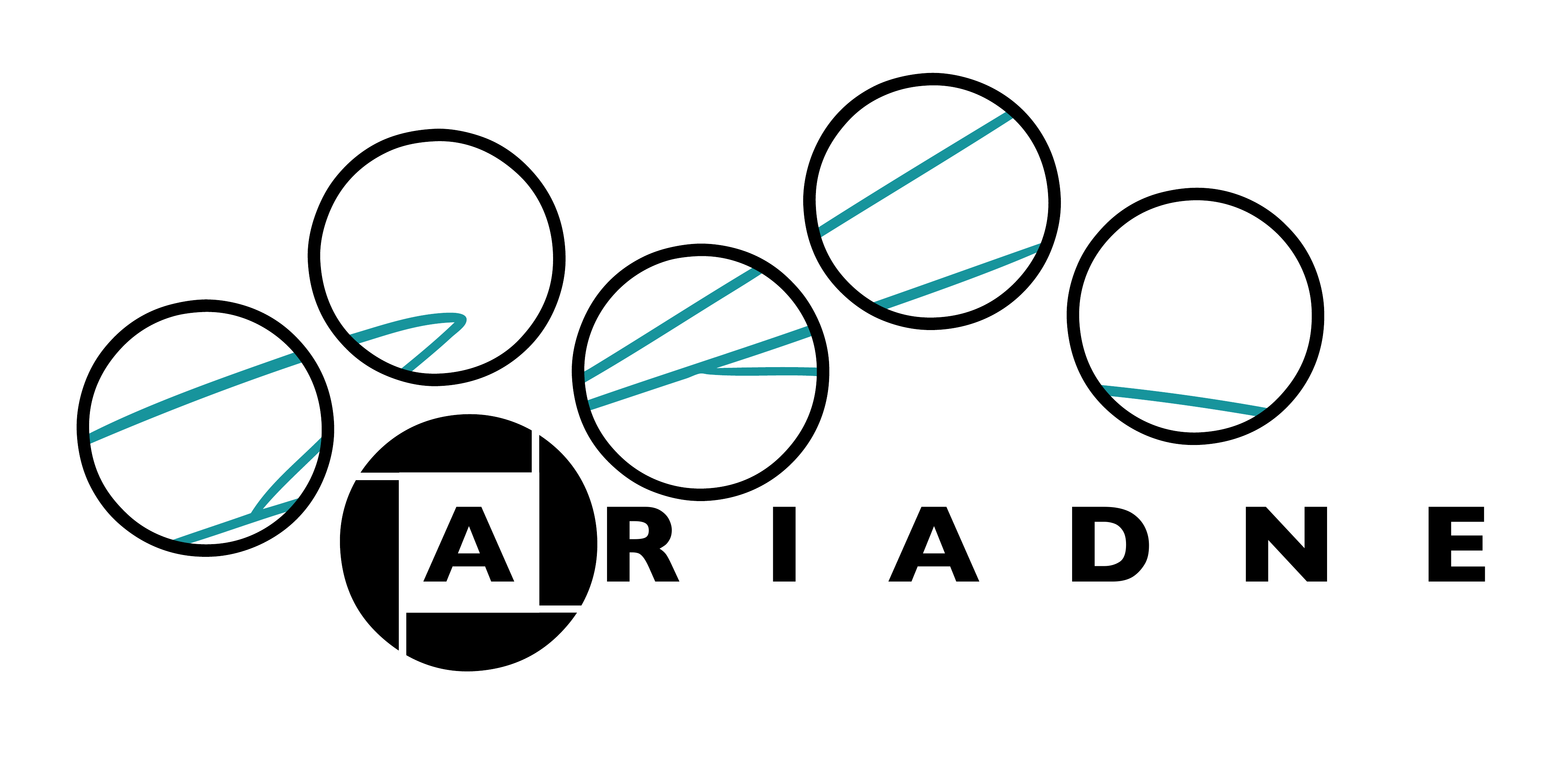}}

\begin{document}

\includepdf[offset=72 -72]{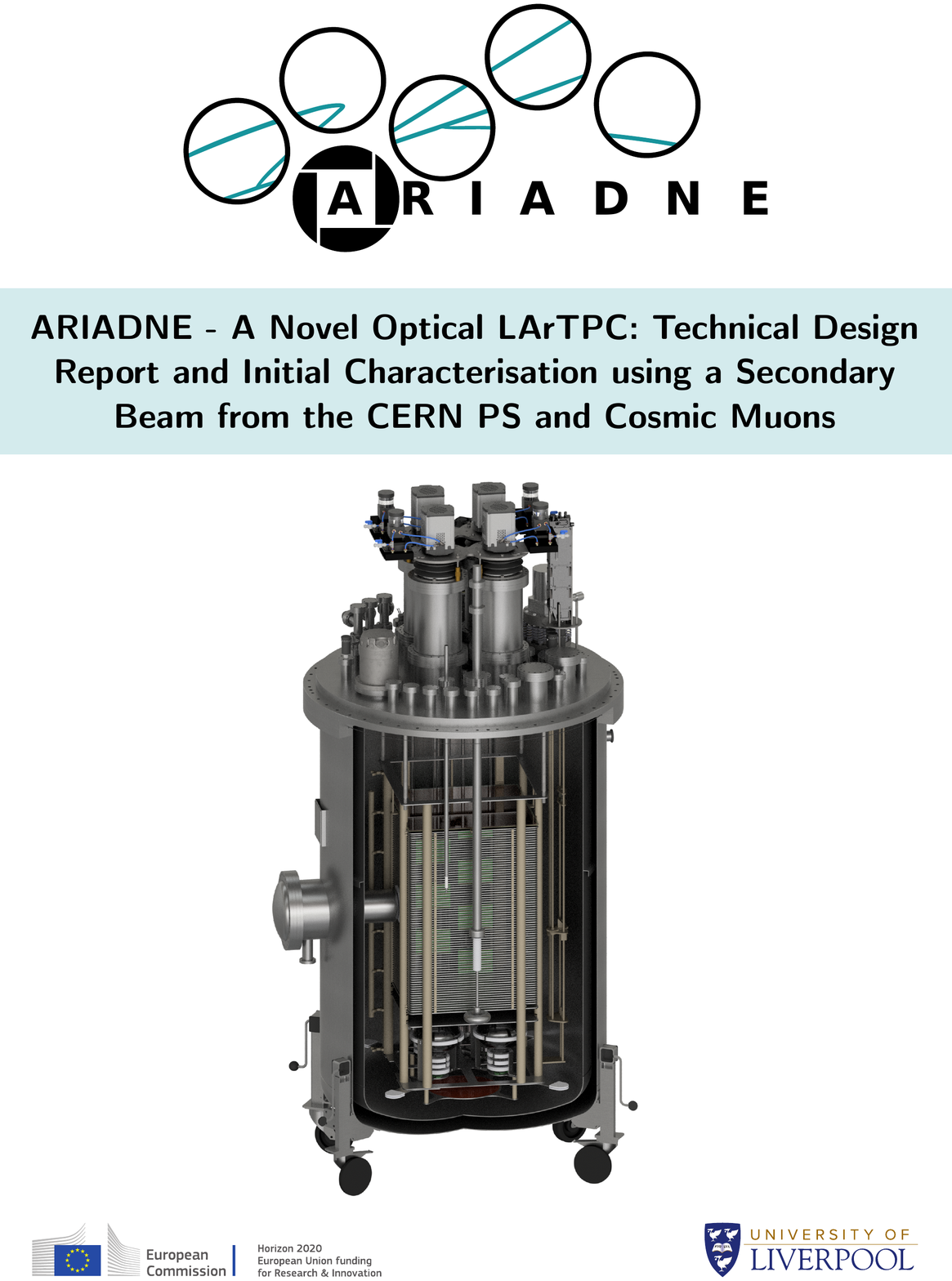}

\maketitle
\flushbottom

\newpage
\FloatBarrier
\section{Introduction}

\subsection{The ARIADNE Experiment}
Liquid Argon Time Projection Chambers (LArTPCs) have continued to grow in size and sophistication since their inception at the end of the 1970s. The current generation of neutrino detectors include the three experiments making up the Short Baseline Neutrino Program~\cite{SBNP}: SBND (112~tons), MicroBooNE (89~tons) and ICARUS-T600 (470~tons), as well as the single- and dual-phase ProtoDUNE experiments \cite{ProtoDUNE-SP, ProtoDUNE-DP} (411 and 300~tons respectively). It is therefore clear that kiloton-scale detectors will soon become a reality. In the Long Baseline Neutrino program, the DUNE project~\cite{DuneCDRVol1, DuneCDRVol2, DuneCDRVol3, DuneCDRVol4, DuneIDRVol1, DuneIDRVol2, DuneIDRVol3} is proposing the use of four 17,000~ton LArTPCs.

Large dual-phase TPCs are also widely used in direct dark matter detection experiments. DarkSide \cite{DARKSIDE20k} is proposing the use of a 20~tonne LArTPC, and LUX-ZEPLIN is based around a TPC containing 7~tonnes of liquid xenon \cite{LUXZEPLIN}.
\\
\\
ARIADNE (\textbf{AR}gon \textbf{I}m\textbf{A}ging \textbf{D}etectio\textbf{N} chamb\textbf{E}r) is a 1-ton (330~kg fiducial mass) dual-phase LArTPC, built to pursue the initial demonstration of optical readout \cite{ccdargon, emccdargon} on a larger scale. ARIADNE additionally serves as a platform for future R\&D and the maturation of optical readout.

In searches for dark matter, the detection principle of dual-phase detectors relies upon the readout of both prompt scintillation light (S1) produced in the liquid phase and secondary electroluminescence light (S2), produced during the extraction of electrons from the liquid phase into the gas phase.

For dual-phase detectors in the neutrino sector, more stringent tracking performance requirements have favoured a different readout approach. Typically, tracking is performed using ionised electron signal, rather than the electroluminescence light signal. By using a Thick Gas Electron Multiplier (THGEM) in the gas phase, the ionised electron signal is amplified, boosting signal-to-noise. This amplified charge signal is then typically collected on a segmented anode, thereby enabling tracking. 

The ARIADNE experiment is exploring novel optical readout techniques, which allow for high resolution tracking using the secondary scintillation light produced by a THGEM. Optical readout of secondary scintillation light produced within a THGEM has previously been reported in dual-phase cryogenic avalanche detectors, using Geiger-mode avalanche photodiodes \cite{BondarGAPD}, silicon photomultipliers \cite{THGEMReadout}, CCD \cite{ccdargon} and EMCCD cameras \cite{emccdargon}. Optical readout has the potential to offer many benefits, including massive simplification and savings in construction and operation costs. Improvements in detector performance may also be realised, particularly in terms of spatial resolution and low energy thresholds. 
\\
\\
The ARIADNE TPC is 54 x 54~cm ($x, y$) with a total drift length of 80~cm ($z$). Four PMTs are installed below the TPC, providing detector triggering via the detection of primary scintillation (S1) light, and also performing auxiliary detection of secondary scintillation (S2) light.

The field cage of the TPC is composed of a total of 79 field-shaping rings, spaced 1~cm apart. The high density of the rings ensures high drift field uniformity. The cathode is installed at the base of the field cage, and can be biased up to 80~kV, providing a drift field of up to  1~kV/cm within the TPC. This bias is provided by a custom high-voltage feedthrough, described in more detail in Section~\ref{subsec:HVFT}. Above the field cage is an extraction grid, with a Thick Gas Electron Multiplier (THGEM) 11~mm above the grid. Both segmented and monolithic THGEM designs have been used on ARIADNE - these are detailed in Section~\ref{subsec:THGEM}. The space between the extraction grid and THGEM forms the extraction region, which is nominally biased with an extraction field of 3~kV/cm.

\begin{figure}
\includegraphics[width=\textwidth]{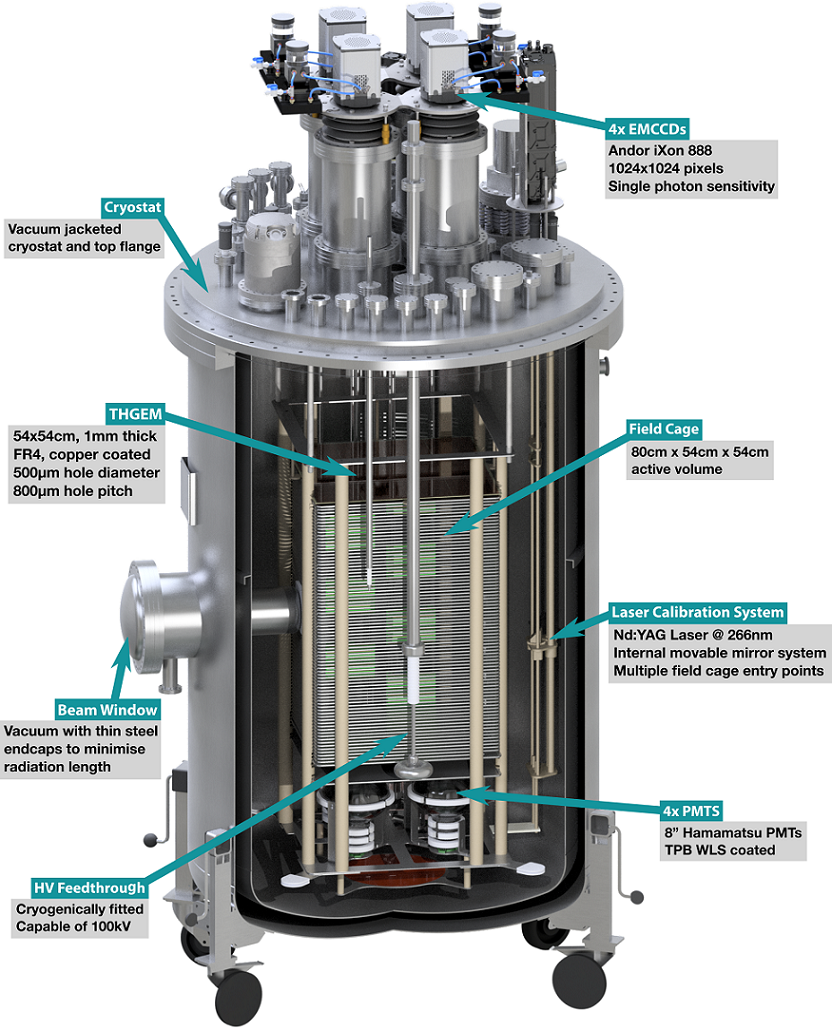}
\caption{The ARIADNE detector. The TPC is housed inside a vacuum-jacketed cryostat. A beam window and beam plug can be seen on the left of the image, penetrating through the cryostat walls and towards the TPC. Four EMCCD cameras are installed on the top flange, looking down into the cryogenic volume through viewports.}
\label{fig:ARIADNE}
\end{figure}

For the purposes of optical readout, a glass plate, with a coating of Tetraphenyl Butadiene (TPB) on its underside, is placed 2~mm above the THGEM. This plate shifts the 128~nm Vacuum Ultra Violet (VUV) light that is produced in the THGEM holes to 420~nm, allowing the EMCCD cameras and PMTs to operate close to their peak quantum efficiencies. Four EMCCD cameras, installed outside of the cryostat, look down at the THGEM through optical viewports and capture the S2 light produced in the THGEM holes.
\\
\\
Figure~\ref{fig:ARIADNE} shows a general overview of the ARIADNE detector.

\vspace{5mm}

\subsection{Operation Principle} 
\begin{figure}
\includegraphics[width=0.89\textwidth]{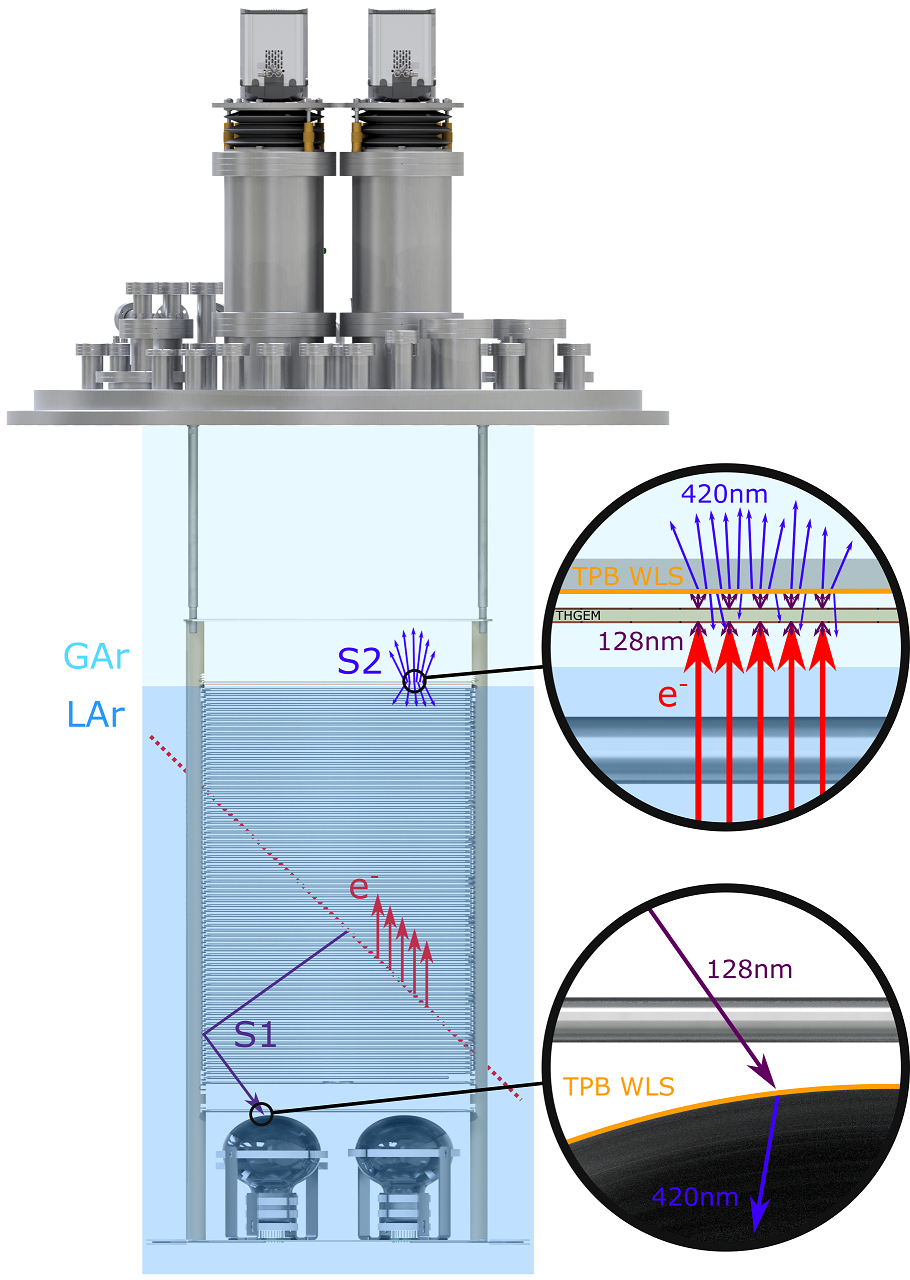}
\caption{The operation principle of ARIADNE. An ionising track in the detector creates free electrons in the liquid argon, and the drift field in the TPC directs these electrons up towards the THGEM. The electrons are extracted out of the liquid phase and into the THGEM holes by the extraction field. Once inside a THGEM hole, the electrons are accelerated by the large electric field across the THGEM, and S2 light is produced by electroluminescence during this process. This S2 light is detected by the external EMCCD cameras and the PMTs.}
\label{fig:DetectionPrinciple}
\end{figure}

Figure~\ref{fig:DetectionPrinciple} shows a schematic of the operation principle of ARIADNE.
\\
\\
As a charged particle passes through LAr, it causes both ionisation and excitation of the argon atoms. The excited argon atoms emit primary scintillation (S1) light at 128~nm as they de-excite, which is immediately detected by the PMTs at the base of the TPC. The timing of this S1 pulse is used as the event trigger. The ionisation electrons are drifted towards the THGEM by a nominal 0.5~kV/cm drift field within the TPC.

Once at the surface of the liquid, the electrons are extracted into the gas phase by a nominal extraction field of 3~kV/cm, and then directed into the THGEM holes. Once inside a THGEM hole, the electrons experience a much larger electric field - nominally of the order of 30~kV/cm - across the THGEM. This field is sufficient to accelerate each electron above the scintillation threshold of gaseous argon. Each electron gains enough energy to excite the gas argon atoms that they collide with, resulting in the production of secondary scintillation (S2) light. The EMCCD cameras capture this S2 light, and reproduce a two-dimensional image of the ionising event in the TPC. In doing so, ARIADNE retains the positional information of the S2 light, with a nominal resolution of 1.1~mm per EMCCD sensor pixel. The high resolution of the EMCCD sensors provides high granularity tracking, and their single photon sensitivity allows for optimisation of detection thresholds.

\newpage
\FloatBarrier
\section{Detector Design}

\subsection{TPC and Drift Field}
\label{subsec:TPC}
\begin{figure}[ht]
\includegraphics[width=\textwidth]{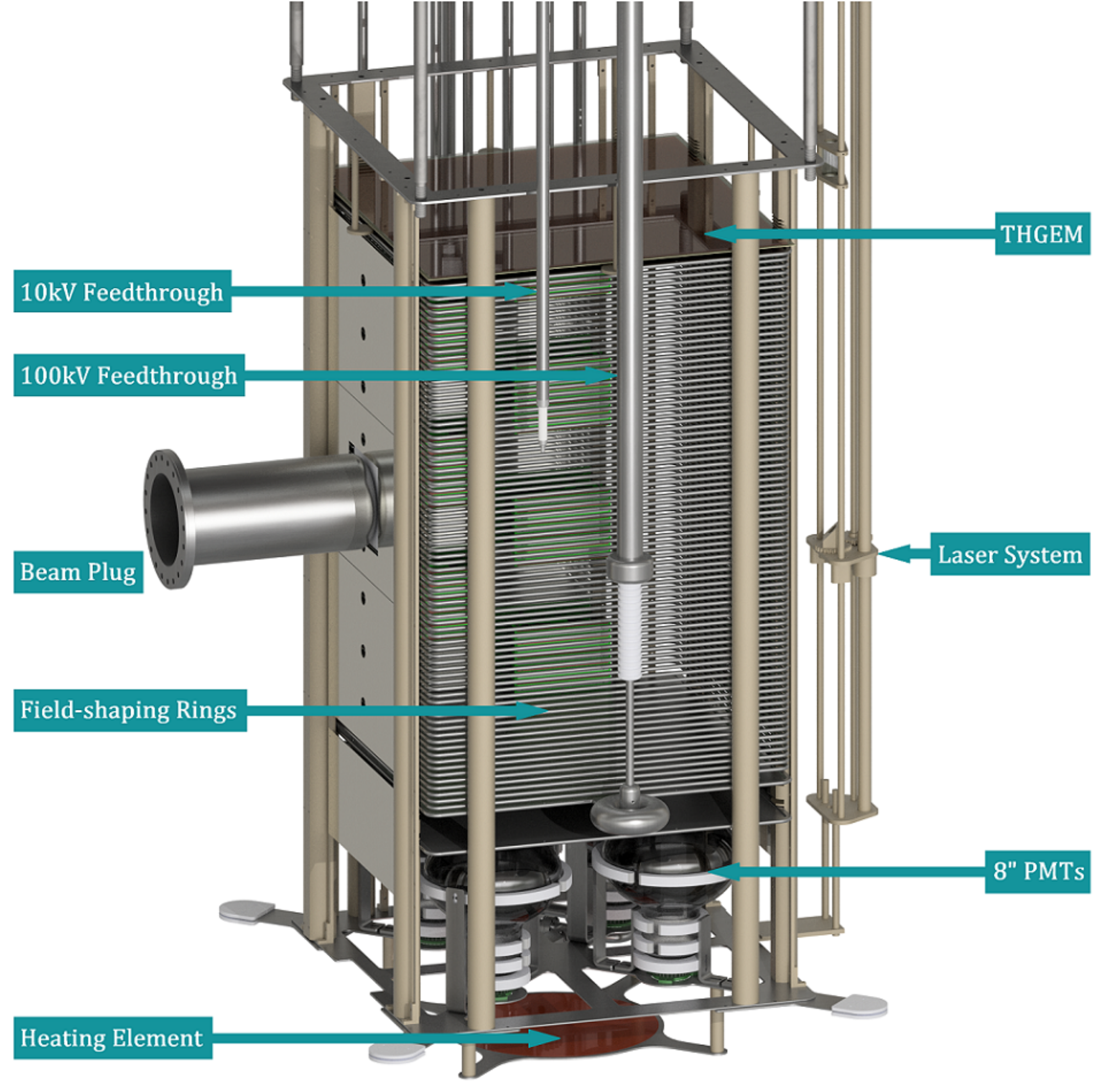}
\caption{The ARIADNE TPC, with key detector components highlighted.}
\label{fig:TPC}
\end{figure}

Figure~\ref{fig:TPC} shows a CAD model of the ARIADNE TPC and other key detector components.\\

\begin{figure}[ht]
\includegraphics[width=0.42\textwidth]{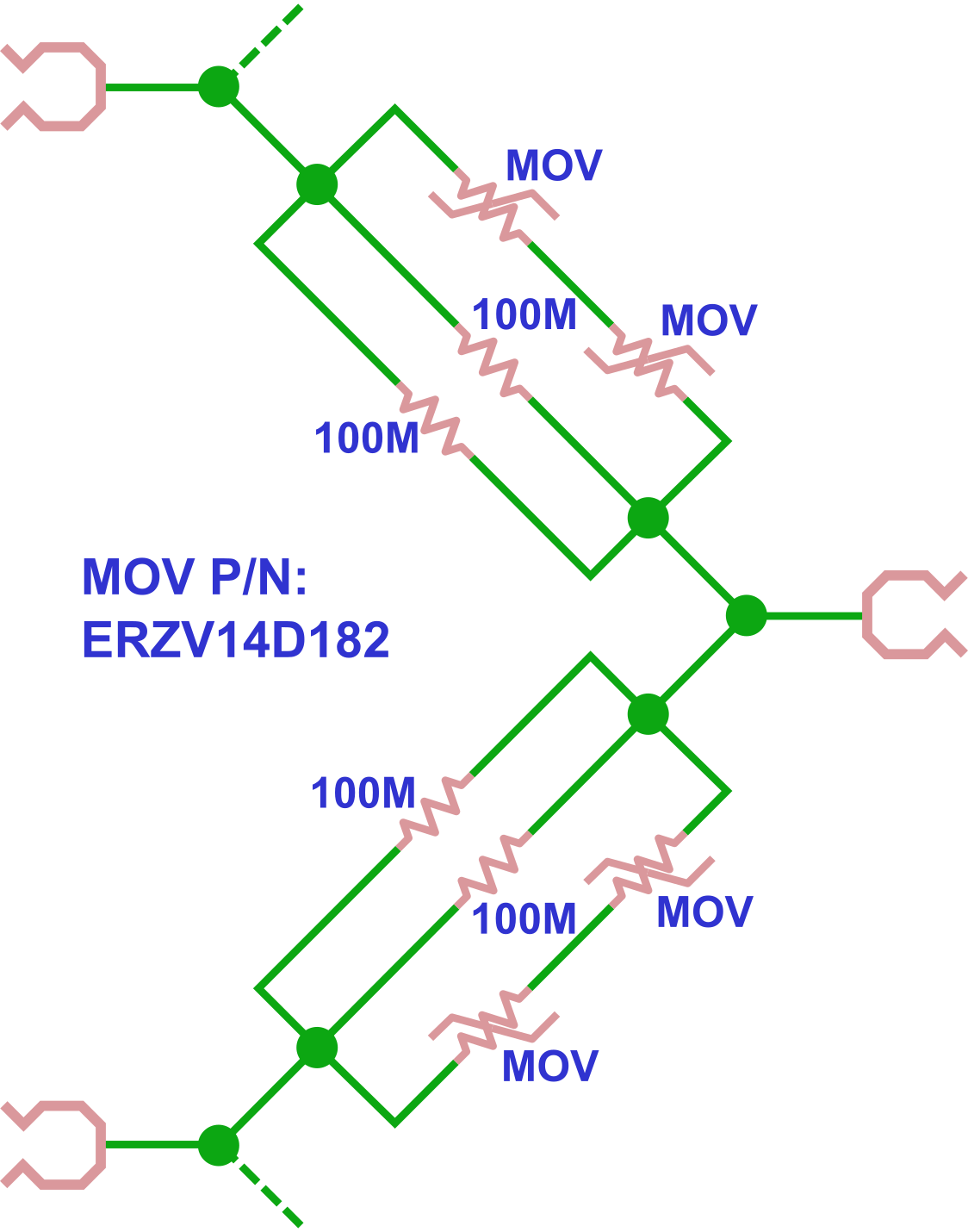}
\caption{Schematic of the resistor chain connections and components between three adjacent field-shaping rings. The fuse clips at top-left, bottom-left and middle-right connect to the rings. Between each pair of clips are two 100~M$\Omega$ resistors in parallel, giving a total of 50~M$\Omega$ resistance between rings. In parallel with this are the metal-oxide varistors (all labelled as ``MOV'') for transient/surge protection.}
\label{fig:ResistorChainSchem}
\end{figure}

\noindent The field cage is defined by 79 stainless steel field-shaping rings, formed using CNC (Computer Numerical Control) wire bending. The ends of these rings are welded together and polished smooth to remove sharp surfaces that might induce local high electric fields. The cage is supported by 8 VICTREX\textsuperscript{\textregistered{}} PEEK 450G  rods, which have slots machined at regular intervals to support the field-shaping rings (one per slot). The rings have a vertical (centre-to-centre) spacing of 1~cm, giving a total TPC drift length of 80~cm, including the spacings between the top ring and the extraction grid, and the bottom ring and the cathode.
\\
\\
The uniform gradient of the drift field is produced by a series of 100~M$\Omega$ resistors connecting pairs of adjacent field-shaping rings, as well as connecting the extraction and cathode grids to the top and bottom rings respectively. This resistor chain therefore dictates the potential difference between one ring and the next. A schematic of the connections between rings is shown in Figure~\ref{fig:ResistorChainSchem}.

In addition to the fixed-value resistors, the chain also includes metal-oxide varistors, which limit the effects of any transients or voltage spikes by acting as an efficient short circuit above 1800~V. Otherwise, a sudden large potential difference between two or more rings - and the resulting high current - could damage the fixed-value resistors or other parts of the detector.\\

\begin{figure}[ht]
\begin{subfigure}{0.49\textwidth}
  \centering
  \includegraphics[height=51mm]{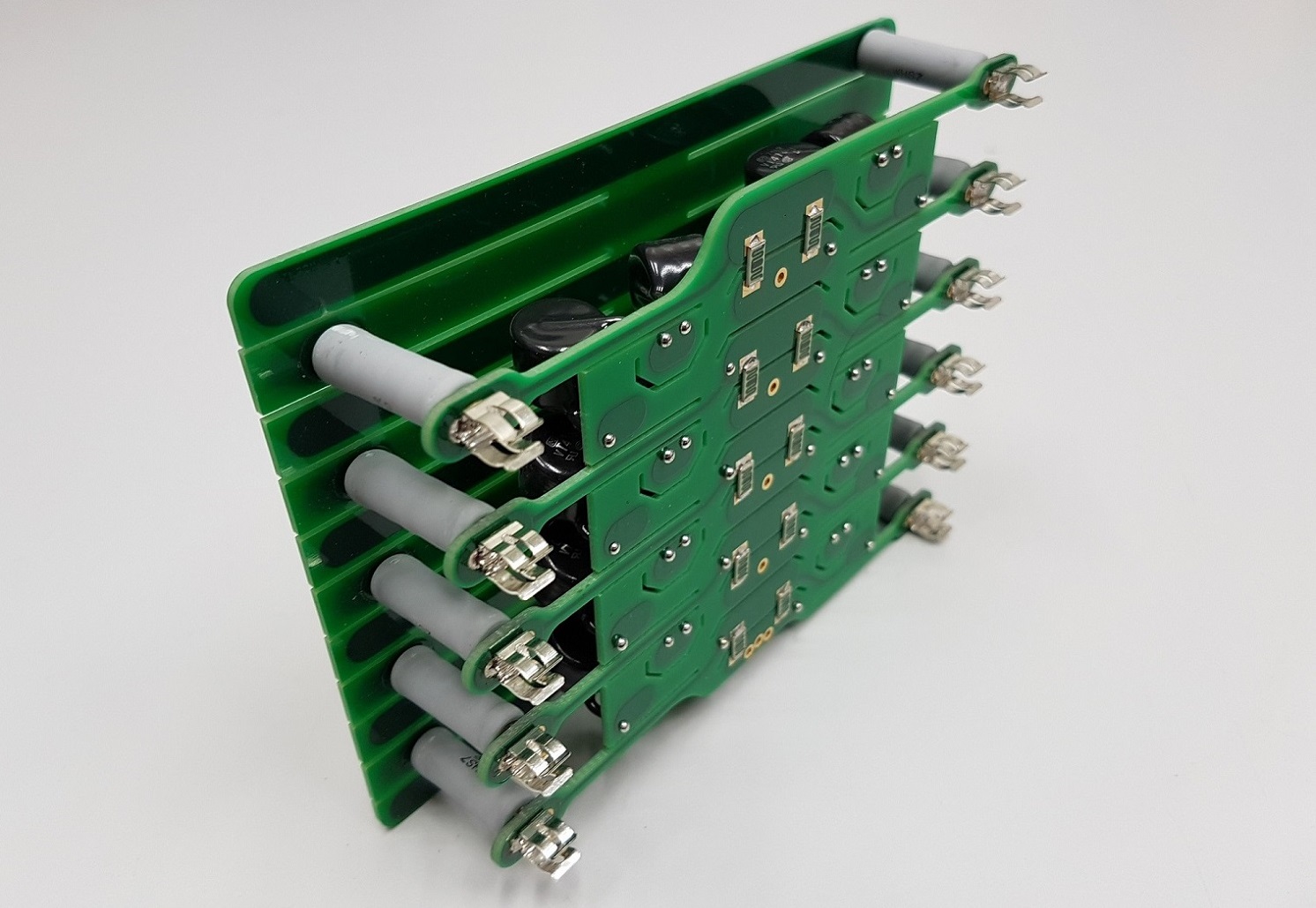}
\end{subfigure}
\hspace{1mm}
\begin{subfigure}{0.49\textwidth}
  \centering
  \includegraphics[height=51mm]{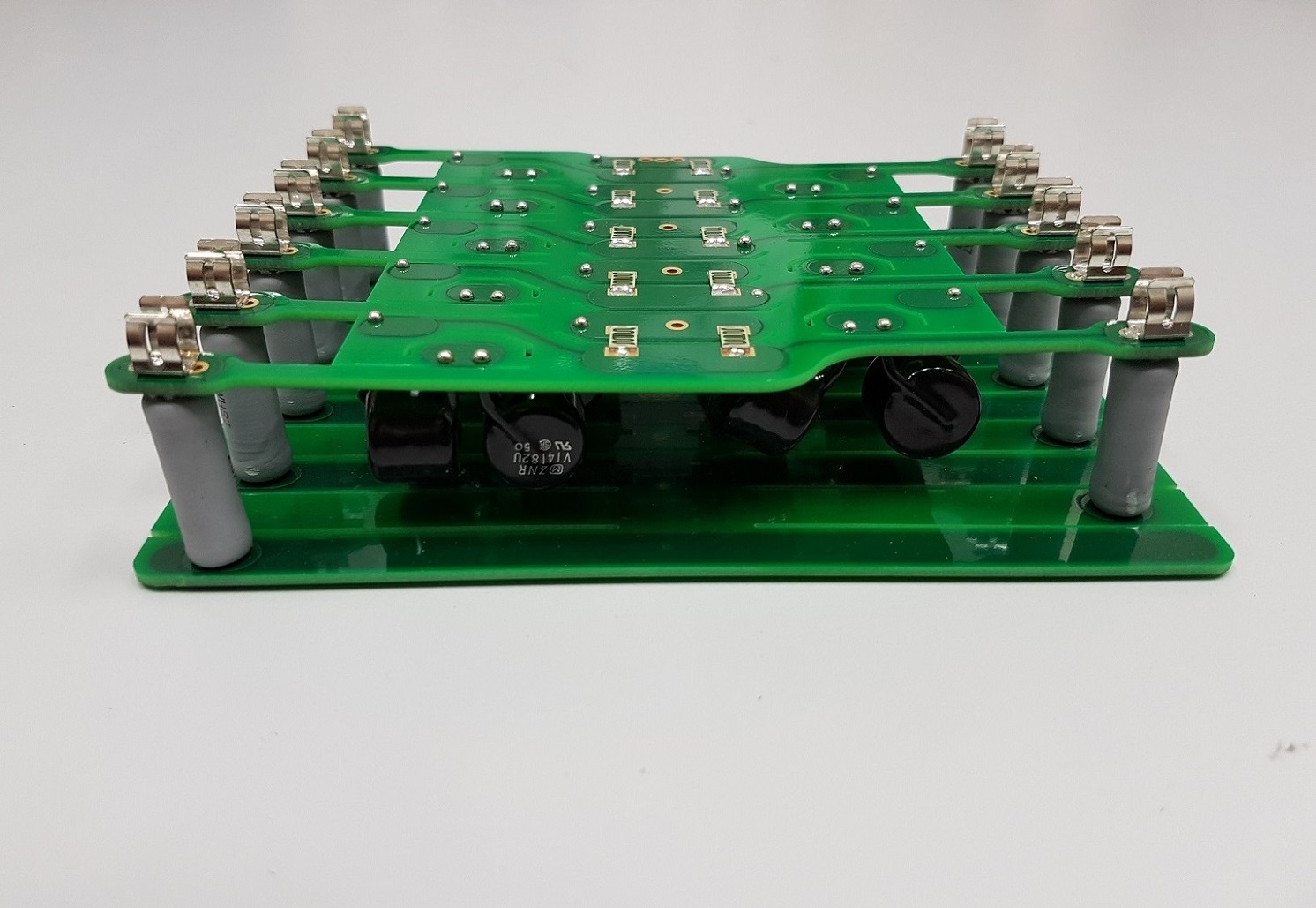}
\end{subfigure}
\caption{One of the resistor chain modules, which covers 9 unique field-shaping rings. The two PCBs can be seen, with the first housing the surface-mounted resistors and varistors, and the second acting as a field shaper. The fuse clips for connecting the module to the field-shaping rings are located on the flexible extensions to the first PCB.}
\label{fig:ResistorChain}
\end{figure}

\noindent The resistors are grouped onto 9 modules, one of which is shown in Figure~\ref{fig:ResistorChain}. Each module consists of two PCBs - the first of which sits closest to the field-shaping rings and houses the resistors and varistors, and the second acting to shape the electric field around the first PCB. This reduces its impact on the field uniformity within the TPC, and is achieved using embedded copper strips sitting parallel to - and biased at the same voltage as - the field-shaping rings.

Each module houses 11 fuse clips for connecting to the field-shaping rings (1 per clip). The middle 7 rings are uniquely connected to each module, with the two top-most and bottom-most rings also being connected to the modules above and below respectively. The top 9 clips of each module are coupled to resistors as in Figure \ref{fig:ResistorChainSchem}, but to ensure that the resistances of the overlapping rings remain consistent, the bottom 2 clips are not. Instead, they act as purely passive structural support. The clips are located on extensions to the first PCB layer, which can flex slightly to account for any small non-uniform contraction between the different materials present under cryogenic conditions. This allows each clip to keep a consistent and tight connection to its ring. The modules are fitted to the inside of the field cage, in order to avoid introducing localised high electric fields outside the TPC.\\

\begin{figure}[ht]
\includegraphics[width=\textwidth]{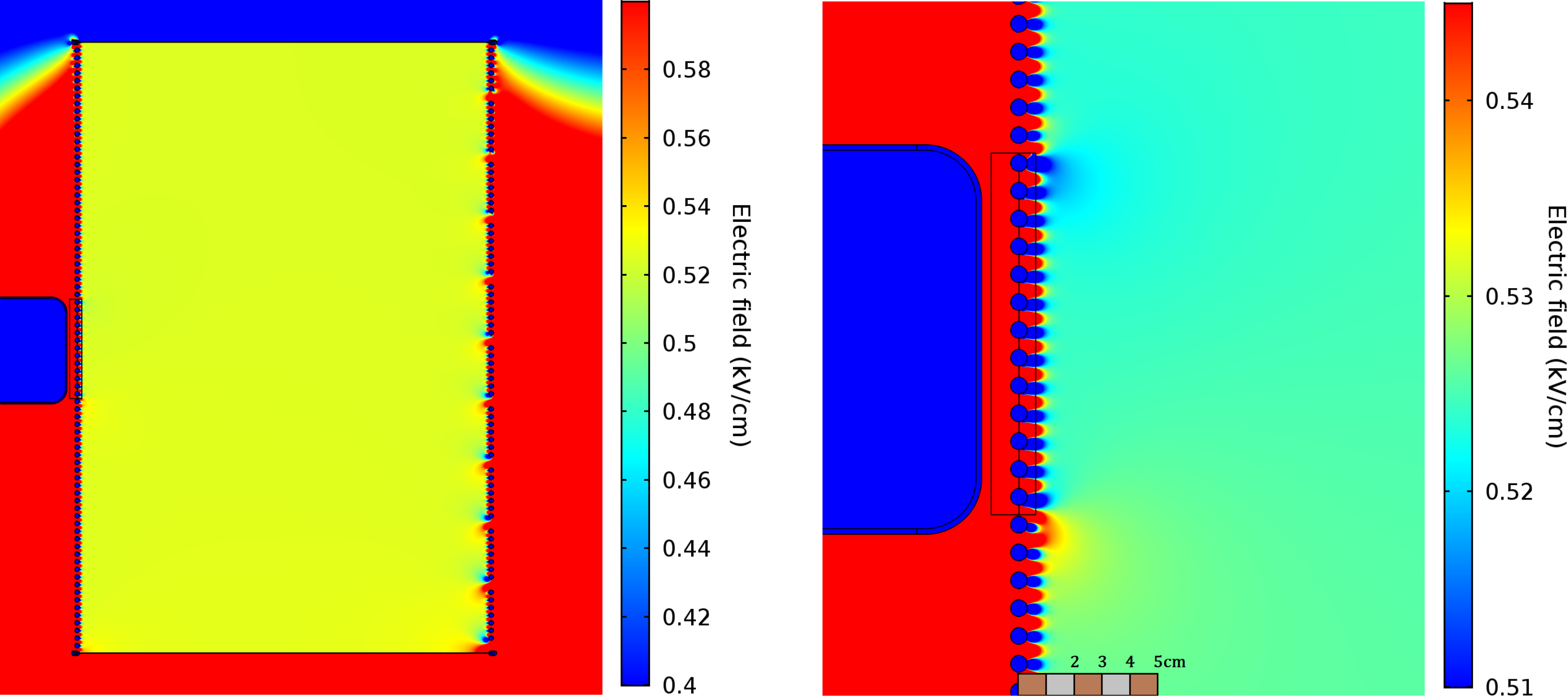}
\caption{The results of COMSOL drift field simulations for the entire field cage (left) and the region within the field cage close to the beam plug (right). Small localised deviations from drift field uniformity of up to 15\% exist in the region near the beam plug, but all non-uniformities subside to a level of no more than 1~\% beyond a distance of $\approx$ 2~cm inwards from the field-shaping rings, as indicated by the scale bar at the bottom.}
\label{fig:DriftFieldTPC}
\end{figure}

\noindent Under normal operating conditions, the drift field has a nominal gradient of 0.5~kV/cm, but it is designed for operation up to 1~kV/cm. The uniformity of the drift field was optimised during the design phase of the experiment using COMSOL\texttrademark{} \cite{COMSOL} simulations of the TPC, as shown in Figure~\ref{fig:DriftFieldTPC}.

At and near the centre of the TPC, deviations from a uniform drift field are lower than 1\%, and increase to $\approx$~10\% near the outermost edges of the field cage. Localised deviations of up to 15\% exist in the regions closest to the beam plug (discussed in Section~\ref{subsec:BeamPlug}) and the gaps in the field-shaping rings for the laser calibration system (discussed in Section~\ref{subsec:Laser}). However, these deviations from uniformity are outside - and therefore do not affect - the (instrumented) active volume, which has an $x-y$ area of 53 $\times$ 53~cm as defined by the active area of the THGEM - i.e. slightly smaller than the 54 $\times$ 54~cm $x-y$ area of the field cage.

\subsection{Energy Containment}
\label{subsec:Containment}
GEANT4 simulations were performed to study the energy containment of the ARIADNE TPC. Energy containment is defined as the fraction of a particle's total ionising energy deposition that occurs within the fiducial volume. A particle that deposits all of its ionising energy within the fiducial volume would have 100$\%$ energy containment. Particles with sufficient momentum to pass through the TPC without stopping would have reduced energy containment. 

\begin{figure}[tp]
\includegraphics[width=0.75\textwidth]{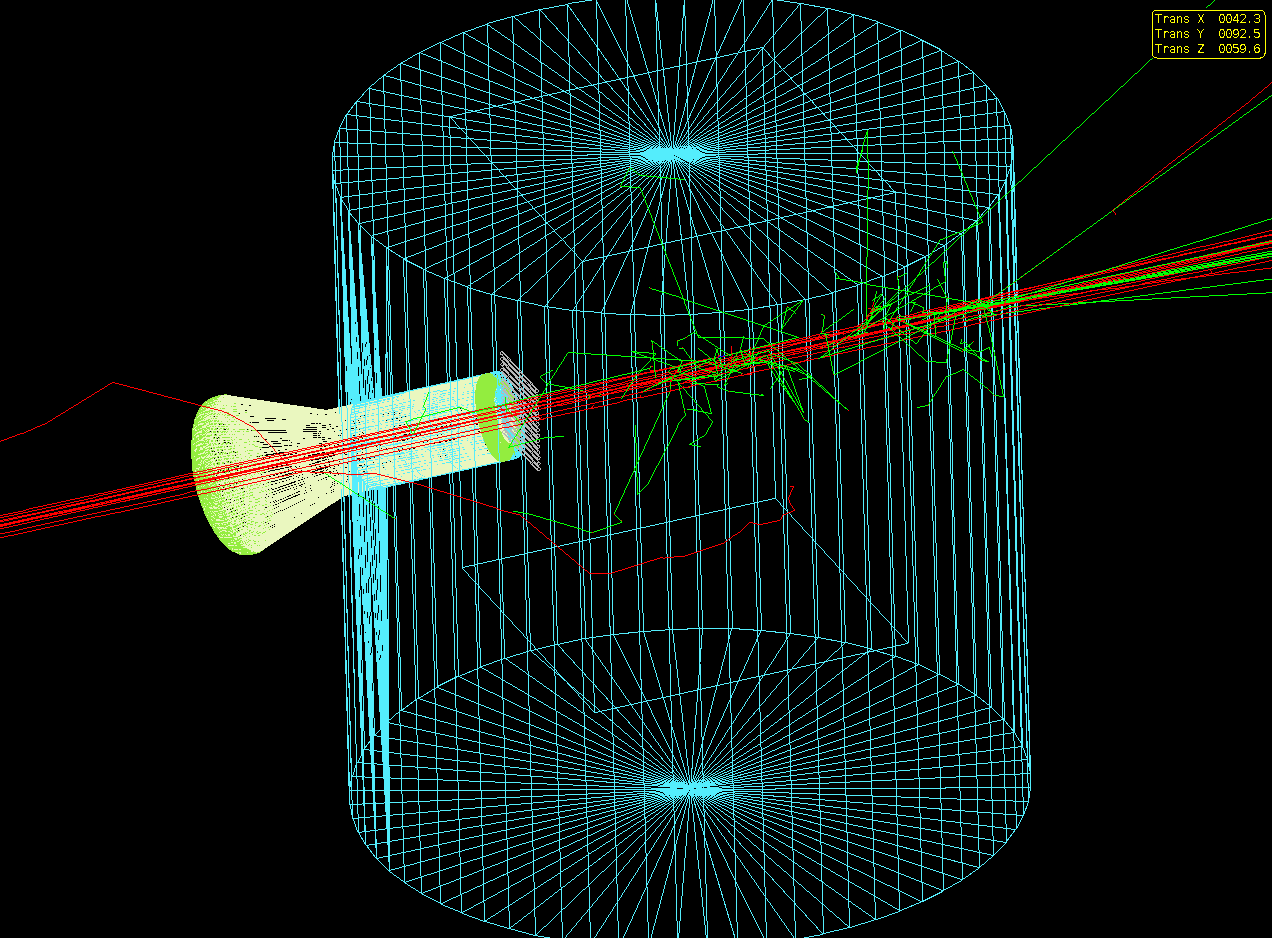}
\caption{An example GEANT4 simulation of a spill of beam particles passing through the ARIADNE detector (from left to right on the image). The fiducial LAr volume is shown by the inner blue box, and the uninstrumented LAr volume by the outer blue cylinder. The beam plug (discussed in Section~\ref{subsec:BeamPlug}) is shown in green, and the grey rectangle indicates a small section of the field-shaping rings.}
\label{fig:DetectorSimulation}
\end{figure}

\begin{figure}[bp]
\centering
  \includegraphics[width=0.86\textwidth]{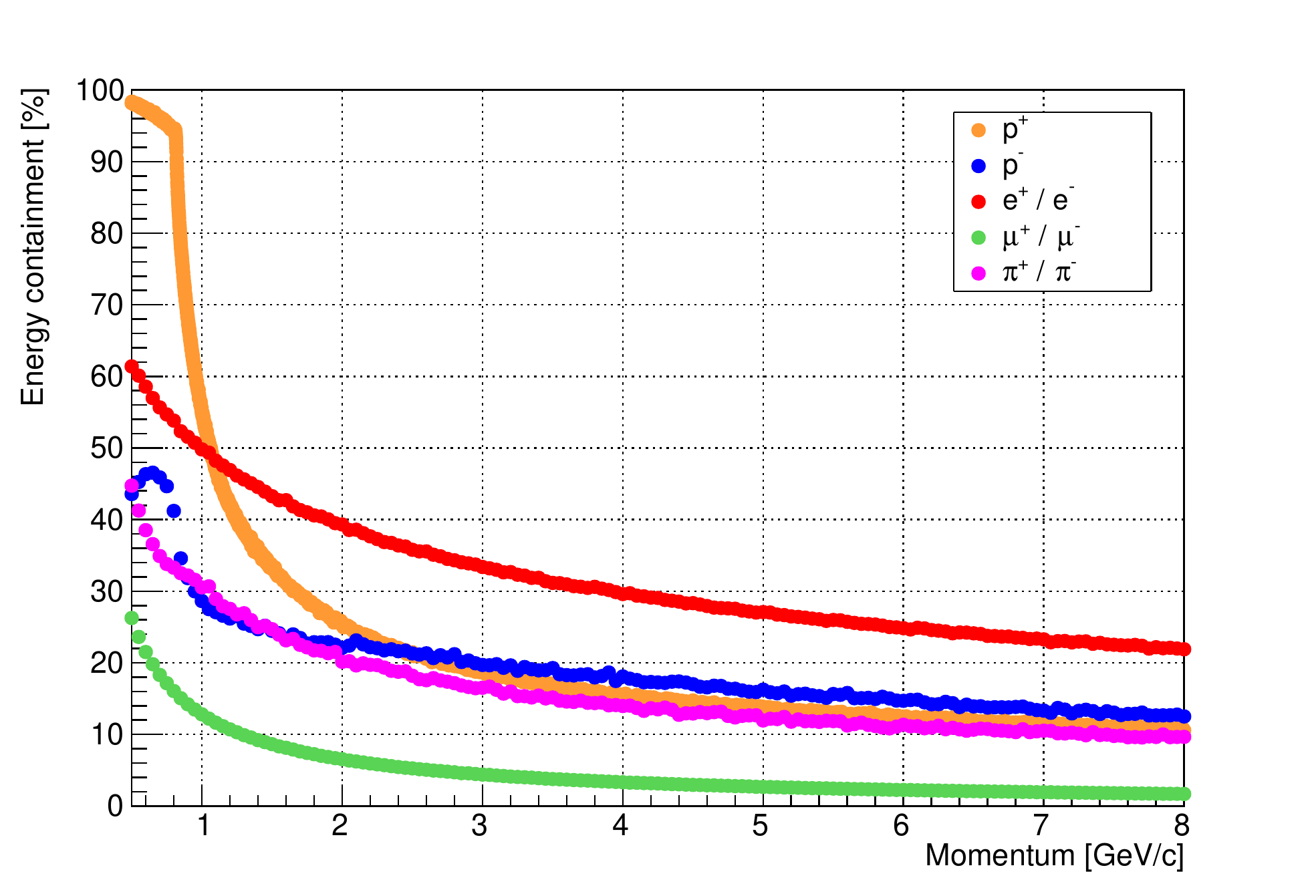}
  \caption{The results of energy containment simulations, showing the fraction of the total ionising energy deposition that is found within the ARIADNE TPC for a given particle and momentum. Per-point statistical uncertainties are comparable in size to the points displayed on the plot.}
\label{fig:containment}
\end{figure}

The simulated detector geometry is shown in Figure~\ref{fig:DetectorSimulation}. The cuboid-shaped fiducial LAr volume is visible, as well as the surrounding uninstrumented LAr. For the energy containment simulations, a particle gun was used to generate particles of various species and momenta on the boundary of the fiducial volume, at the approximate location where beam particles enter the TPC. The effects of the beam window, beam plug and field-shaping rings were not modelled. Each primary particle was tracked from the particle gun position until the point at which it stopped. The total ionising energy depositions both within and outside the TPC were measured. The results of the simulations are shown in Figure~\ref{fig:containment}.\\

\noindent At low momenta, the energy containment is relatively high - 60\% for electrons and positrons at 0.6~GeV/c - but a small momentum increase results in a significant reduction in containment - with electron / positron containment dropping by 10\% as the momentum increases by 0.4~GeV/c. However, the relation between containment and momentum shallows as the latter continues to increase - with electron / positron containment decreasing to 30\% at 4.0~GeV/c, and by a further 10\% as the momentum then doubles to 8.0~GeV/c. Muons and pions both exhibit a similar trend, but with an overall lower absolute energy containment due to their specific interaction rates, as do protons and anti-protons at high momenta. However, at low momenta, protons and anti-protons exhibit some unique behaviour.

Low momentum protons are essentially fully contained within the TPC. A sharp drop in containment is seen as momentum increases from $\approx$ 0.8 to 2.0~GeV/c. This is explained by the nature of proton energy deposition. Protons exhibit a very sharp Bragg peak, which contains a large fraction of the total ionising energy deposit of the proton track. As the momentum of the incident protons increases, the location of the Bragg peak moves across the TPC. Beyond $\approx$ 0.8~GeV/c, The Bragg peak is no longer contained within the active volume. After this point, the ionising deposit of the Bragg peak is lost and a sudden decrease in energy containment is seen. 

Anti-proton annihilation often results in products that travel both forwards (in the same direction as the incident anti-proton) and backwards (towards the source of the anti-proton). Low-momentum anti-protons annihilate close to the incident edge the fiducial volume, resulting in the backwards-travelling products quickly leaving the TPC, reducing containment. As the anti-proton momentum increases up to $\approx$ 0.5~GeV/c, the average annihilation vertex moves further into the active volume, with the backwards-travelling products being more contained within the TPC. This explains the initial rise of containment seen in the simulation. Beyond $\approx$ 0.5~GeV/c, the average anti-proton annihilation vertex moves closer to the far edge of the fiducial volume. The forward-travelling interaction products are more likely to escape through the far edge of the TPC, resulting in the sudden containment drop between $\approx$ 0.5 and 1.0~GeV/c.
\\
\\
These containment studies were used to inform the most useful momentum range over which the ARIADNE detector should be operated: from a lower limit of 0.5~GeV/c (in order to fully contain the low-momentum protons and anti-protons) to an upper limit of 8~GeV/c (beyond which the containment of the majority of particle species decreases below $\approx$ 10\%). This momentum range was therefore used during operation at the CERN T9 beamline (discussed in Chapter~\ref{chp:CERN}).

\subsection{Cathode and High Voltage Feedthrough}
\label{subsec:HVFT}
Positioned 1~cm below the bottom field-shaping ring, the ARIADNE cathode grid consists of a stainless steel frame containing a tensioned stainless steel mesh with 70\% optical transparency. A 4.5~inch diameter field-shaping torus is welded to one side of the frame, and this is where the connection to the high voltage (HV) feedthrough is made. This torus shields the electric field where the HV connection is made, thus minimising corona effects. An external Heinzinger PNChp series 100~kV negative polarity power supply \cite{Heinzinger} provides the cathode bias. 
\\
\\
Due to the high bias required on the cathode, the HV feedthrough of LArTPCs in general poses a significant engineering challenge. Care and attention is required to design and manufacture a feedthrough that is capable of providing both UHV leak tightness as well as discharge-free operation during extended periods of time in demanding cryogenic conditions.

The central core of the ARIADNE HV feedthrough makes connection with the cathode using a spring-loaded tip, which ensures a good electrical connection even if the cathode changes position slightly in cryogenic conditions. In order to make the feedthough vacuum tight, the central core is enclosed in a thick ultra-high-molecular-weight polyethylene (UHMWPE) jacket, which has been cryofitted onto the central core. This process ensures a leak-tight seal through the cold flow of UHMWPE as it warms and compresses tightly against the central core.

To provide mechanical stiffness to the core assembly, the UHMWPE jacket is itself enclosed in an electrically-grounded stainless steel jacket. This jacket extends from the Conflat flange on the air side of the feedthrough down to approximately 30~cm above the HV connection at the cathode. The large diameter of this outer stainless steel jacket provides substantial mechanical strength to the feedthrough, and its lower end terminates in a smaller, 68~mm diameter torus for field-shaping, thereby minimising the field between the central core and the end of the jacket.

\begin{figure}[ht]
\centering
\vspace{5mm}
\includegraphics[width=\textwidth]{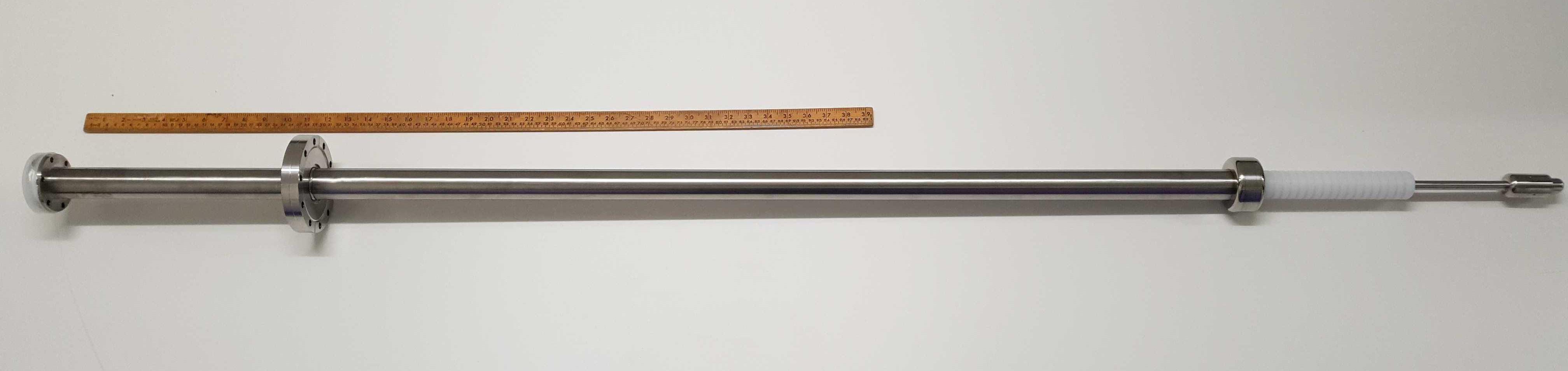}
\caption{The ARIADNE high voltage feedthrough. A metrestick is shown for scale.}
\label{fig:HVFT2}
\end{figure}

\begin{figure}[ht]
\begin{subfigure}{0.48\textwidth}
  \centering
  \includegraphics[width=\textwidth]{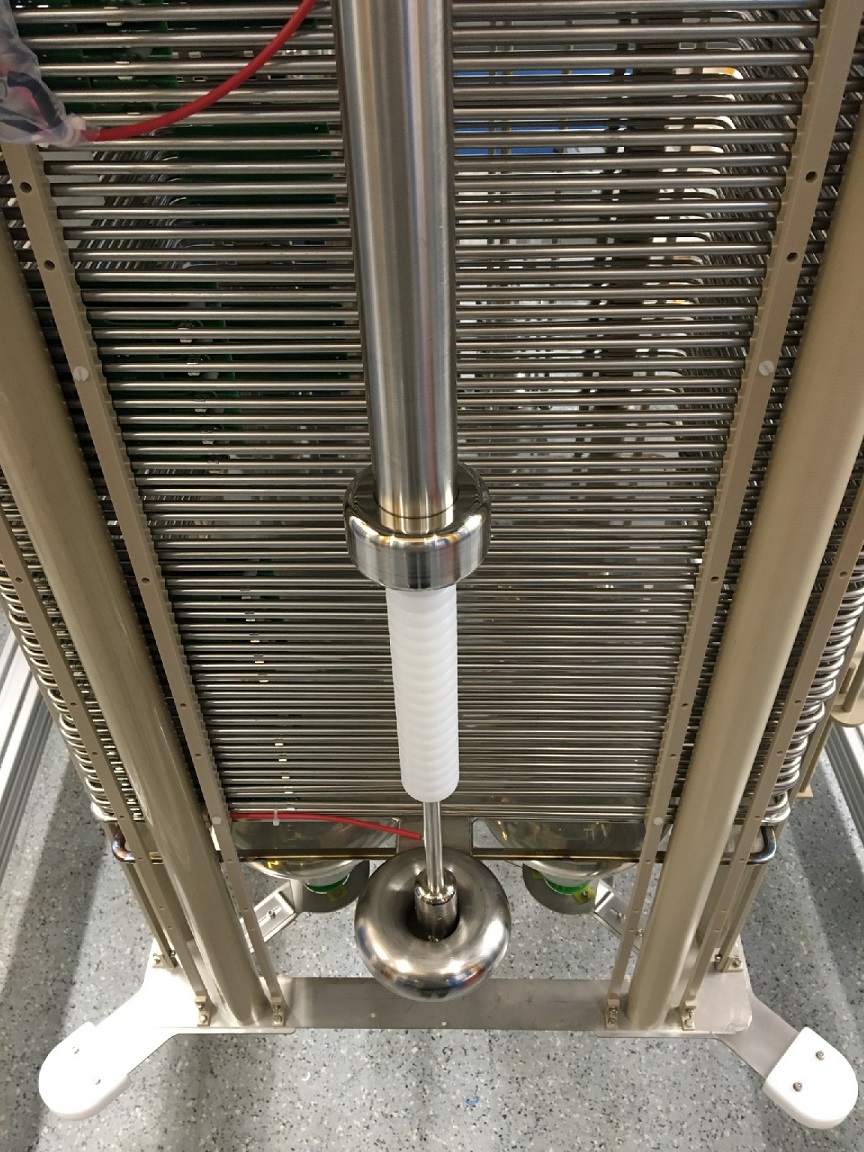}
\end{subfigure}
\hspace{1mm}
\begin{subfigure}{0.48\textwidth}
  \centering
  \includegraphics[scale=0.237]{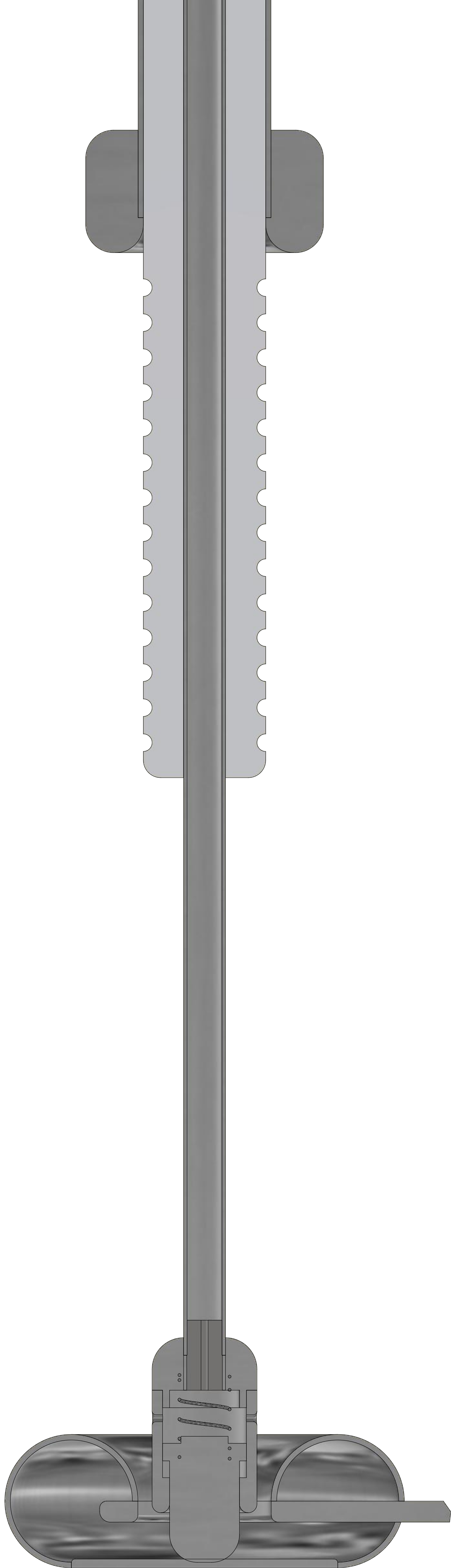}
\end{subfigure}
\caption{(Left) The ARIADNE HV feedthrough in-situ on the TPC, with the spring-loaded tip sitting within the cathode grid torus. (Right) A cross section model of the lower end of the HV feedthrough.}
\label{fig:HVFT}
\vspace{5mm}
\end{figure}

\begin{figure}[tp]
\includegraphics[width=\textwidth]{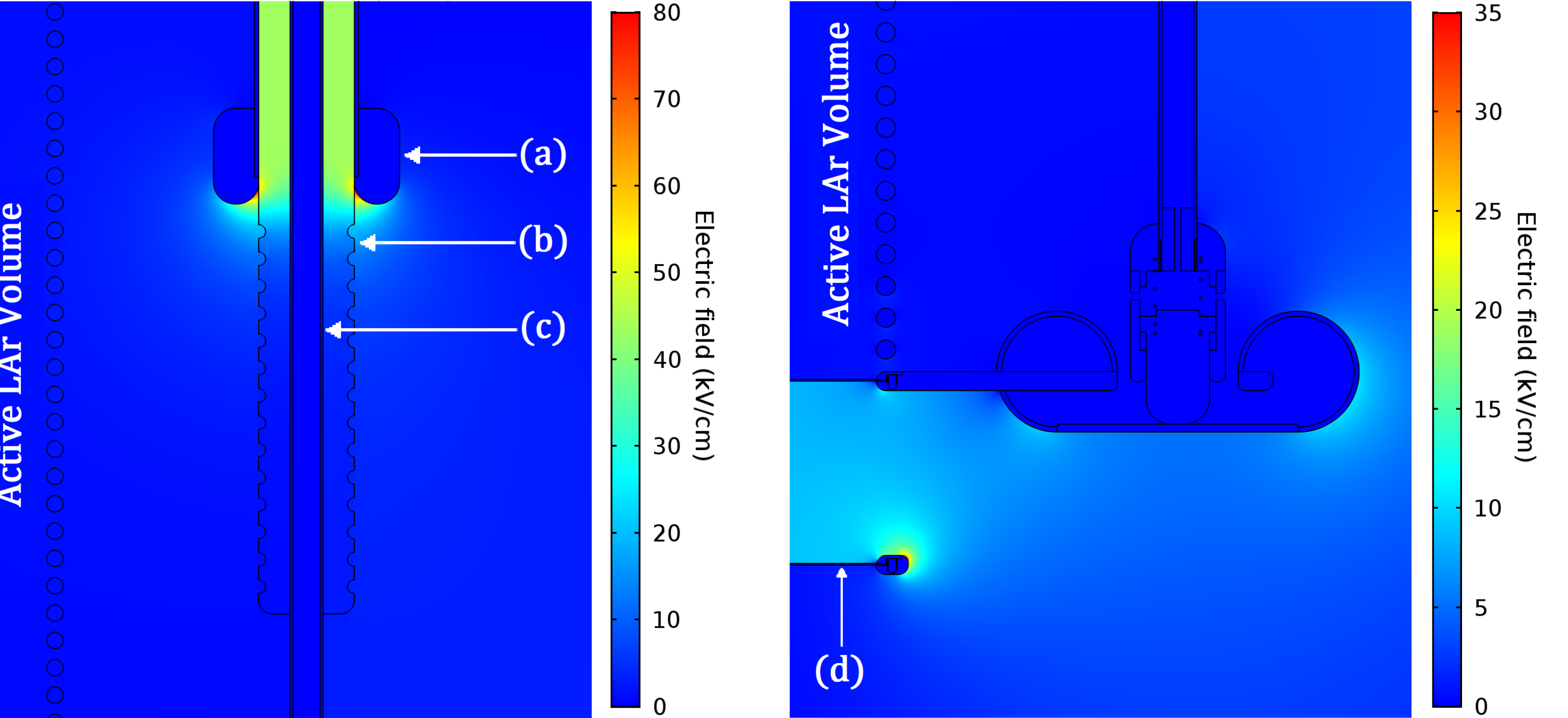}
\caption{The results of COMSOL simulations of the HV feedthrough. (Left) A high electric field of 80~kV/cm is present at the field-shaping torus that terminates the feedthrough's grounded outer jacket (a). However, the high dielectric strength of the inner UHMWPE jacket (b) prevents any discharges between the central core (c) and the outer jacket. (Right) The effectiveness of the torus surrounding the connection to the cathode is visible, with the largest field of 35~kV/cm occurring between the cathode and PMT protection (d) grids.}
\label{fig:FieldFeedThrough}
\end{figure}

The ARIADNE HV feedthrough has been successfully operated up to a bias of 80~kV, and is leak-tight to better than 10$^{-8}$~mbar~l/s. Figures~\ref{fig:HVFT2} and \ref{fig:HVFT} show the feedthrough in isolation and in-situ on the TPC respectively, along with a cross section model of the lower end. COMSOL simulations, shown in Figure~\ref{fig:FieldFeedThrough}, were performed to validate the design of the HV feedthrough as a whole.

\subsection{THGEM}
\label{subsec:THGEM} 
A Thick Gaseous Electron Multiplier (THGEM) \cite{THGEMReview} is a type of gas micropattern detector, broadly consisting of two conducting copper planes separated by an insulating layer (typically made of FR4), with holes passing directly through all three layers. This type of device is also known as a Large Electron Multiplier (LEM) in the context of dual-phase neutrino detectors. By applying separate biases to the top and bottom planes of the THGEM, an electric potential is set up across the device, and a strong electric field is created within each hole. For the following discussions, we define the nominal THGEM field (typically expressed in kV/cm) as the potential difference applied across the THGEM, divided by the THGEM thickness.
\\
\\
Different regimes of operation are possible depending on the THGEM field.

At a low field, electrons that enter the THGEM holes will simply be drifted through and collected on the top plane. As the field increases, the incident electrons gain enough energy to overcome the scintillation threshold - that is, they have enough energy to excite gas argon atoms as they travel through the holes of the THGEM. When these excited argon atoms de-excite, they release photons with wavelengths centred around 128~nm - the peak scintillation wavelength of argon. This regime of operation is known as the proportional electroluminescence regime.

As the THGEM field continues to increase, the incident electrons may gain enough energy to ionise the gas argon atoms, liberating additional electrons which can themselves be accelerated and gain enough energy to liberate further electrons (i.e. electron multiplication). The resulting electron avalanche is known as a Townsend discharge.

The ratio of the number of electrons leaving each hole to the number of incident electrons defines the gain of the THGEM.
\\
\\
Traditionally, high gain - and therefore a high THGEM field - is required to overcome the noise of the THGEM readout electronics. However, if both the gain and the number of incident electrons are large enough, sufficient ionisation may occur in a hole that a plasma is formed. This in turn may be electrically conductive enough to induce a discharge between the top and bottom THGEM planes. This dependence of discharge probability on the incident electron population means that the THGEM field must be carefully chosen: it should be low enough that a discharge is not induced during high incident electron population events, but high enough to give a good signal-to-noise ratio for low incident electron populations.

\begin{figure}[h]
\begin{subfigure}{0.49\textwidth}
  \centering
  \includegraphics[width=\textwidth]{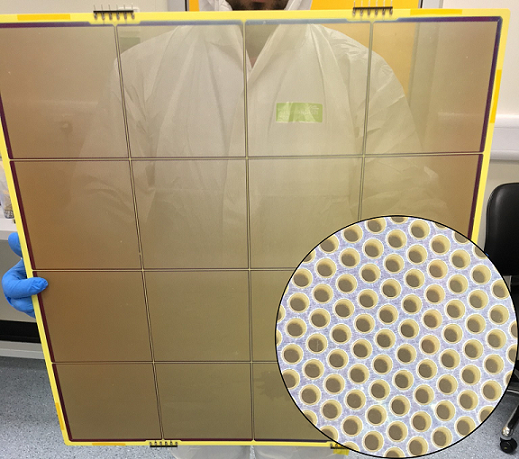}
\end{subfigure}
\hspace{1mm}
\begin{subfigure}{0.49\textwidth}
  \centering
  \includegraphics[width=\textwidth]{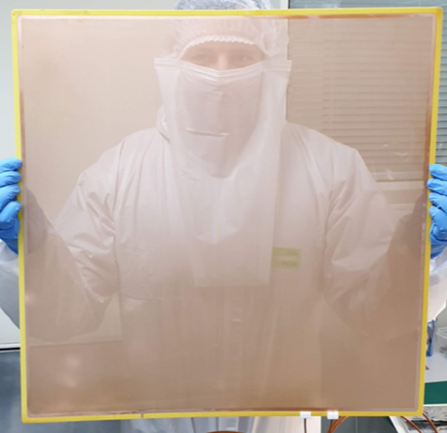}
\end{subfigure}
\caption{(Left) The 16-pad segmented ARIADNE THGEM. Inset: the THGEM hole structure (diameter: 500~$\mu$m, pitch: 800~$\mu$m). The 50~$\mu$m dielectric rim is visible around each hole. (Right) The monolithic ARIADNE THGEM. The hole structure and spacing is identical to that of the segmented THGEM.}
\label{fig:THGEM}
\end{figure}

Optical readout allows the THGEM to operate more reliably in the proportional electroluminescence regime compared to charge readout, as the optical signal is more stable to variations in the number of incident electrons, and can be produced at a lower THGEM field. Optical readout also still benefits from the electron-multiplication electroluminescence regime of the THGEM, as the charge gain also results in exponentially enhanced light production. Each incident electron can result in the production of many hundreds of photons, depending on the exact THGEM field \cite{THGEMLightYield, THGEMReadout, CryogenicAvalanche}, and so optical readout may allow for improved detection thresholds compared to charge readout, as well as improved stability by allowing the THGEM to operate at a lower field.
\\
\\
Two different THGEM designs have been used in the ARIADNE detector. Both were built by the CERN PCB Workshop, and are shown in Figure~\ref{fig:THGEM}. The ``segmented'' design has the top plane segmented into 16 charge-independent pads, and the ``monolithic'' design has a single entity top plane.

\noindent The two designs share a common underlying structure: a total area of 55.6 $\times$ 55.6~cm, an active area (the area covered by holes) of 53 $\times$ 53~cm, and a total thickness of 1~mm. The holes have a diameter of 500~$\mu$m, with a 50~$\mu$m dielectric rim and 800~$\mu$m hole-to-hole pitch. The bottom plane of both designs is a single monolithic pad, with a bias provided by a separate HV supply to that used for the top plane.

When using the segmented THGEM, each of the pads is biased independently, but to the same potential, using an external pre-amp and bias distribution board. The pre-amp also collects the charge on each of the pads, at both high and low pre-amp gains simultaneously, for a total of 32 charge readout channels. These are then passed to a digitiser for recording. Two pre-amp gains are used so as to make use of as much of the digitiser's dynamic range as possible across both small and large signals.

When using the monolithic THGEM, the top plane is biased and read out through a single external pre-amp.
\\
\\
The segmented THGEM was used for the studies conducted at the CERN T9 beamline (discussed in Chapter~\ref{chp:CERN}), while the monolithic design was used for cosmic muon operation at the University of Liverpool (discussed in Chapter~\ref{chp:Liverpool}).

\subsection{PMTs and Light Collection} 
\begin{figure}[ht]
\centering
\includegraphics[width=0.75\textwidth]{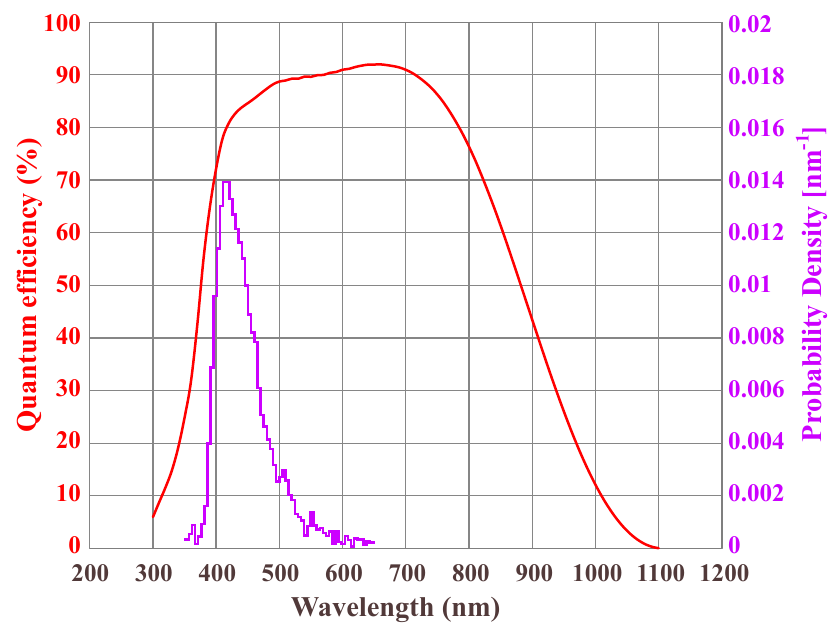}
\caption{The emission spectrum of TPB (purple), with the quantum efficiency of the EMCCD cameras (red) for comparison. This figure has been adapted from \cite{Andor} and \cite{TPBSpectrum}.}
\label{fig:TPBQE}
\end{figure}

Argon emits scintillation light with a peak wavelength of 128~nm. However, the spectral response range of the Hamamatsu R5912-20 PMTs lies between 300 and 650~nm, with a peak quantum efficiency at 420~nm \cite{Hamamatsu}. The EMCCD cameras (discussed in more detail in Section~\ref{subsec:EMCCDs}) have a spectral response range between 300 and 1000~nm, with a quantum efficiency of >~80~$\%$ between 420 and 780~nm, as shown in Figure~\ref{fig:TPBQE}. Therefore, since neither the PMTs nor the EMCCDs are directly sensitive to the VUV 128~nm photons, wavelength shifting is required.

As well as considerations relating to the emission spectrum of the wavelength shifter, it is necessary for it to be cryogenically stable. Tetraphenyl Butadiene (TPB) is an obvious choice in both regards: it has a peak emission wavelength of 420~nm and an emission spectrum tail out to 550~nm (as shown in Figure~\ref{fig:TPBQE}) and a successful history of use in many cryogenic LArTPC dark matter and neutrino experiments \cite{DEAP-3600, ICARUS}.\\

\noindent In total, it was necessary to coat the four PMTs, the 54 $\times$ 54~cm wavelength shifting glass sheet positioned above the THGEM, and all of the 3M\texttrademark{} ESR (Enhanced Specular Reflector) Vikuiti\texttrademark{} polyester foil panels. These reflector panels surround the TPC, improving the light collection efficiency of the PMTs. The chosen method for this coating was vacuum evaporation, which produces a uniform, repeatable coating that ensures predictable wavelength shifting performance of the applied coatings.

When considering the TPB coating on the PMTs and glass, it is important to strike a balance between the TPB thickness and optical transparency. If the coating is too thick, then a photon which is absorbed and re-emitted by the TPB may not itself escape, but if the coating is too thin, the efficiency of shifting the 128~nm VUV photons is poor. The PMTs and glass plate were coated with a  0.05~mg$/$cm$^2$ layer, which was chosen based on various studies \cite{ICARUS, Spanu, ArDM_wavelength_shifter} which have demonstrated the response of PMTs to various TPB coating thicknesses. Optical transparency is not a concern for the reflectors, but instead, a balance is required between the increasing light collection efficiency and the increasing brittleness of the TPB coating as its thickness increases. A coating of 0.22~mg$/$cm$^2$ thickness was found to be a reasonable compromise.

\begin{figure}[ht]
\begin{center}
\includegraphics[width=0.75\textwidth]{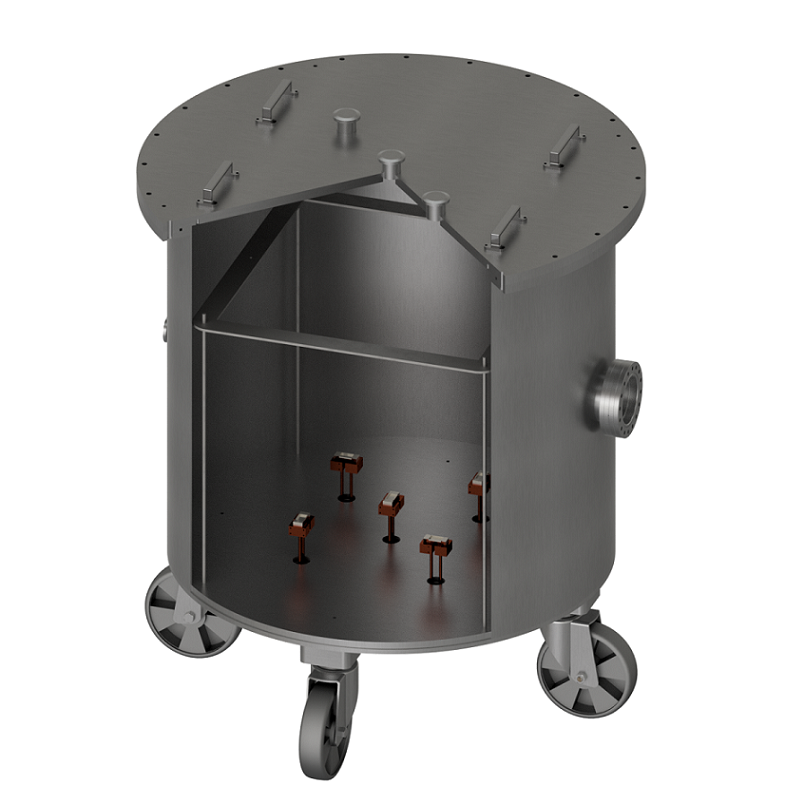}
\caption{A CAD model of the vacuum evaporation chamber, which was used to apply a TPB coating to the PMTs, the glass plate and the reflector panels. The five crucibles for holding the TPB powder can be seen at the bottom of the chamber, along with their associated copper blocks for heating and the frame for suspending the component(s) to be coated. The viewport is shown on the right of the chamber.}
\label{fig:VacuumEvaporationChamber}
\end{center}
\end{figure}

\begin{figure}
\begin{subfigure}{0.585\textwidth}
  \begin{subfigure}{\textwidth}
    \centering
    \includegraphics[width=\textwidth]{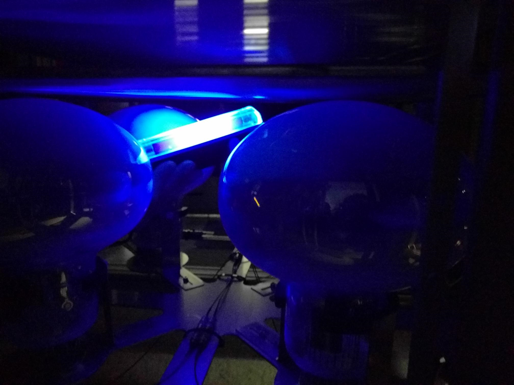}
  \end{subfigure}
  \begin{subfigure}{\textwidth}
    \vspace{2mm}
    \centering
    \includegraphics[width=\textwidth]{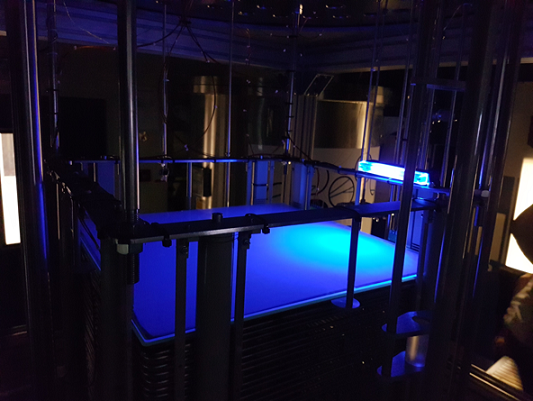}
  \end{subfigure}
\end{subfigure}
\hspace{1mm}
\begin{subfigure}{0.395\textwidth}
  \centering
  \includegraphics[width=\textwidth]{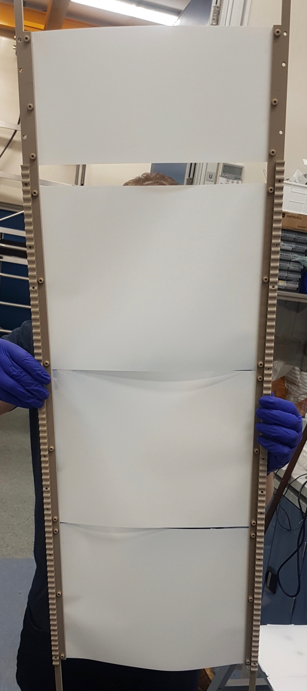}
\end{subfigure}
\caption{(Top-Left) The PMTs and (Bottom-Left) the glass sheet, viewed under UV light illumination to show the effect of the TPB coating. (Right) One of the TPC reflector assemblies after TPB coating.}
\label{fig:TPBCoatingImages}
\end{figure}

A bespoke vacuum evaporation chamber - a model of which is shown in Figure~\ref{fig:VacuumEvaporationChamber} - was used for coating the listed ARIADNE components with TPB. During the coating procedure, the chamber is evacuated, and five crucibles (which hold the solid TPB powder) are slowly heated over a period of several hours via a high current passed through copper blocks positioned underneath each one. The component(s) to be coated are suspended from a frame positioned above the crucibles, and the chamber as a whole is capable of coating a maximum single area of 60 $\times$ 60~cm. It is also equipped with a viewport, so that the coating process can be monitored. Figure~\ref{fig:TPBCoatingImages} shows the PMTs, wavelength shifting glass sheet and reflector panels coated in TPB.

\subsection{LAr Purification and Cryogenics}
\label{subsec:Cryogenics}
In order to most efficiently drift electrons the full 80~cm drift length of the TPC, it is desirable to maximise the electron lifetime, denoted by $\tau$. This lifetime is heavily dependent on purity, and can be approximated by: $\tau$~[$\mu$s] $\approx$ 300~/~$\rho$, where $\rho$ is the O$_{2}$ equivalent impurity concentration in parts per billion (ppb) \cite{ElectronLifetime1, ElectronLifetime2}.

The fraction of electrons lost during drift by attachment to impurities is given by $Q=Q_0 e^{-t\textsubscript{drift}/\tau}$, where $Q$ is the number of electrons surviving a drift time $t$\textsubscript{drift}, $Q_0$ is the initial number of ionised electrons and $\tau$ is the electron lifetime as before. When $\tau = t$\textsubscript{drift}, only $\approx$~27\% of electrons can be expected to survive. The drift velocity of electrons in LAr is 1.5~mm/$\mu$s at the nominal ARIADNE drift field of 0.5~kV/cm \cite{LArDriftVelocity}, and so the maximum drift time is 530~$\mu$s - approximately corresponding to 0.6~ppb O$_{2}$ equivalent impurities in the detector volume. In practice, it is always preferable to have an electron lifetime longer than the maximum drift time, since the probability of an electron surviving the drift process is statistical, and the survival rate continues to be enhanced for $\tau$ much larger than $t$\textsubscript{drift}. A lifetime of $\approx$~5000~$\mu$s is required for 90\% of electrons to survive drifting the full ARIADNE detector, and for 99\% electron survival probability a lifetime of over 0.5~s is required.
\\
\\
The purity can also be measured by directly utilising the slow component of argon scintillation emission. This component - the characteristic decay time of which is denoted by $\tau_2$ - is caused by the decay of triplet excimers. (This is in contrast to the decay of singlet excimers, which produce the fast emission component with a decay time denoted by $\tau_1$.) The exact value of $\tau_2$ is dependent on the source of the excitation, with measurements of 1540~ns (1100~ns) \cite{argonPurity} and 1590~ns (1660~ns) \cite{argonPurity2} being reported in ultra-pure LAr for excitation via electrons (alpha particles). However, regardless of the excitation source, impurities in the argon can quench the production of triplet excimers more than that of singlet excimers \cite{excimerQuenching}, leading to a smaller slow emission component, and thus a $\tau_2$ value smaller than the cited results.

The value of $\tau_2$ is calculated by fitting a double-exponential function to the shape of S1 pulses recorded by a single PMT. The sensitivity of this method is limited to a few ppb equivalent impurity concentration \cite{argonPurification}, but it is still a useful tool during the initial stages of the purification process - for example, when the TPC is not fully submerged in LAr and application of a drift field is therefore not possible.
\\
\\
The combined internal recirculation and purification system has been generously sized in an attempt to ensure the best possible electron lifetimes even if the initial argon fill purity is poor.

The system is suspended from the underside of the top flange, and is completely submerged in the LAr volume to minimise heat losses that may have otherwise been introduced with an external purification system. A bespoke positive displacement pump, designed and built by the authors specifically for ARIADNE, provides a nominal flow rate of 300~L/h through a 13.5~L filtration cartridge. This cartridge is filled with a mixture of Engelhard Cu-0226~S~14~$\times$~20 Mesh (85.6\%) working alongside 3~\AA ~and 4~\AA ~molecular sieves (both 7.2\%), to provide efficient filtration of O$_2$ and H$_2$O impurities. A similar, smaller recirculation and purification system is described in \cite{argonPurification}.

\begin{figure}[ht]
\centering
\includegraphics[width=0.75\textwidth]{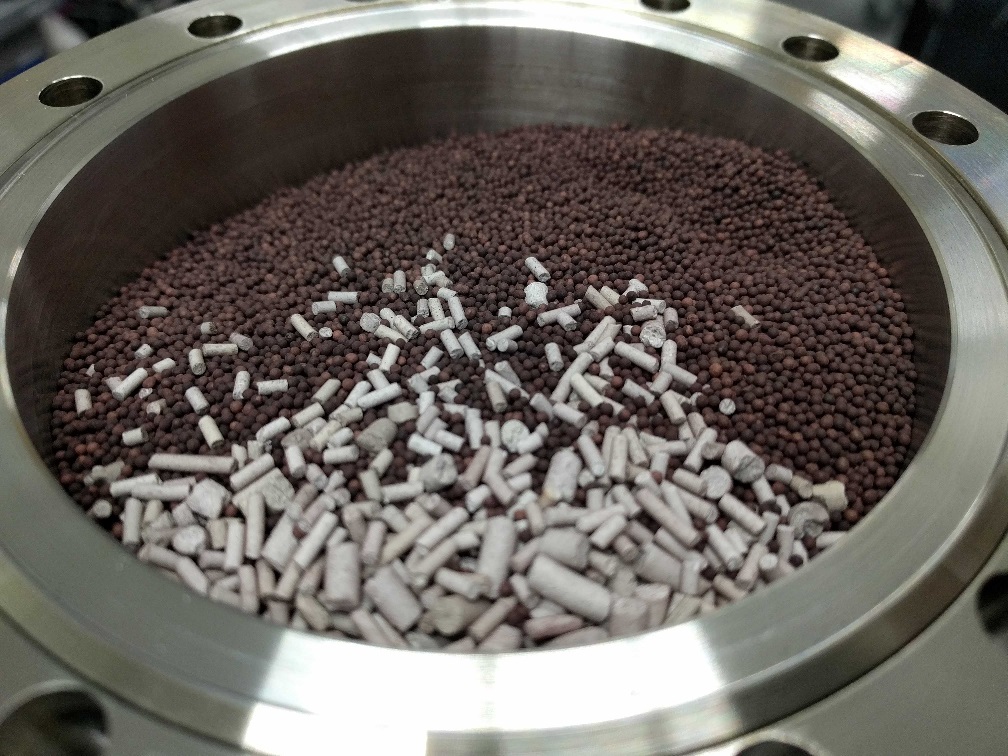}
\caption{The external 30~L filtration cartridge, showing the mixture of 85.6\% Engelhard Cu-0226~S~14~$\times$~20 (brown balls) and 7.2\% of both 3~\AA ~and 4~\AA ~molecular sieves (white cylinders in two different sizes). The same mixture is present in the internal 13.5~L cartridge.}
\label{fig:externalCartridge}
\end{figure}

The same mixture of chemicals is also present in a larger, 30~L external filtration cartridge, shown in Figure~\ref{fig:externalCartridge}, which provides initial purification of the LAr immediately after it leaves the supply dewar.  While not essential, this extra purification step reduces the burden on the internal purification system.\\

\begin{figure}[ht]
\centering
\includegraphics[width=\textwidth]{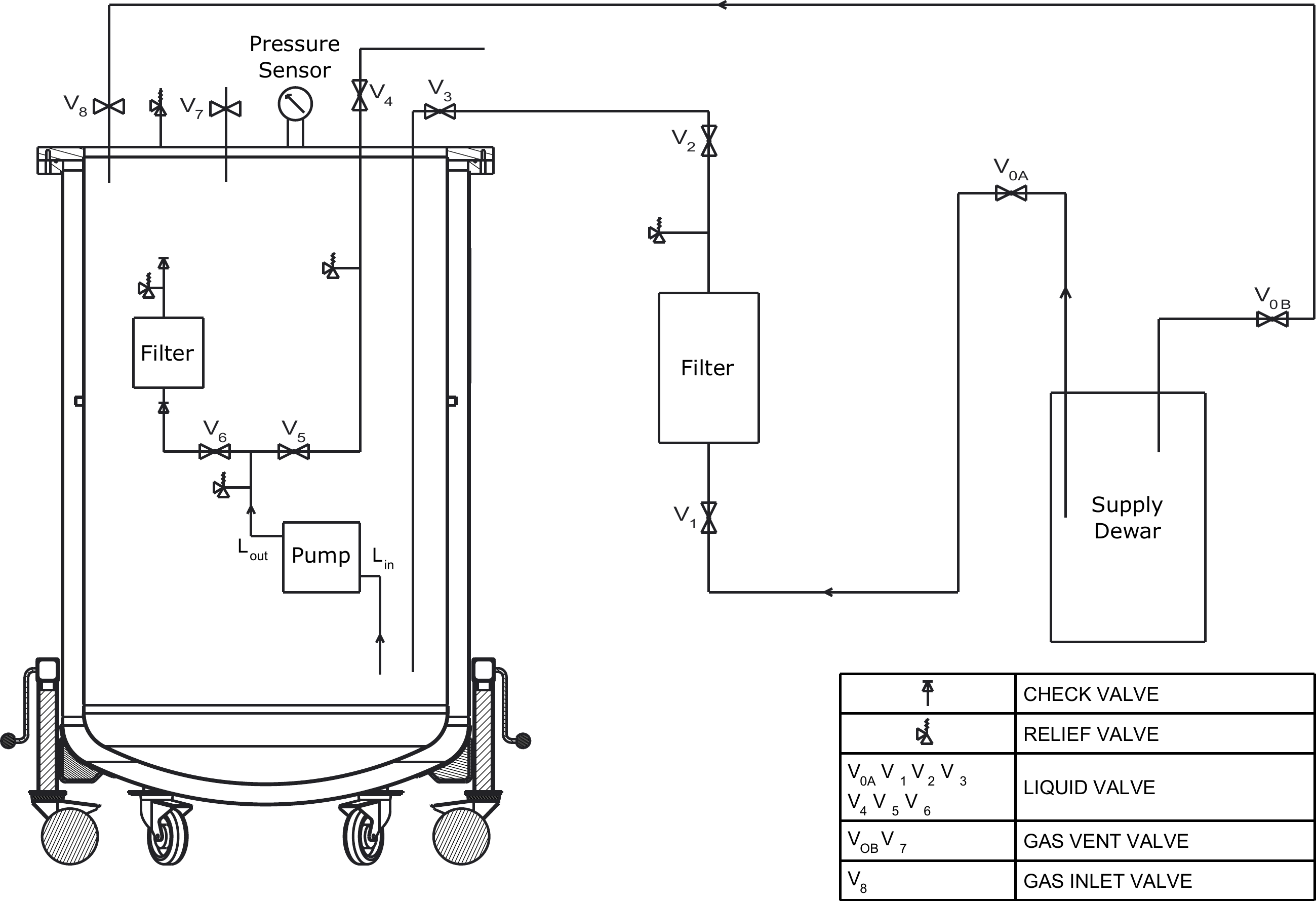}
\caption{Schematic showing the various lines and valves used to fill, recirculate, purify and evacuate LAr to and from the ARIADNE detector. See main text for a detailed explanation.}
\label{fig:ValveDiagram}
\end{figure}

\noindent Figure~\ref{fig:ValveDiagram} shows a schematic of the system that is used to fill, recirculate, purify and evacuate LAr to and from the ARIADNE detector.

The LAr is initially stored in a supply dewar, and passes into the external filtration cartridge via an inlet line controlled with two valves ($V_{0A}$ and $V_{1}$). The LAr leaves the filtration cartridge (which is equipped with a 1.5~bar(g) safety relief valve), and enters the main detector volume via valves $V_2$ and $V_3$. This ``liquid-in'' line deposits the LAr at the bottom of the cryostat.

The cryostat is equipped with multiple pressure sensors (discussed in more detail in Section~\ref{subsec:Monitoring}, and only one of which is shown in Figure~\ref{fig:ValveDiagram} for clarity), a 0.6~bar(g) relief valve, and two exhaust lines for releasing excess pressure if required. These lines are independently controlled using valves $V_7$ and $V_8$. Gas can also be let into the detector volume via $V_8$.

During purification, LAr is drawn from the lowest part of the volume into the pump (shown by `$L_{in}$' on Figure~\ref{fig:ValveDiagram}), and then passes into the internal filtration cartridge via valve  $V_6$. This cartridge is equipped with check valves (rated to 10~mbar) that only allow one-way flow, therefore ensuring that there is no contamination of the chemicals within the cartridge when it is at atmospheric pressure. Once purified, the LAr then empties back into the detector at a sufficient distance from the $L_{in}$ line to ensure well-distributed purity within the volume. An alternate route out of the pump - controlled by valves $V_5$ and $V_4$ - is used for pumping the LAr out of the cryostat.

The $V_5$ and $V_6$ valves are mutually exclusive - that is, when one is fully closed, the other is fully open. This is done via a gearing system that rotates the valves in opposite directions, with user control being provided by a single rotation feedthrough.\\

\begin{figure}[ht]
\centering
\includegraphics[width=\textwidth]{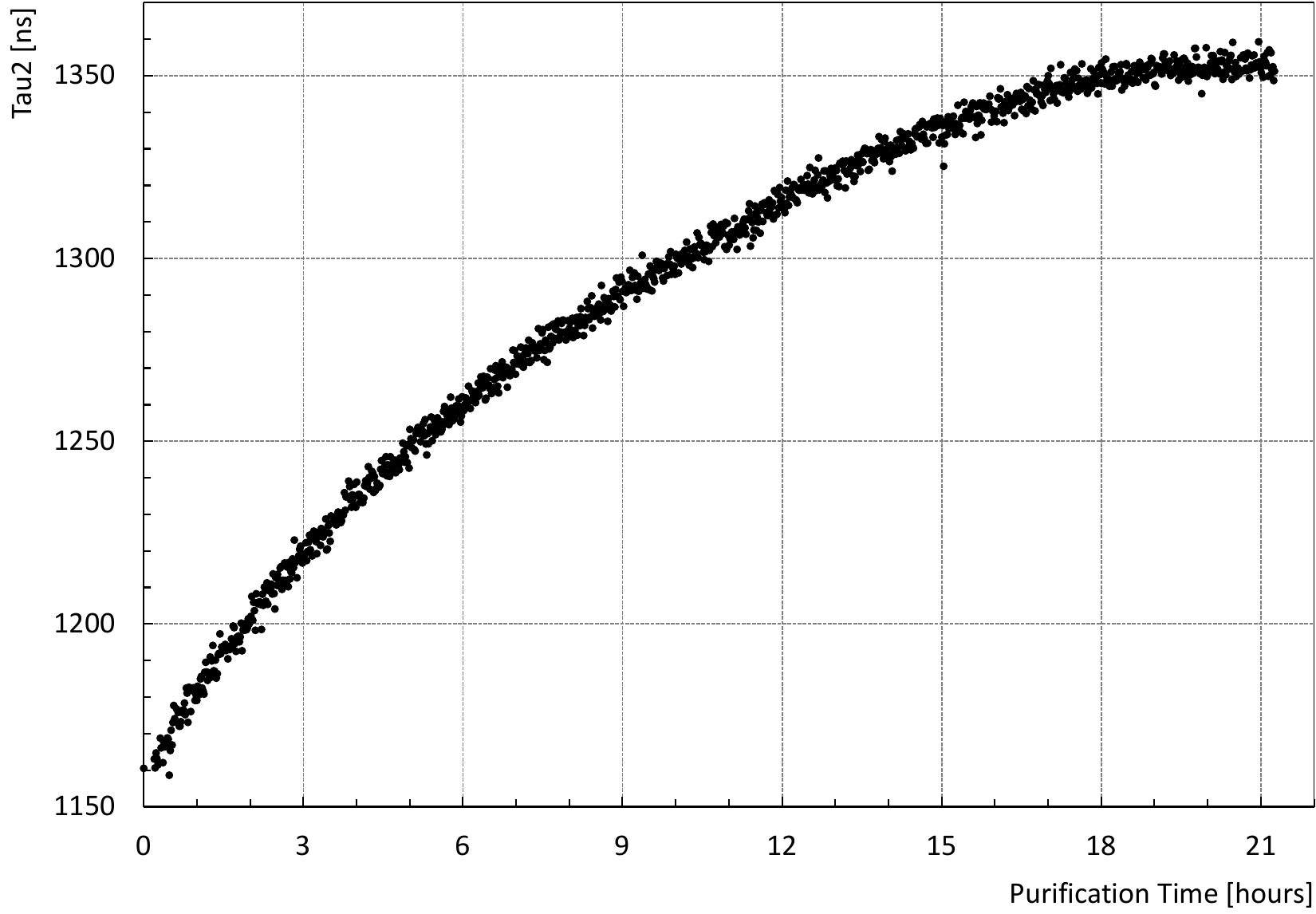}
\caption{Progression of $\tau_2$ - the decay time of the slow scintillation emission component - as a function of purification time.}
\label{fig:tau2Plot}
\end{figure}

\noindent Figure~\ref{fig:tau2Plot} shows the progression of $\tau_2$ - the decay time of the slow scintillation emission component - as a function of purification time. The starting $\tau_2$ value of $\approx$ 1160~ns relates directly to the initial LAr fill purity, and it can be seen that after only 20 hours of purification, $\tau_2$ has plateaued at a value of 1350~ns. This is somewhat less than the reported values previously discussed, but may be attributed to the inefficiency of the purification system at removing $N_2$. This is considered to be a light-quenching impurity, but has little effect on the electron lifetime. Further purification will therefore continue to remove O$_{2}$ contamination - thus increasing the electron lifetime - with little additional change in $\tau_2$. For the results discussed in Sections \ref{subsec:CERNResults} and \ref{subsec:LivResults}, the electron lifetime was $\approx$ 250~$\mu$s (as determined from analysis of the PMT S2 pulse lengths), which corresponds to an impurity level of 1.2~ppb O$_{2}$ equivalent.\\

\noindent In a dual-phase TPC, the LAr level must be accurately controlled within the extraction region, since any variation of the level within this region can modify the effective extraction field, and may therefore effect the charge extraction efficiency.

\begin{figure}[ht]
\includegraphics[width=0.60\textwidth]{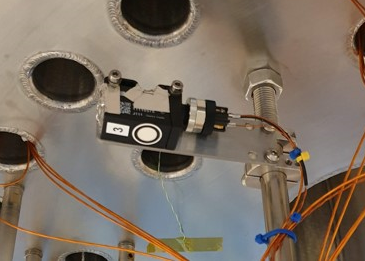}
\caption{One of the three Baumer Ultrasonic sensors that are used to measure the flatness and fill level of LAr within the ARIADNE cryostat. All three are mounted at the same height - a few centimetres below the underside of the top flange.}
\label{fig:ultrasonics}
\end{figure}

ARIADNE has a variety of dedicated sensors to precisely monitor the LAr fill level. An internal cryogenic (Microsoft HD-3000 \cite{ccdargon}) webcam provides video footage of the extraction region for easy - albeit qualitative - visual inspection. A more quantitative approach is to use the capacitance between the extraction grid and the bottom plane of the THGEM. However, due to the large area of the THGEM, this method was found to be extremely sensitive to small fluctuations in the LAr level (which naturally occur during filling) and also to wetting of the THGEM. In addition, neither of these methods provide information about the flatness of the LAr fill level. To remedy this, ARIADNE is equipped with 3 Baumer U500 series ultrasonic distance measuring sensors \cite{Baumer}, shown in Figure~\ref{fig:ultrasonics} and mounted just below the underside of the top flange. These precisely and quantitatively monitor the fill level in three locations.

Calibration of the sensors was performed prior to closing the detector, and involved first confirming (using an external testing setup) that each sensor has the same response over their common operating range. The sensors were then mounted at their respective positions above the TPC, with fine adjustments to the height ensuring that they all give the same readout value with respect to a common reference point - the TPB-coated glass plane. The sensors were then rotated to their final positions facing over the sides of the TPC, giving an unhindered view of the liquid.

\noindent These sensors have been successfully operated to temperatures as low as -40{\textdegree}C, which is considerably below the minimum temperature of the underside of the top flange when the cryostat is filled. The nature of ultrasonic sensors in general means that they are sensitive to changes in pressure and temperature, but relative measurements between the sensors will still be reliable. Therefore, by combining measurements from the sensors, it is possible to both measure the absolute level and also ensure that the level is uniformly flat across the entire extraction region. The flatness of the fill level relative to the THGEM and extraction grid is adjusted by small adjustments to the six cryostat wheels.\\

\begin{figure}[ht]
\vspace{5mm}
\includegraphics[width=\textwidth]{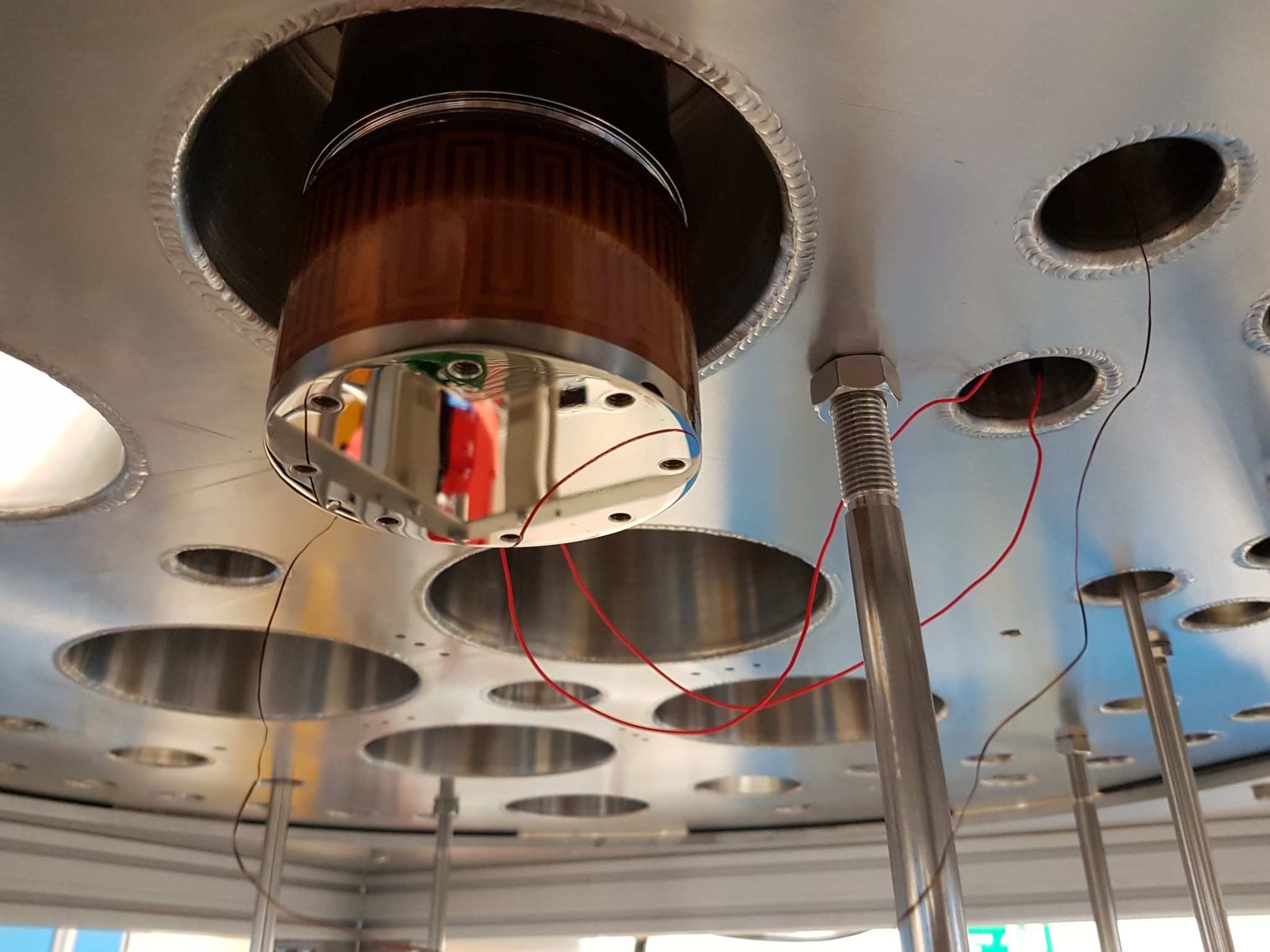}
\caption{The Cryomech AL300 cryorefrigerator cold head extending down from the top flange of ARIADNE. A flexible kapton heater element wrapped around the cold head regulates cooling, and thus allows the pressure in the cryostat to be accurately controlled.}
\label{cryorefrigerator}
\end{figure}

\noindent Once the LAr filling process is complete, heat losses of the cryostat will cause the LAr to slowly boil off, thus increasing the internal pressure and affecting the LAr fill level. A Cryomech AL300 cryorefrigerator unit \cite{Cryomech}, installed on the underside of the top flange and shown in Figure~\ref{cryorefrigerator}, is used to ensure that the LAr remains liquid inside the cryostat. When operating at 50Hz and 80K, this provides a peak of 266~W of cooling capacity to the internal volume of the cryostat to overcome the aforementioned heat losses. Therefore, by adjusting the heat input to the cryorefrigerator's cold head such that the cryostat pressure remains fixed, a stable LAr level is ensured.

\subsection{Laser Calibration System}
\label{subsec:Laser} 
The use of lasers for calibration in LAr detectors - via the three photon excitation process that allows the ionisation of argon atoms \cite{Rossi:2011zzc} - has been previously been demonstrated by multiple groups \cite{Ereditato:2014jj}. In the same vein, the ARIADNE detector features a calibration system that utilises a pulsed Nd:YAG (neodymium-doped yttrium aluminium garnet) UV laser.

The main purpose of the laser calibration system is to allow measurements of the overall field uniformity to be taken, in both the vertical and horizontal planes. These measurements can then be used to make corrections for any significant non-uniformities in the field. Such non-uniformities can be identified using the principle that, upon entering the active volume, the laser photons ionise the argon atoms, and the free electrons are drifted to the gas phase of the TPC in the same way as ionisation electrons from a particle track would be. Therefore, the resulting EMCCD images will be affected by any non-uniformities in the drift field. 

\begin{figure}[h]
\centering
\vspace{5mm}
\includegraphics[width=\textwidth]{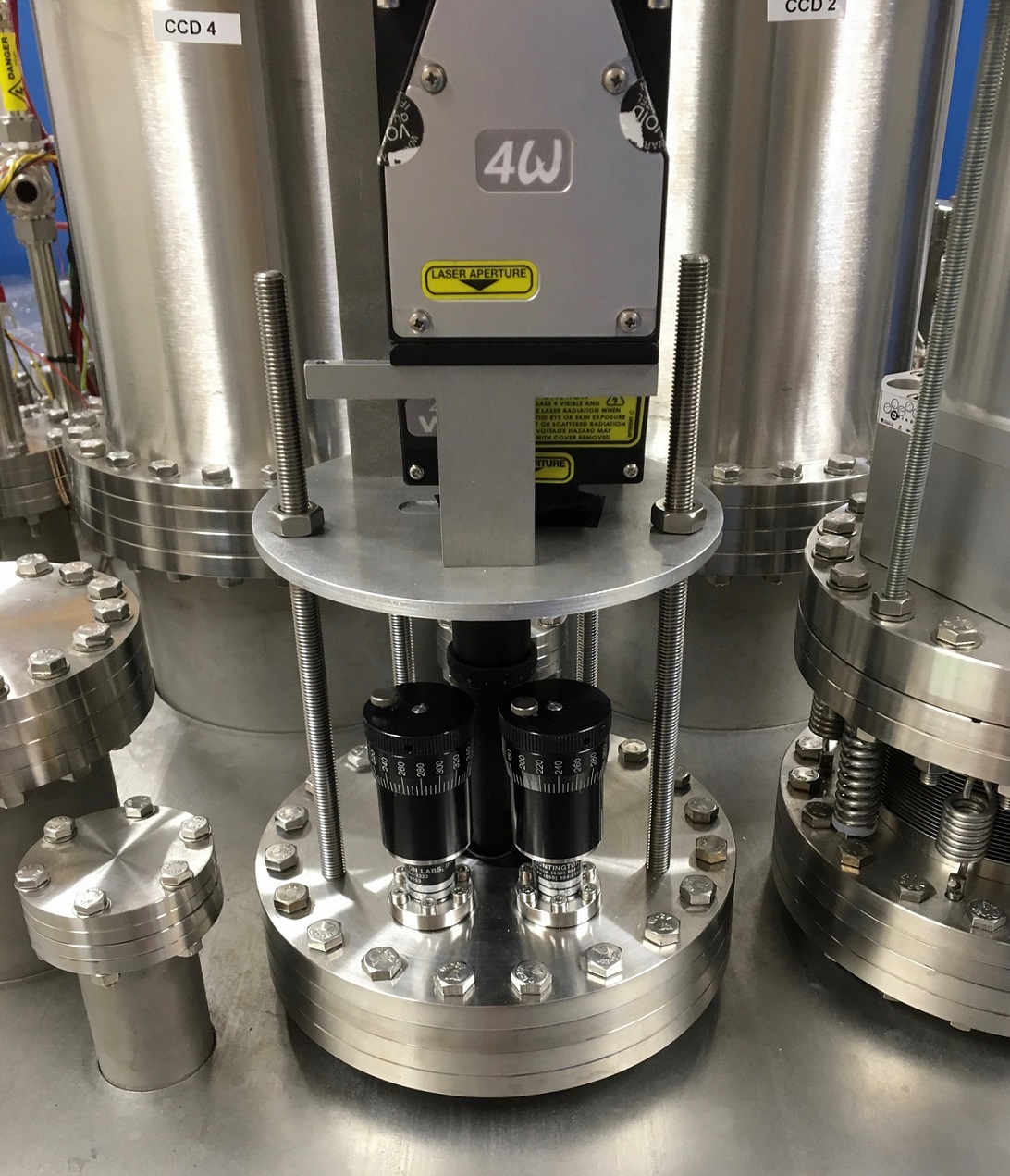}
\caption{The Quantel Q-Smart 100 pulsed Nd:YAG laser, mounted on the top flange of the ARIADNE cryostat. Two rotary feedthroughs are used to control the position and orientation of the UV mirror.}
\label{fig:laser-controls}
\end{figure}

The specific laser used on the ARIADNE detector is a Quantel Q-Smart 100 \cite{QSmart} pulsed Nd:YAG laser emitting 1064~nm light at 100~mJ per pulse and shown in Figure~\ref{fig:laser-controls}. Harmonic modules are used to change the wavelength to 266~nm, in the process reducing the pulse energy to 20~mJ. The compact dimensions of the laser allow for it to be vertically mounted directly onto the top flange of the cryostat, meaning that the beam travels directly into the cryostat through an optical feedthrough, without the need to align the detector to an external optical table.

\begin{figure}
\includegraphics[width=\textwidth]{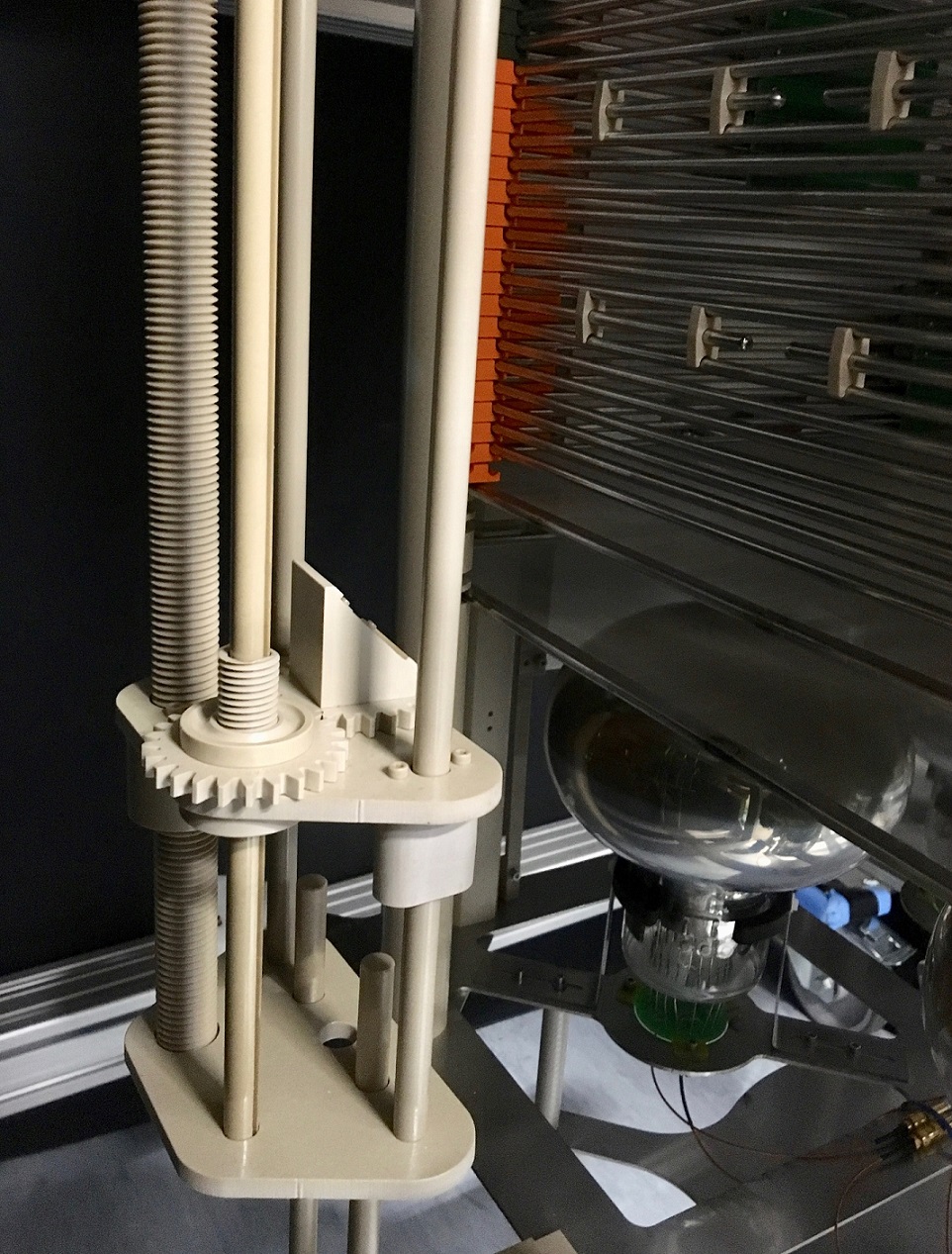}
\caption{The lower section of the laser support structure, showing the platform with the angled mirror facing towards the TPC. Also visible are the gaps in specific field-shaping rings that allow the laser beam to enter the active volume. For clarity, the reflector panel for this side of the TPC is not shown.}
\label{fig:laser-platform}
\end{figure}

Suspended from the underside of the top flange is a platform, shown in Figure~\ref{fig:laser-platform}, that holds a UV mirror at a 45\textdegree{} angle, which acts to redirect the vertical laser beam into the horizontal plane. The entire structure is constructed from VICTREX PEEK 450G - the same material that the TPC support rods are constructed from, to minimise any differences between the thermal expansion of the laser support structure and the TPC. This reduces vertical misalignments between the relative position of the laser beam and the field cage.\\

\vspace{10mm}

\begin{figure}[ht]
\centering
\includegraphics[width=0.80\textwidth]{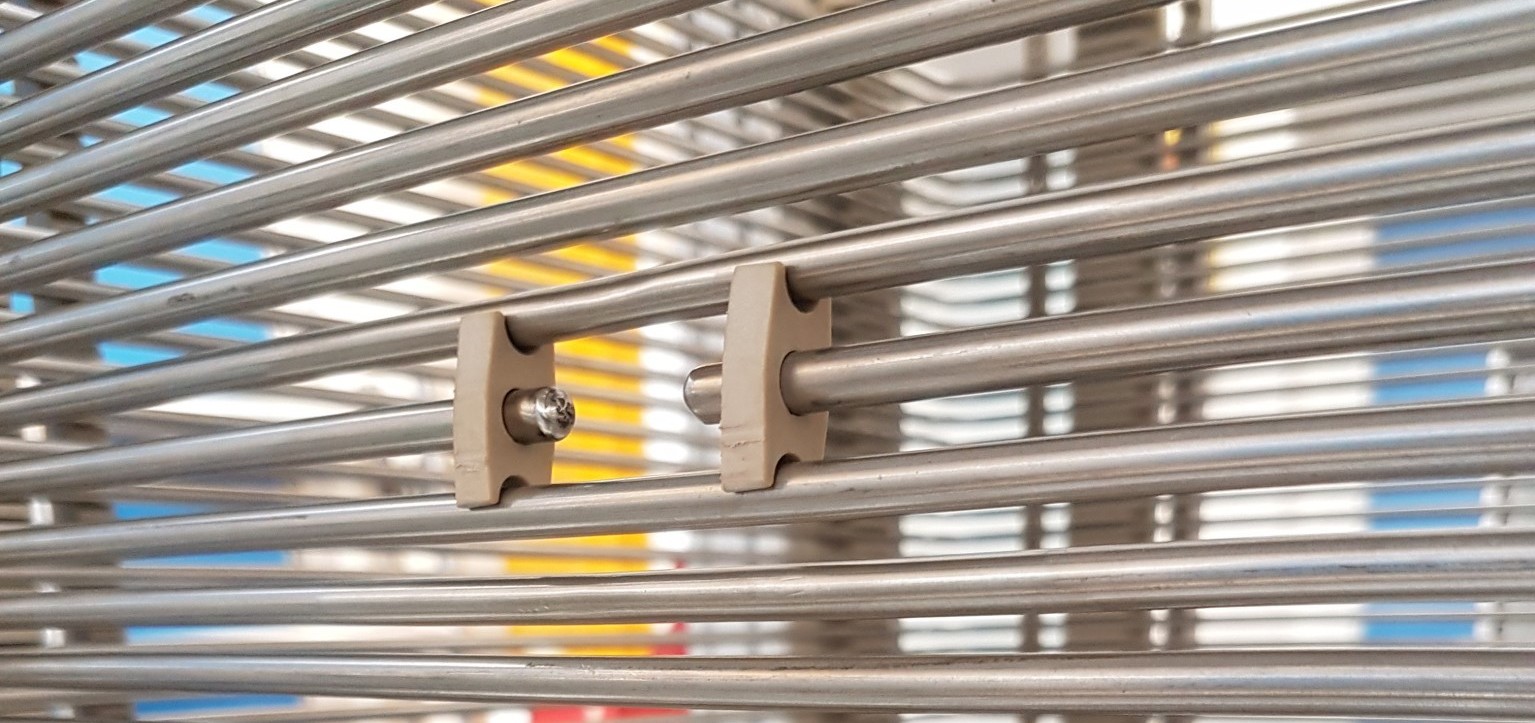}
\caption{One of the eight small-width gaps in the TPC. The smoothed and polished ends of the field-shaping ring are supported by PEEK elements.}
\label{fig:LaserGaps}
\end{figure}

\noindent The laser support structure has mechanical elements which allow for manipulation of the position and orientation of the mirror, and hence the horizontal beam. These elements are controlled using two independently-operated rotary feedthroughs on the top flange, as shown in Figure~\ref{fig:laser-controls}. One of the rotary feedthroughs is connected to a lead screw that, when rotated, moves the mirror platform up and down via a lead screw nut rigidly connected to the platform. This therefore allows the beam to be traversed vertically along the entire height of the TPC. Connected to the second rotary feedthrough is a keyed rod which is connected to a corresponding toothed hollow rod on the platform. Rotating this rod adjusts the angle of the mirror - and thus the direction of the laser beam - in the horizontal plane.
\\
\\
The TPC has been designed with gaps in the field-shaping rings, shown in Figure~\ref{fig:LaserGaps}, which allow the laser beam to pass into the active volume at specific heights. There are eight small-width gaps, evenly spaced in a vertical line spanning the full height of the TPC, which together permit a vertical scan of the TPC by the laser. There is additionally a single wider gap at the bottom of the TPC that allows for a full horizontal scan. Corresponding holes have been cut into the reflector panels on the near side of the TPC.\\

\begin{figure}
\centering
\includegraphics[width=0.65\textwidth]{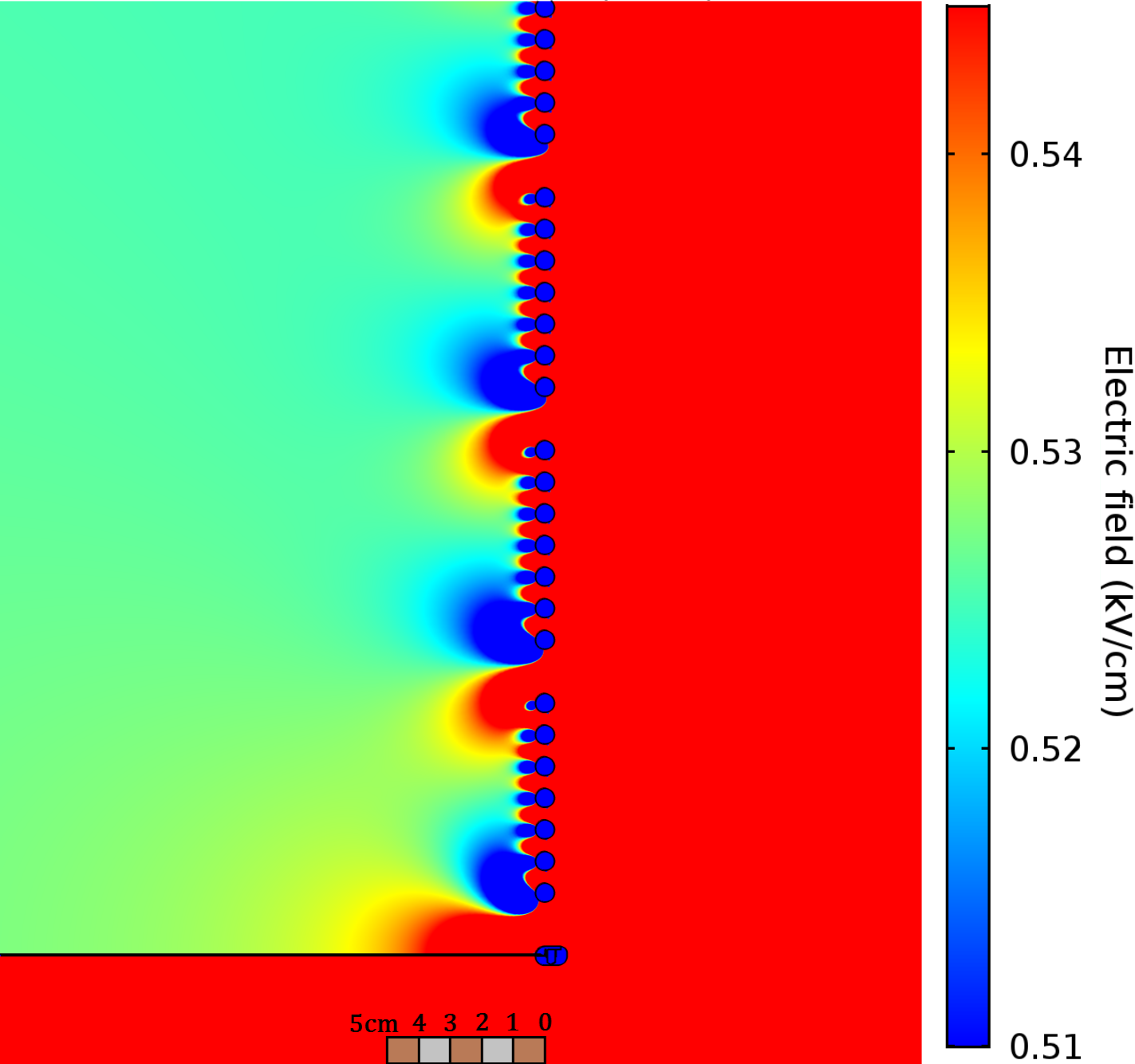}
\caption{The results of a COMSOL simulation of the effect of the gaps in the field-shaping rings on the drift field. The non-uniformity is $\approx$ 15\% - similar in magnitude to the effect of the beam plug (discussed in Section~\ref{subsec:BeamPlug}) - and extends up to $\approx$ 3~cm into the TPC, as indicated by the scale bar at the bottom.}
\label{fig:FieldLaserGaps}
\end{figure}

\noindent A COMSOL simulation was performed to investigate the effects of these ring gaps on the TPC drift field, and the results are shown in Figure~\ref{fig:FieldLaserGaps}. There is a $\approx$ 15\% non-uniformity of the drift field extending into the TPC for several centimetres, which is similar in magnitude to the non-uniformity induced by the beam plug (as discussed in Section~\ref{subsec:BeamPlug}). As noted previously in Section~\ref{subsec:TPC}, these non-uniformities subside to an acceptable level within just a few centimetres. The uniform drift field on that face of the TPC can be restored in the future by replacing the gap-containing rings with standard ones, if these ring gaps are no longer required.

\subsection{Cryostat, Top Flange and Beam Plug}
\label{subsec:BeamPlug}
The ARIADNE cryostat contains all of the previously discussed components of the detector, and consists of a vacuum-jacketed body and top flange, with mylar super-insulation installed inside the jacket to minimise heat losses. The cryostat is certified for operating pressures of up to 4~bar(g), and has a rated capacity of 1530 litres at 95\% full.

A DN200 conflat (CF) flange is welded to the outer wall of the cryostat, and is used to mount the external 0.9~mm thick stainless steel beam window. Correspondingly, the inner wall of the cryostat has a DN160 CF flange axially aligned to the DN200 flange on the outer wall, and the beam plug - which extends into the cryostat - is mounted onto this inner-wall flange.

\begin{figure}[ht]
\centering
\vspace{5mm}
\includegraphics[width=\textwidth]{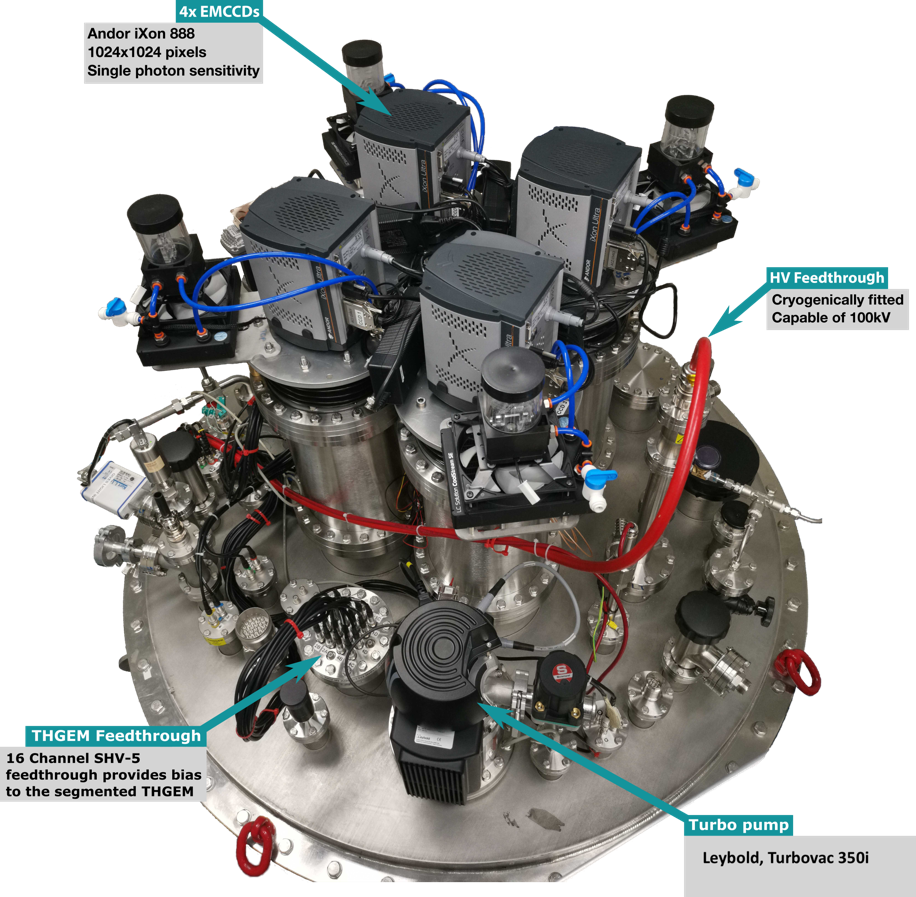}
\caption{The top flange of the ARIADNE cryostat. The 4 EMCCDs are mounted on the 2 $\times$ 2 array of DN200 CF ports, and various other ports are occupied by feedthroughs for the internal detector components, pressure and temperature sensors, viewports, valves and other external components.}
\label{fig:TopFlange}
\end{figure}

The cryostat is supported by six height-adjustable wheels. In the case of beamline operation, these wheels allow the beam window and plug to be aligned to the height of the beam.
\\
\\
The top flange, shown in Figure~\ref{fig:TopFlange}, houses the various ports and feedthroughs for mounting the external components and providing power, readout and other information to and from the internal components.

Some of these ports have been positioned in specific locations for specific purposes. Of particular note, the EMCCDs (described in more detail in Section~\ref{subsec:EMCCDs}) are mounted on 4 DN200 CF ports with additional DN200-to-100 CF zero-length reducer viewports. These four ports are arranged in a 2 $\times$ 2 array centred around the vertical axis of the cryostat. The previously discussed cryorefrigerator unit is mounted on a specific DN160 CF port, the aforementioned HV feedthrough is located on a DN75 CF port located directly above the cathode grid torus, and the turbo pump - which aids in evacuating the cryostat to ultra-high vacuum prior to introducing LAr - also has a dedicated location. There is also a custom DN63 CF port positioned centrally on the top flange, which could be used for future optical readout R\&D. Other ports are populated by the various non-location-specific sensors, viewports and feedthroughs necessary for operating the detector.

\noindent The seal between the top flange and the cryostat body is made using a Viton\textsuperscript{\textregistered} Grade ``A'' gasket. This material was chosen for its ability to maintain a flexible and stable seal at a temperatures as low as -20{\textdegree}C \cite{vitonSpecs}, which is well below that of the underside of the flange even when the cryostat is filled. The cryostat additionally has a machined groove which would allow the placement of a 2~mm diameter indium seal, if so required.\\

\begin{figure}
\begin{subfigure}{0.84\textwidth}
  \centering
  \includegraphics[width=\textwidth]{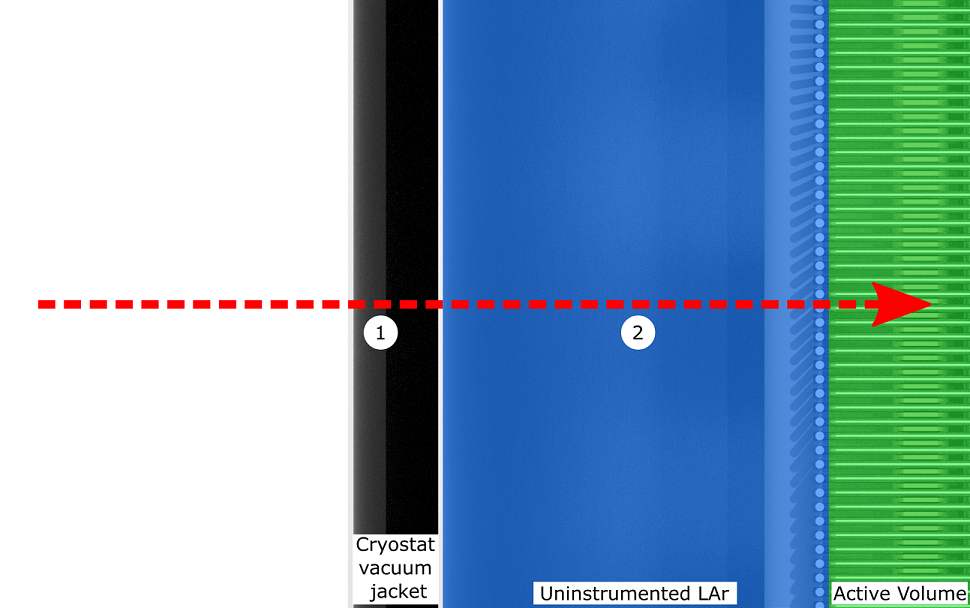}
  \caption{Beam transport through an unmodified ARIADNE cryostat, where the incoming beam passes through both cryostat walls (1) and a large length of uninstrumented LAr (2). In this configuration, the total material budget is 2.29 X$_0$.}
  \label{fig:WithoutBeamWindow}
\end{subfigure}
\begin{subfigure}{0.84\textwidth}
  \centering
  \vspace{5mm}
  \includegraphics[width=\textwidth]{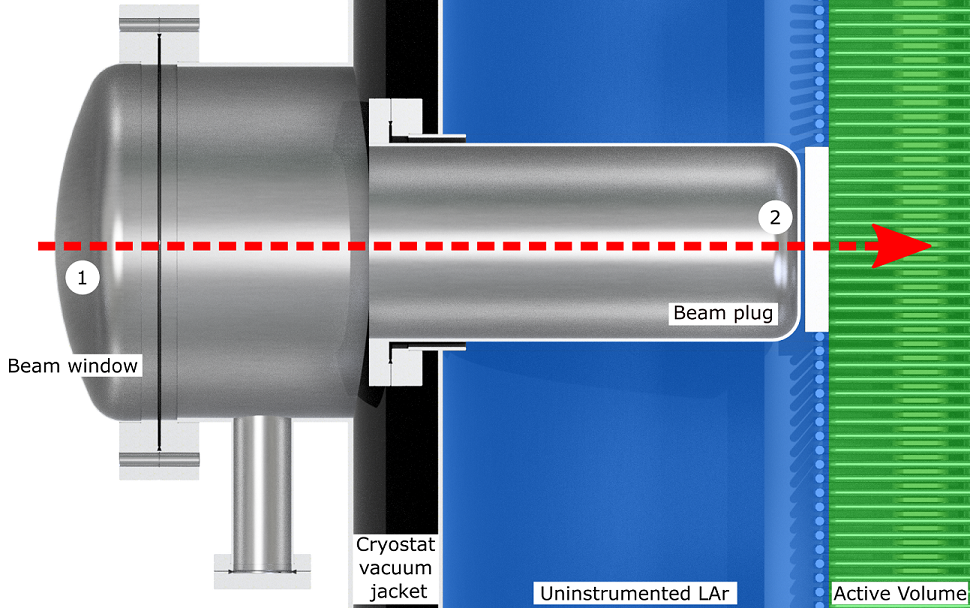}
  \caption{Beam transport through the ARIADNE beam window, plug and UHMWPE element. The incoming beam first passes through a 0.9~mm thick stainless steel beam window (1), and then vacuum before exiting the beam plug through a 2~mm thick stainless steel wall (2). Particles then pass through a 3.5~mm thick slice of uninstrumented LAr followed by a 16~mm UHMWPE element before reaching the active volume. The total material budget for this design is 0.21 X$_0$.}
  \label{fig:WithBeamWindow}
\end{subfigure}
\caption{Scenarios for beam transport from outside the ARIADNE cryostat into the active volume of the TPC.}
\label{fig:BeamWindowOptions}
\end{figure}

\noindent Without the presence of the beam window and plug penetrating through the cryostat body, incoming particles would encounter a significant amount of material before entering the active volume of the TPC - this scenario is shown in Figure~\ref{fig:WithoutBeamWindow}. The overall effect of this material on the quality of the incoming beam can be characterised using the radiation length, $X_0$ - the distance that a particle can travel through a particular material before its energy is reduced to $1/e$ of its original value.  Table~\ref{tab:radiationLengths1} shows the thicknesses and corresponding radiation lengths for the materials that an incoming particle would encounter in the scenario of there being no beam window or plug.

The beam window and plug together form a continuous extension to the cryostat body's vacuum jacket, allowing incoming beam particles to travel much deeper into the detector without impedance from any material. This in turn minimises losses and deviations to the incoming particle momenta and direction. The addition of the beam window and plug is shown schematically in Figure~\ref{fig:WithBeamWindow}, and Table~\ref{tab:radiationLengths2} shows the material budget for the scenario of beam transport into the TPC via the beam plug. (The additional UHMWPE displacement element is discussed below.) The use of the beam window and plug therefore reduces the energy loss of particles entering the TPC significantly - from 90\% to less than 20\%.

\begin{figure}[p]
\centering
\vspace{5mm}
\includegraphics[width=0.65\textwidth]{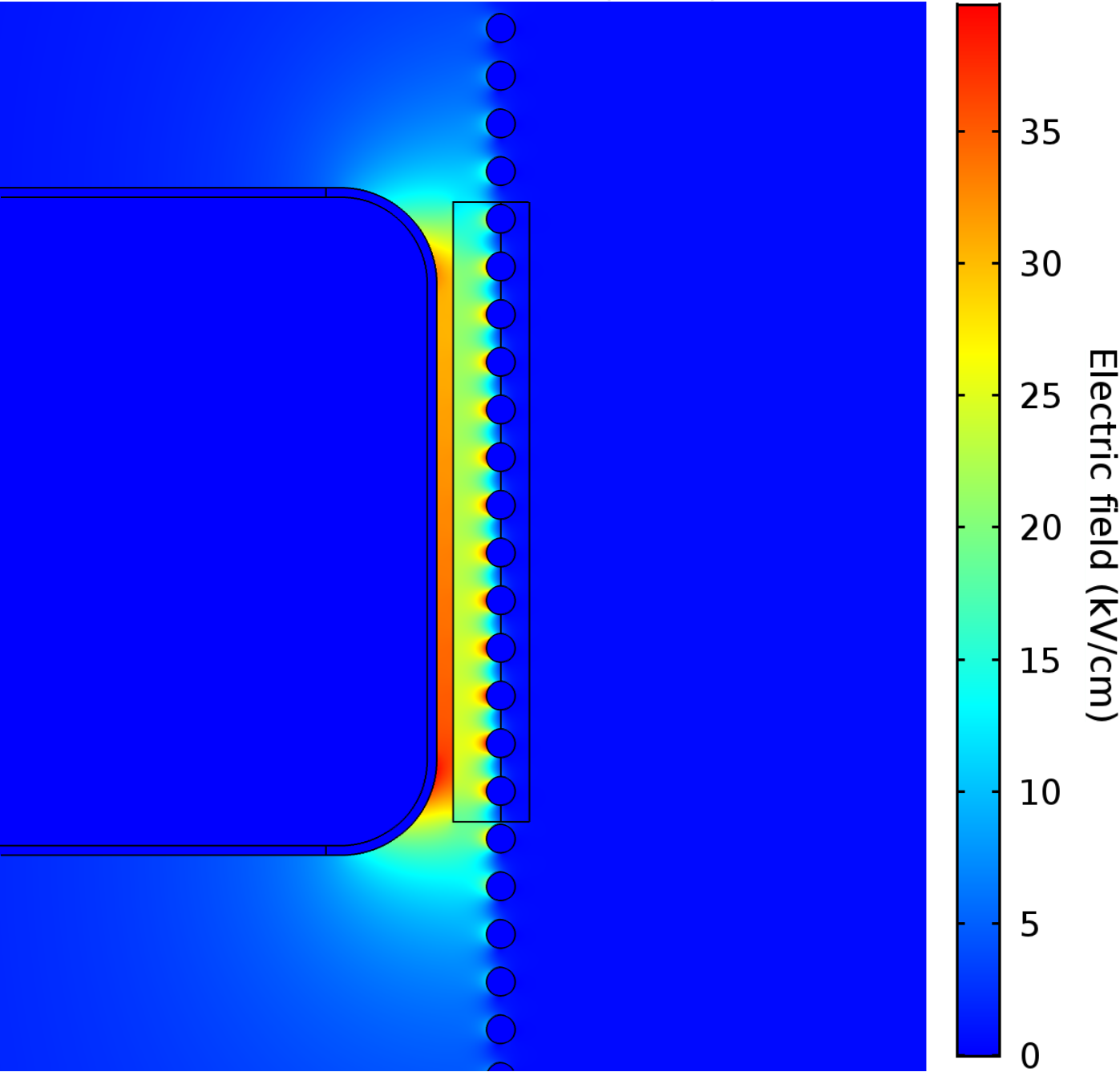}
\caption{Results of a COMSOL simulation of the electric field in the region outside the field cage near the end of the beam plug. The length of the beam plug was carefully chosen to ensure that the field between the beam plug and the field cage does not exceed 40~kV/cm - a safe lower limit for the dielectric strength of LAr. The UHMWPE element, discussed in the main text, is also shown.}
\label{fig:FieldBeamPlug}
\end{figure}

Since the beam plug is electrically grounded to the cryostat, its distance from the field cage must be carefully controlled. COMSOL simulations were performed to determine the minimal spacing required between the field cage and the beam plug to avoid LAr breakdown. This has previously been observed for fields as low as 40~kV/cm \cite{LArBreakdown}, and so this value was used as the upper limit on the electric field between the beam plug and the field cage. The COMSOL simulation with the beam plug installed at $\approx$~10.5~mm from the field cage - the results of which are shown in Figure~\ref{fig:FieldBeamPlug} - indicates that the electric field does not exceed the limit of 40~kV/cm.\\

\begin{table}[p]
\begin{subtable}{\textwidth}
  \centering
  \begin{tabularx}{0.97\textwidth}{|p{6.5cm}|c|c|c|}
    \cline{1-4}
    \textbf{Material} & \textbf{$X_0$ (cm)} & \textbf{Thickness (cm)} & \textbf{Thickness ($X_0$)} \\
    \cline{1-4}
    \cline{1-4}
    Stainless Steel (outer cryostat wall) & 1.76 & 0.4 & 0.23 \\
    \cline{1-4}
    Stainless Steel (inner cryostat wall) & 1.76 & 0.3 & 0.17 \\
    \cline{1-4}
    LAr (uninstrumented) & 14.0 & 26.4 & 1.89 \\
    \cline{1-4}
    \textbf{Total} &  &  & \textbf{2.29} \\
    \cline{1-4}
  \end{tabularx}
  \caption{The material budget when the beam transport occurs through an unmodified ARIADNE cryostat.}
  \label{tab:radiationLengths1}
\end{subtable}
\begin{subtable}{\textwidth}
  \vspace{5mm}
  \centering
  \begin{tabularx}{0.97\textwidth}{|p{6.5cm}|c|c|c|}
    \cline{1-4}
    \textbf{Material} & \textbf{$X_0$ (cm)} & \textbf{Thickness (cm)} & \textbf{Thickness ($X_0$)} \\
    \cline{1-4}
    \cline{1-4}
    Stainless Steel (outer beam window face) & 1.76 & 0.09 & 0.05 \\
    \cline{1-4}
    Stainless Steel (inner beam window wall) & 1.76 & 0.2 & 0.11 \\
    \cline{1-4}
    LAr (uninstrumented) & 14.0 & 0.35 & 0.02 \\
    \cline{1-4}
    UHMWPE (displacement element) & 50.0 & 1.6 & 0.03 \\
    \cline{1-4}
    \textbf{Total} &  &  & \textbf{0.21} \\
    \cline{1-4}
  \end{tabularx}
  \caption{The material budget when the beam transport occurs through the ARIADNE beam window and plug.}
  \label{tab:radiationLengths2}
\end{subtable}
\caption{Material budgets for beam transport from outside the ARIADNE cryostat into the active volume of the TPC.}
\label{tab:MaterialBudgetOptions}
\end{table}

\noindent With the beam plug extending to its minimum safe distance from the field cage, the beam must still be transported through the remaining 19~mm of uninstrumented LAr to reach the active volume of the TPC. Even this thickness would affect the quality of any incoming beam, and so to displace this material, an UHMWPE element is attached to the field cage. This element fills almost the entire space between the end of the beam plug and the active volume. Since UHMWPE has a much longer radiation length of $\approx$~50~cm, this helps to maintain the incoming beam quality, while not affecting the shape of the electric field near and around the beam plug.

\subsection{EMCCD Cameras}
\label{subsec:EMCCDs}
\begin{figure}[ht]
\begin{subfigure}{0.49\textwidth}
  \centering
  \includegraphics[width=\textwidth]{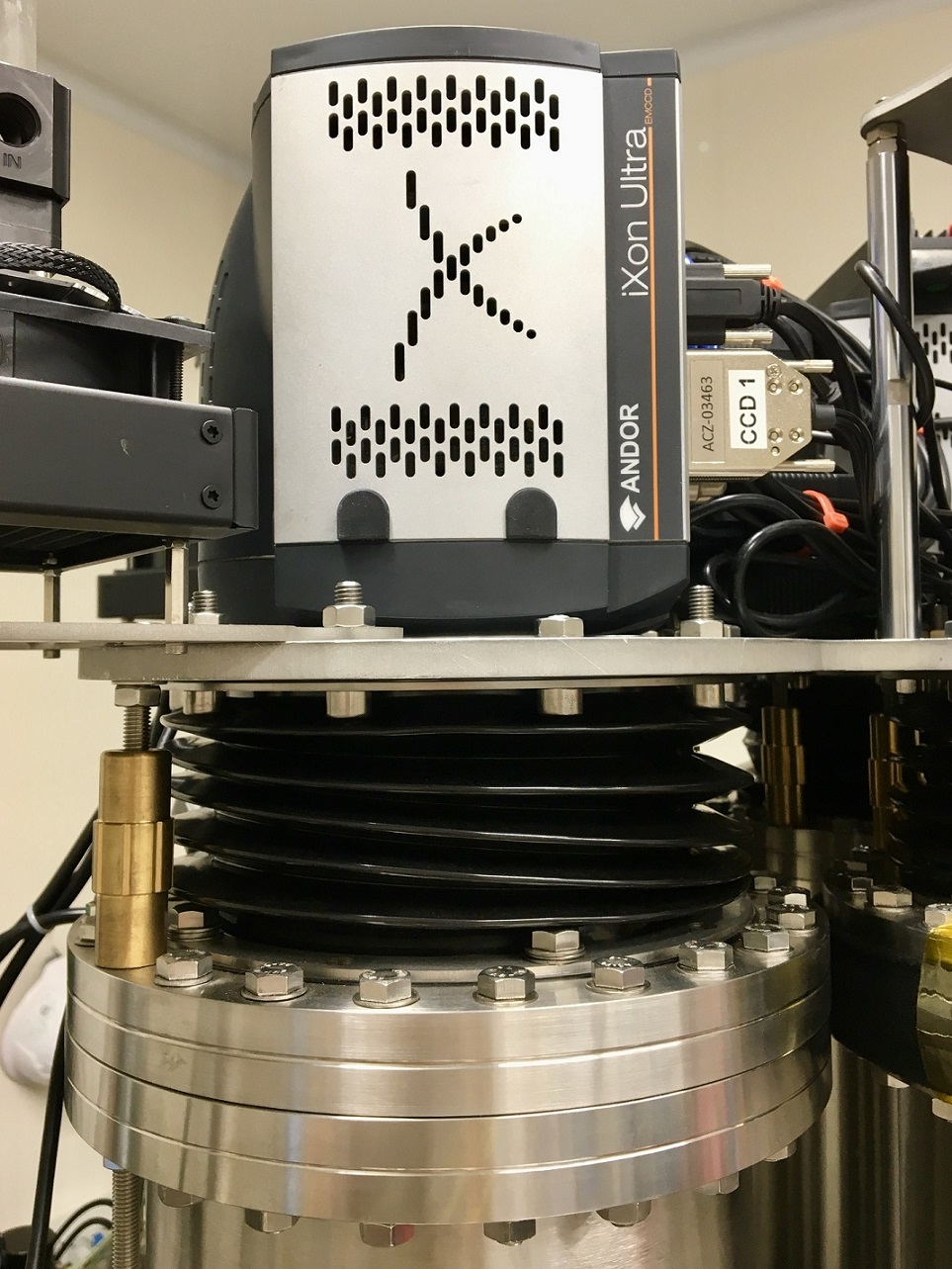}
  \caption{With the rubber bellow secured in the ``up'' position, ensuring light tightness.}
\end{subfigure}
\hspace{1mm}
\begin{subfigure}{0.49\textwidth}
  \centering
  \includegraphics[width=\textwidth]{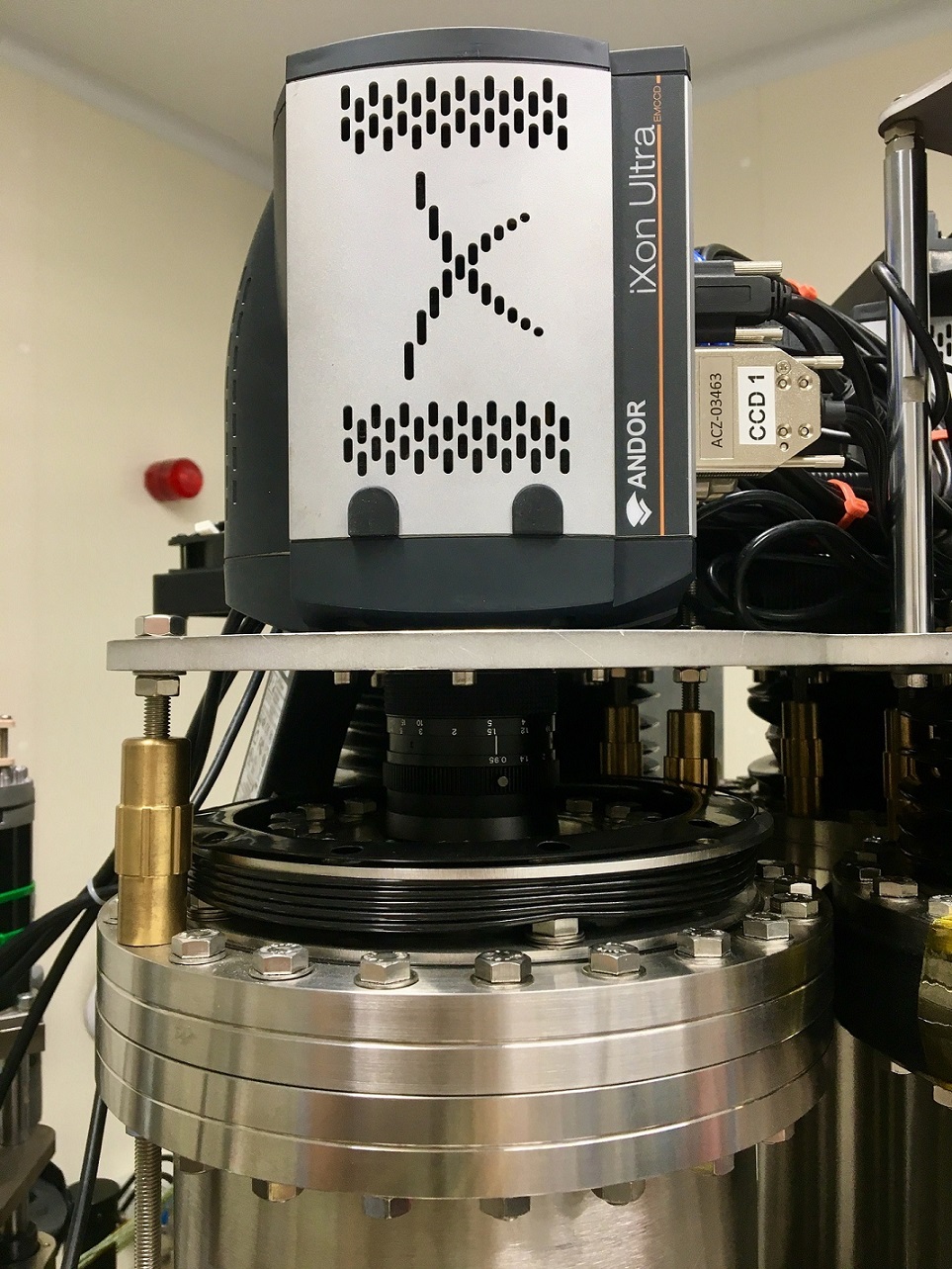}
  \caption{With the rubber bellow in the ``down'' position, allowing access to the Spacecom lens.}
\end{subfigure}
\caption{Side views of one of the Andor EMCCDs in position on the common mounting plate, which is in turn secured to the 4 DN200 nipples on the top flange of the ARIADNE cryostat.}
\label{fig:EMCCD-side-view}
\end{figure}

In order to achieve the best possible energy resolution and lowest possible energy threshold, it is important to maximise the collection efficiency of the S2 light produced at the THGEM. To that end, ARIADNE uses four Andor iXon Ultra 888 EMCCD cameras \cite{Andor} to capture this light - these were previously shown in Figure~\ref{fig:TopFlange}, and additional side-on views are shown in Figure~\ref{fig:EMCCD-side-view}.

Each EMCCD is coupled to a Spacecom VF50095M lens (with a speed of f/0.95 and focal length of 50~mm), and all four devices are securely fixed to a common mounting plate positioned above the 4 DN200 nipples on the top flange of the cryostat. This plate has been finely machined to ensure that the EMCCDs are all flat relative to each other, and its tilt can be adjusted to a very fine level using threaded rods - allowing for precise alignment of the EMCCDs' flatness relative to the THGEM inside the cryostat. This has been achieved by matching the plate's tilt to that of the THGEM, which has been measured to the order of 0.1\textdegree{} using a high-precision level finder. The lenses and top of each DN200 nipple are fully enclosed by rubber bellows, which ensure light tightness while still allowing access to the lenses' focus and iris controls when required.  The external location of the mounting assembly gives excellent access to the EMCCDs themselves, allowing them to easily be maintained and upgraded. Additionally, the optical isolation of the TPC from the external EMCCD cameras provides protection against potential TPC grounding issues and/or detector high voltage transients.\\

\noindent The sensor used in each of the EMCCD cameras has a quantum efficiency of approximately 80$\%$ at 420~nm (the peak wavelength of the TPB emission spectrum), as previously shown in Figure~\ref{fig:TPBQE}, and the cameras as a whole are single photon sensitive. This is achieved through the use of EM gain - applied on a linear multiplicative scale up to a maximum factor of 1000 - to amplify the incoming photon signal above the readout noise of the charge amplifier. The EMCCD sensor pixels have a full well depth of up to 80,000 $e^-$, which is sampled at 16-bit digitisation. This gives relative intensity measurements of up to 65,356 ADU per pixel, allowing for high resolution calorimetry.

During readout, the pixels can optionally be grouped into square clusters of 2 $\times$ 2, 4 $\times$ 4, 8 $\times$ 8 or 16 $\times$ 16 pixels per cluster - this is referred to as the ``binning''. A readout rate of approximately 60Hz can be achieved when running the cameras in 4 $\times$ 4 binning - this setting was found to give a good balance between track resolution and collected light per binned pixel. Each camera's binning settings are easily changed, which allows for quick detector reconfiguration and switching between high resolution tracking of high intensity signals or coarse resolution tracking of faint light signals using 8 $\times$ 8 or even 16 $\times$ 16 binning.
\\
\\
This model of EMCCD camera has a built-in thermo-electric (TE) air-cooling system that reduces the sensor temperature to between -60 and -80\textdegree{}C. (The exact value is dependent on ambient conditions and the pixel readout rate.) The TE cooling can be augmented by a liquid-cooling circuit, which requires an external liquid source but allows the sensor temperature to be further reduced by up to 20\textdegree{}C.

\newpage
\FloatBarrier
\section{Detector Operation}

\subsection{Data Acquisition}
\label{subsec:DAQ}
The ARIADNE Data Acquisition (DAQ) system governs the synchronised collection of data from the various detector readouts. It has been designed to be straightforward for both experienced and untrained individuals to use, and also allows for small changes to be made to the acquisition settings as needed during detector operation.

The device readouts are as follows:

\begin{itemize}
    \item EMCCDs: these have built-in readout and digitisation electronics, and data is transferred over high-speed USB 3.0.  Each EMCCD also can output a veto signal, which is active during the process of acquiring a frame (i.e. the time during which the device is not yet ready to start a new acquisition).
    \item PMTs: the signals are read out by a CAEN V1720 (12~bit, 250~MS/s) digitiser, which has 8 independent channels, each with a time resolution of 4~ns per sample.
    \item THGEM: the signal from each of the 16 pads is split by a pre-amp into high- and low-gain channels, giving a total of 32 charge readouts. These are read out by a 64-channel CAEN V1740 (12~bit, 62.5~MS/s) digitiser, with a per-sample time resolution of 16~ns.
\end{itemize}

\noindent The maximum event rate is limited to approximately 60Hz (constrained by the EMCCD readout rate previously discussed in Section~\ref{subsec:EMCCDs}), and so the technical requirements for full detector DAQ are not demanding. The relatively low event rate also means that all incoming data can be stored directly to disk, without the need for any online filtering of events.
\\
\\
The central point for the DAQ is a dedicated desktop computer - the ``DAQ PC'', which handles all parts of the DAQ (control, operation, settings and data storage), as well as slow control and monitoring (discussed in Section~\ref{subsec:Monitoring} below). The DAQ Graphical User Interface (GUI) software package - written specifically for ARIADNE by the authors, and run from the DAQ PC - allows full control of the DAQ system. This includes selection of the particular readout device(s), configuration settings for each device, and parameters particular to the run such as the number of events to record and what trigger mode to use.

Alongside the GUI, an Event Viewer has been developed which can be operated in either ``Live'' (showing data as it is processed during a run) or ``Offline'' (showing data previously saved to disk) mode. As well as making use of GPU acceleration to produced high quality output of PMT and THGEM data, there are several basic tools built into the Viewer - for example, manipulation of level thresholds in the EMCCD event display, and automated basic statistical calculations (mean, maximum, etc.) on the data. The Event Viewer can therefore be used at all stages of detector operation - during commissioning, for data quality checks during data-taking, and for basic visual analysis after data-taking is complete. The results presented in Sections \ref{subsec:CERNResults} and \ref{subsec:LivResults} are depicted using the Event Viewer.

\subsection{Trigger System}
\label{subsec:Trigger}
\begin{figure}
\centering
\includegraphics[width=\textwidth]{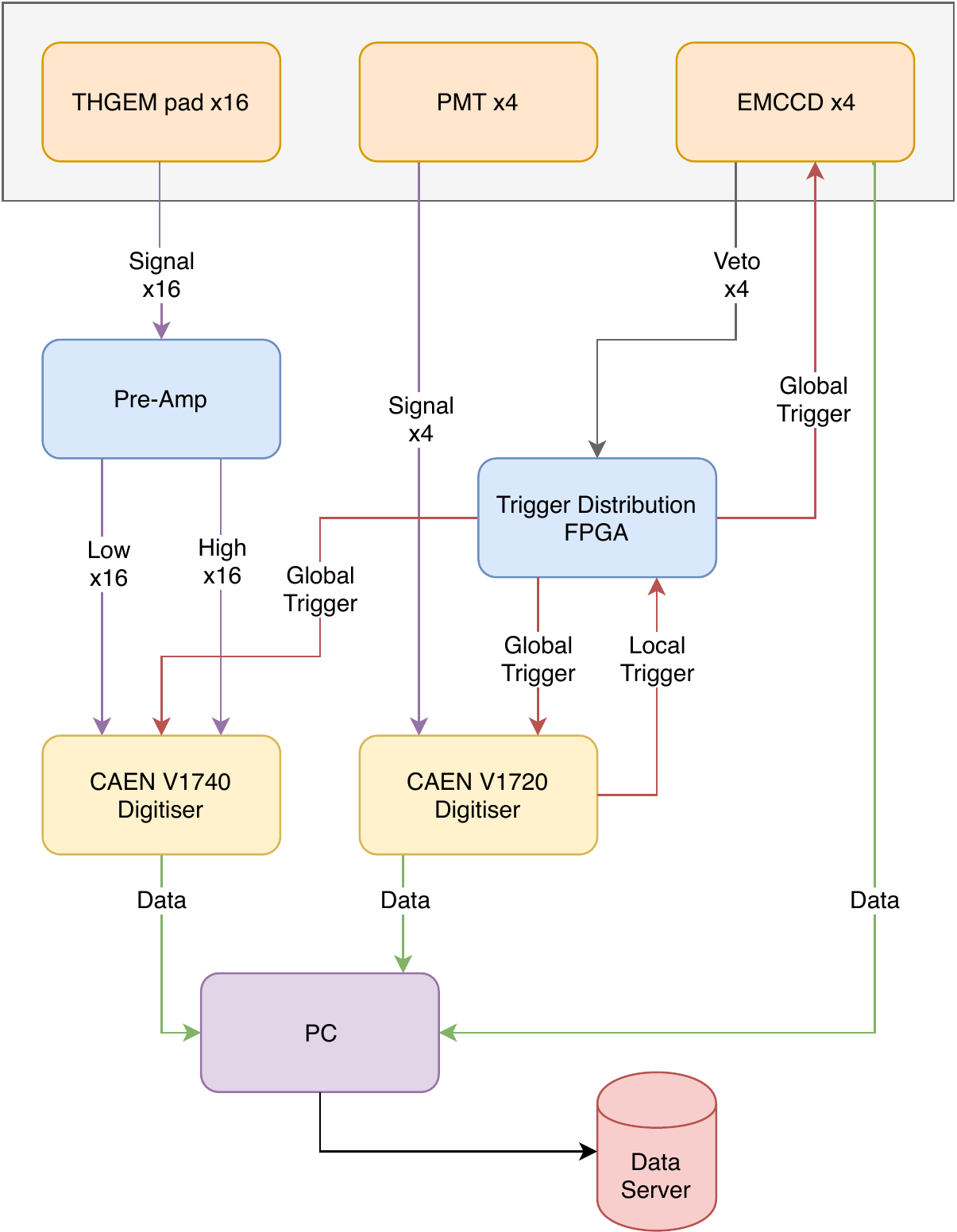}
\caption{Schematic of the ARIADNE trigger and readout system, as implemented only for the detector itself.}
\label{fig:daq-trigger}
\end{figure}

Figure~\ref{fig:daq-trigger} gives a schematic overview of the global ARIADNE detector trigger system.  (Note that this only covers the implementation for the detector itself - additional off-detector trigger sources were used during operation at the CERN T9 beamline, but they are discussed in Chapter~\ref{subsec:CERNOpConfig}.)
\\
\\
The trigger system is implemented in NIM logic.  Each of the PMTs issues a ``local'' (i.e. device-level) trigger via the V1720 digitiser, and these are all passed to a ``trigger distribution'' FPGA, which identifies an internal combined PMT trigger for the entire detector if it finds that any one of the individual PMT triggers is above a pre-defined threshold.

If the combined PMT trigger is found to be in coincidence with a veto signal from any of the EMCCDs, no further triggers are issued, and the device readouts are reset.  If no veto signal is present, i.e. all four EMCCDs are ready to read out, a ``global'' trigger is issued by the FPGA and distributed to the PMT and THGEM digitisers and the EMCCDs, signalling them all to begin data readout.

\subsection{Detector Monitoring and Slow Control}
\label{subsec:Monitoring}
The detector has a number of real-time monitoring and slow control systems.

Three internally-positioned thermocouple temperature sensors - at the bottom, middle and top of the field cage - are used to measure the temperature within the cryostat, and also provide information about the LAr level during filling, operation and emptying of the detector. A fourth internal thermocouple is used to monitor the temperature of the underside of the top flange. The external temperature of the cryostat - and therefore the performance of the vacuum jacket - is monitored using a further two thermocouple sensors, mounted on the outside of the cryostat body and the outer surface of the top flange.

\begin{figure}[p]
\centering
\includegraphics[width=\textwidth]{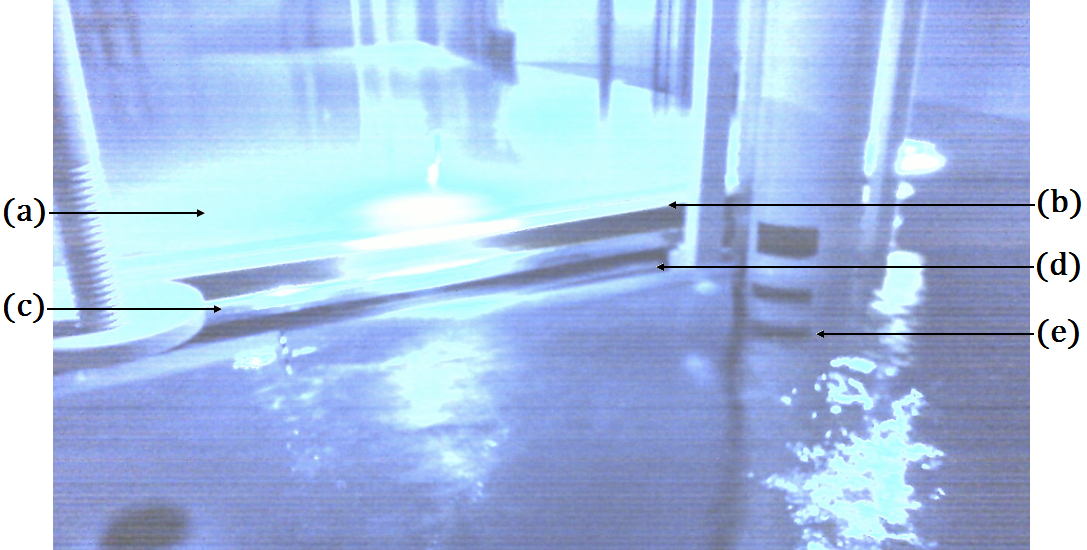}
\caption{An internal view of the cryostat, taken using one of the cryogenic webcams while the detector was being filled.  The TPB-coated glass (a), THGEM (b), extraction grid (c) and first field ring (d) can be seen, with the LAr level just below the ring.  The markings on the PEEK support rod (e) are used to help determine the LAr level.}
\label{fig:CryoWebcam}
\end{figure}

As previously noted in Section~\ref{subsec:Cryogenics}, the LAr level can also be monitored via a set of ultrasonic sensors as well as a webcam. This is positioned such that it has a view of the gap between the THGEM and extraction grid, as seen in Figure~\ref{fig:CryoWebcam}. A second webcam provides a view of the top of the internal filtration cartridge, which allows for the performance of the recirculation system to be monitored. Externally-controlled LEDs are positioned around the top of the TPC to provide illumination for the webcams.

\begin{figure}[p]
\vspace{7mm}
\centering
\includegraphics[width=\textwidth]{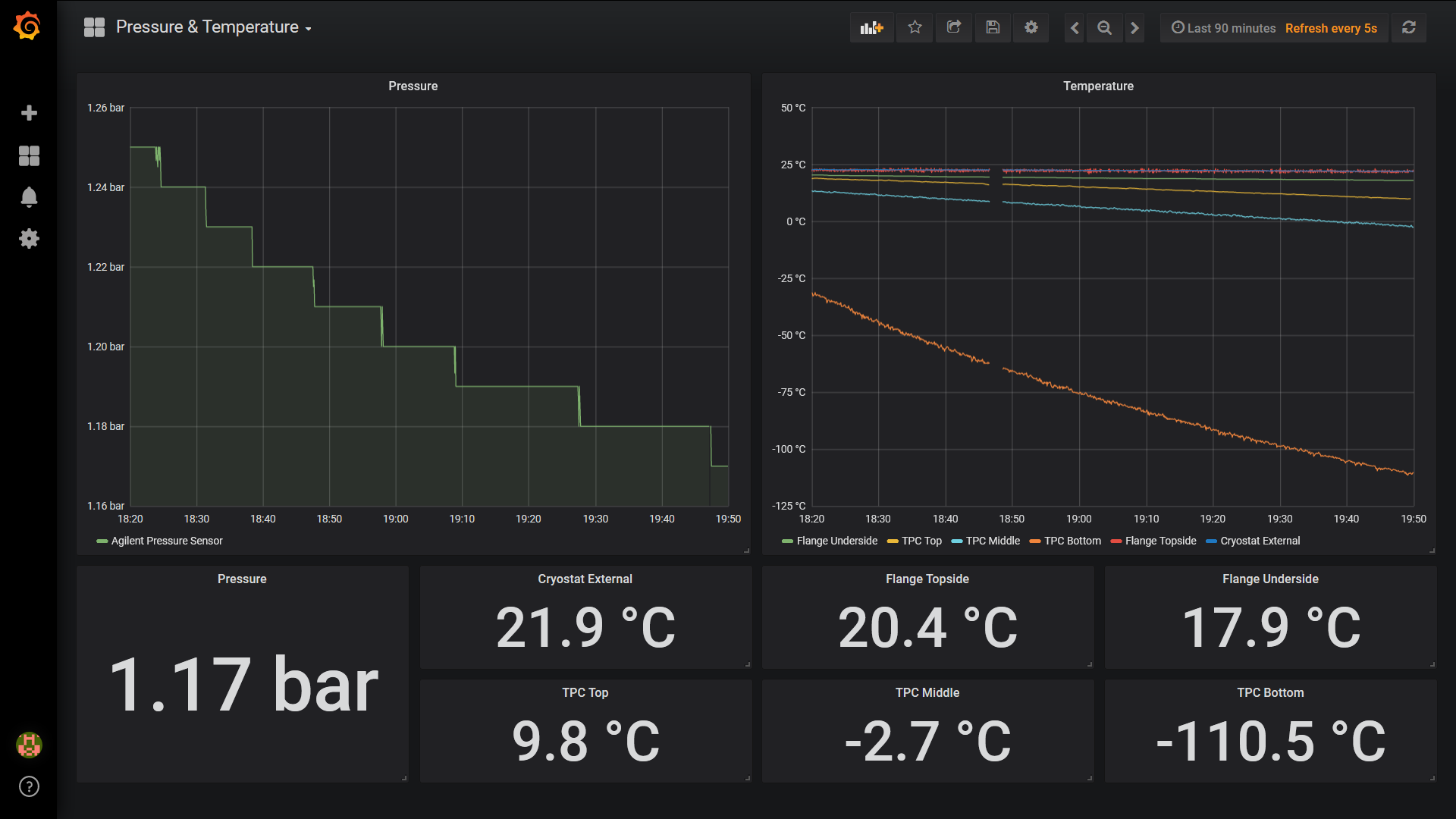}
\caption{The Grafana monitoring display, showing the pressure and temperature sensor readouts.  The numerical values are the immediate values, and the graphs allow longer-term trends to be observed. When this image was taken, the detector was being filled with LAr.}
\label{fig:daq-monitoring}
\end{figure}

Finally, three pressure sensors are available to monitor the internal cryostat pressure: a Kurt Lesker Series 910 covering the range from $1 \times 10^{-5}$ to $2000$~mbar, an Agilent PCG-750 for the range between $5 \times 10^{-5}$ and $1500$~mbar, and a Kurt Lesker HPT-100 operating from $1 \times 10^{-9}$ to $1000$~mbar.  The normal operating pressure of the detector is no more than 1.2~bar, and so the Kurt Lesker 910 and Agilent sensors allow monitoring even in an over-pressure situation.
\\
\\
The data from all of the monitoring devices are passed to a Grafana server running on the DAQ PC, which provides a centralised way to monitor the immediate values, as well as longer-term trends.  This server can be accessed locally (i.e. on the DAQ PC itself), or remotely from any internet-connected browser using the correct user credentials.  An example of the monitoring display on the DAQ PC is shown in Figure~\ref{fig:daq-monitoring}.\\

\noindent The monitoring system is linked to an automated alert system, which can send messages to the ARIADNE collaboration's internal messaging service in the event of any temperature or pressure readings crossing preset alert-triggering thresholds.  This provides a level of protection in the case of problems such as a cooling system failure or power outage, and gives detector operators time to physically access the detector and resolve any issues.

\subsection{Nearline Analysis - Image Stitching}
\label{subsec:Stitching}
\begin{figure}[ht]
\centering
\includegraphics[width=0.70\textwidth]{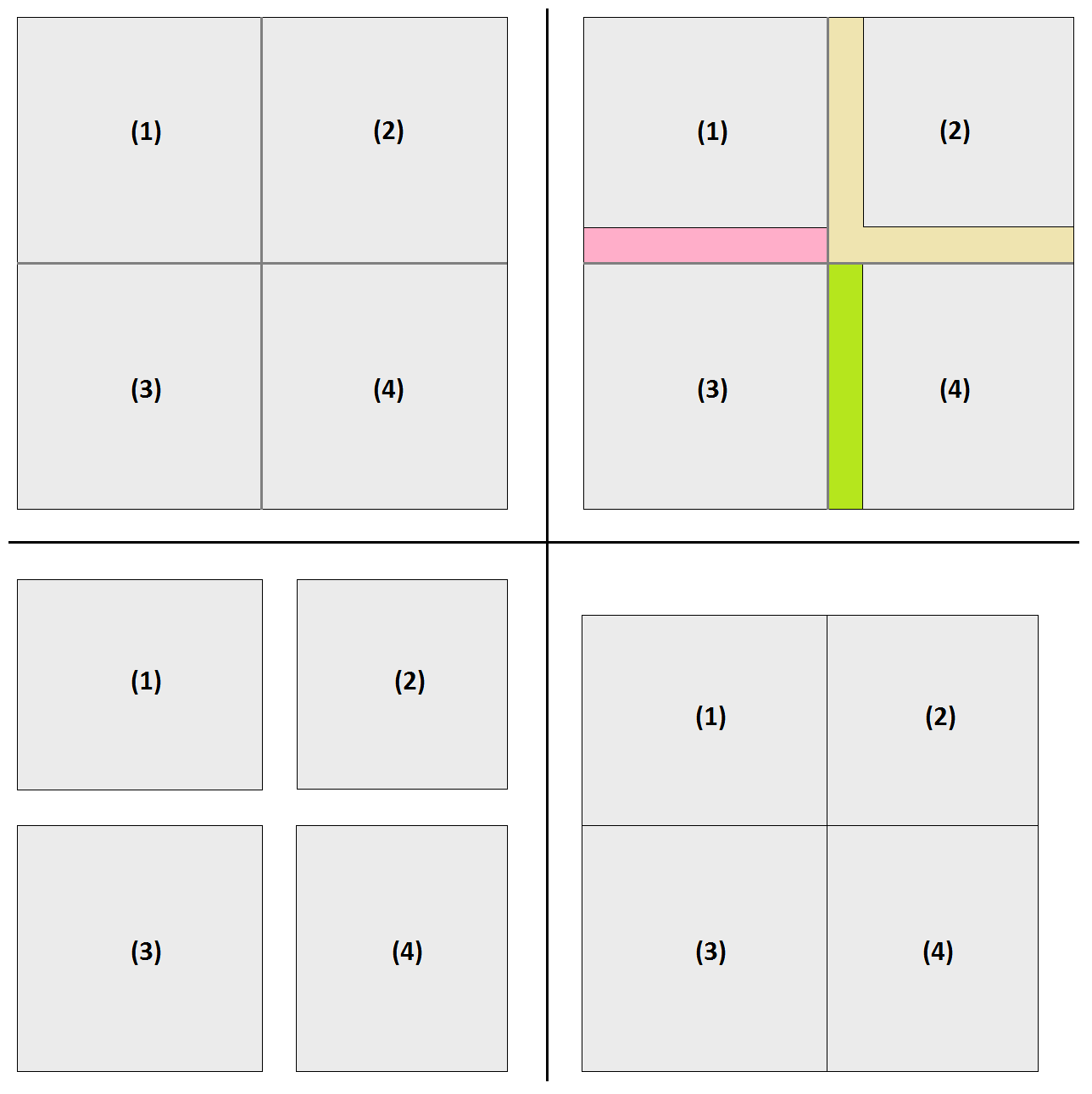}
\caption{The image stitching process for producing a single EMCCD image per event.  The raw image (top-left) is square, as is each EMCCD's individual FoV.  The overlaps between adjacent FoVs are first identified (top-right), depicted in pink, yellow and green for EMCCDs 1, 2 and 4 respectively, and then removed (bottom-left), leaving the EMCCDs with non-square but unique FoVs.  The sizes of these overlaps have been found empirically across a large number of raw images.  The final step is to translate the FoVs such that they align to the same outer edges. This results in an overall square image - slightly smaller than the original raw image - containing no overlaps or empty space (bottom-right).}
\label{fig:stitching_process}
\end{figure}

When the EMCCDs are correctly focused to the plane of the THGEM, the field of view (FoV) of each one is larger than one quarter of the THGEM's area. This results in some overlap between adjacent EMCCDs' FoVs, as well as a view of the region outside the THGEM's outer edges. To correct for these overlapping FoVs, a simple ``image stitching'' algorithm is applied to the raw EMCCD data.  This involves cropping each individual image according to empirically calculated margins, followed by translating the images to remove the now-empty space between them.  The process is shown schematically in Figure~\ref{fig:stitching_process}.

Since the EMCCD positions and FoVs do not change over the detector's operational lifetime, the same stitching algorithm is applicable to all raw EMCCD data. In the discussions of results in Sections~\ref{subsec:CERNResults} and \ref{subsec:LivResults} below, image stitching is always implicitly performed.
\\
\\
Using the 4 $\times$ 4 binning scenario as an example, the dimensions of each single EMCCD image after stitching is approximately 480 $\times$ 480 pixels. Given that this corresponds to the physical 53 $\times$ 53~cm area of the THGEM, the spatial resolution at this binning is therefore $\approx$ 1.1~mm per pixel. Similar calculations show that the resolution is 2.2 and 4.4~mm per pixel for 8 $\times$ 8 and 16 $\times$ 16 binning respectively.

\clearpage

\clearpage
\newpage
\FloatBarrier
\section{Mixed Particle Detection at the CERN T9 Beamline} 
\label{chp:CERN}

\subsection{T9 Beamline} 
\begin{figure}[ht]
\centering
\includegraphics[width=\textwidth]{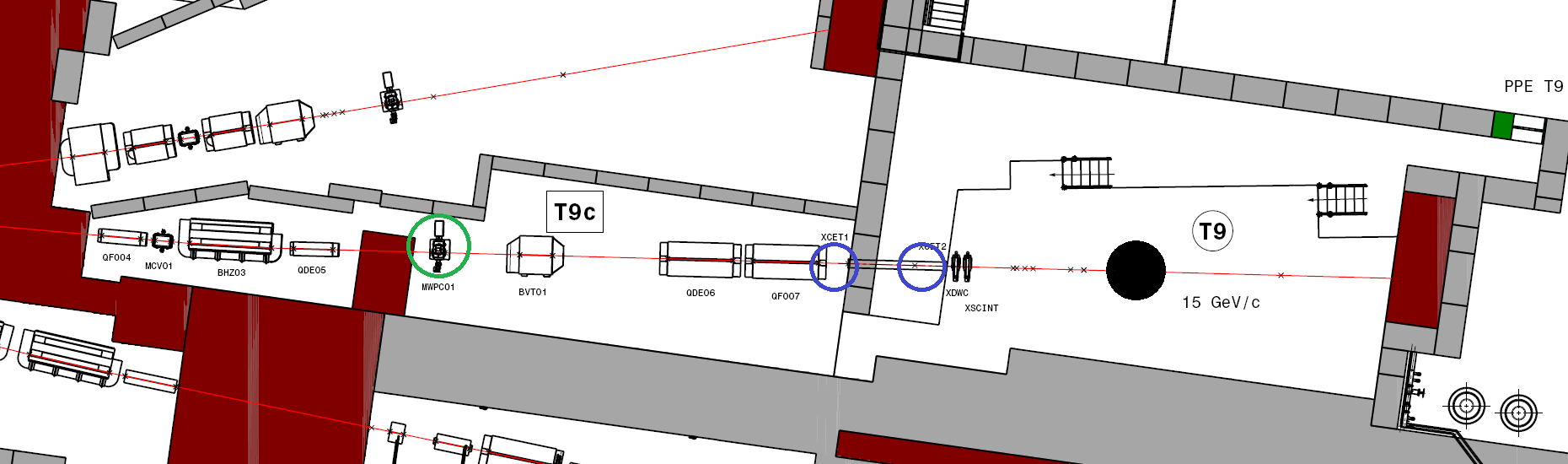}
\caption{Schematic of the T9 beamline and experimental area within the East Hall (adapted from \cite{CERNEastArea}). The air gap is indicated by the green circle, the two in-line Cherenkov counters by blue circles, and the approximate size and position of the ARIADNE detector by the filled black circle.  This view excludes the highly irradiated area closer to the target (to the left of the image), and includes parts of the adjacent T8 and T10 beamlines (below and above T9 respectively).}
\label{fig:t9Schematic}
\end{figure}

In order to characterise the response of the ARIADNE detector and its various sub-systems, a well-described source of particles spanning a range of energies is required. The T9 beamline at CERN's East Hall \cite{CERNEastArea} was chosen for this task. This is one of the Proton Synchrotron (PS) secondary beams, initiated by a proton beam from the PS striking a stationary aluminium or beryllium target located 55~m from the T9 experimental area. The resulting mixed beam of electrons, muons, pions, kaons and protons (along with their respective anti-particles) is produced with momenta between 0.5 and  15~GeV/c and a production angle of 0\textdegree{}.  The beamline is pre-equipped with two in-line Gas Cherenkov counters and an air gap located approximately 20~m from the experimental area, which allows for the placement of a Time Of Flight (ToF) system for independent particle identification (PID). A schematic of the T9 beamline is shown in Figure~\ref{fig:t9Schematic}.\\

\noindent The ARIADNE detector was operated at the T9 beamline for 18 days during March and April of 2018, and data from electrons, muons, pions and protons (as well as their anti-particles) were collected at momenta between 0.5 and 8.0~GeV/c (the range suggested by the energy containment studies discussed previously in Section \ref{subsec:Containment}). No kaon data were collected, as their short lifetimes mean that the vast majority of them will decay within the beamline before reaching the detector.  A total of just under 800,000 events were collected during this period, split evenly between the positive and negative polarities. (Information concerning the relative fractions of each particle species, as measured at the T9 source, can be found in \cite{CERNEastArea}.)

\subsection{Beamline Hardware} \label{ToFSection}
\subsubsection*{Collimators}
The beamline is equipped with a set of collimators, which are used to limit the physical size of the beam, and therefore control the number of particles per spill that reach the experimental area.  The collimators act as 'shutters', with their left, right, top and bottom positions all independently controllable.

\subsubsection*{Cherenkov Counters}
The T9 beamline is also pre-equipped with two in-line Cherenkov counters. Each counter consists of a volume of gas connected to a PMT through an optical viewport. Any particle passing through the counter will emit Cherenkov radiation if it exceeds the phase velocity of light in the gas, the value of which is dependent on the type of gas and its pressure.  Therefore, by changing these two properties of the counter, it is possible to use it to tag different particle species travelling at different velocities, i.e. momenta.

When operating the ARIADNE detector at T9, one of the Cherenkov counters was filled with carbon dioxide at 2.2~bar. Due to their low rest mass, only electrons have a high enough velocity at momenta below 8~GeV/c to emit Cherenkov light in 2.2~bar CO$_2$, and so this Cherenkov counter was used as an electron tagger. The second Cherenkov counter was emptied to vacuum, so as to minimise its impact on the through-going beam (the counters cannot be physically removed from the beamline). 

\subsubsection*{Time of Flight (ToF) System}
\begin{figure}[ht]
\centering
\includegraphics[height=135mm]{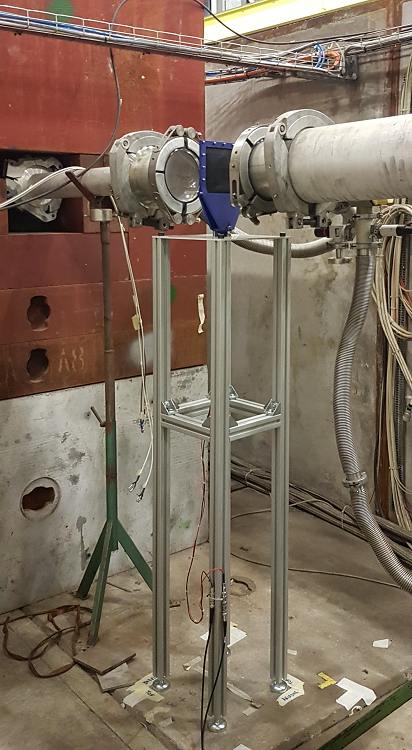}
\hskip 2mm
\includegraphics[height=135mm]{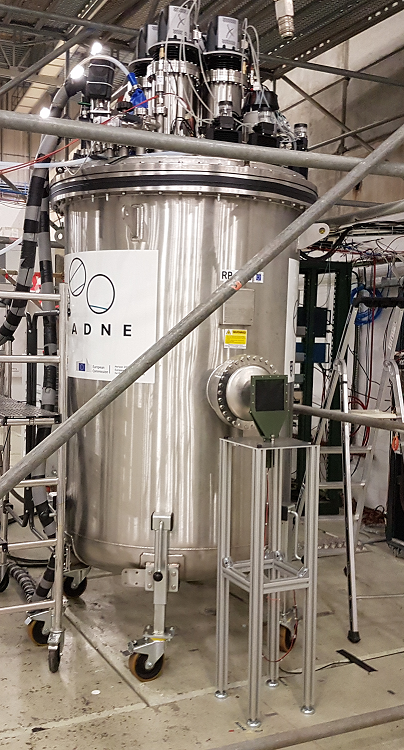}
\caption{The far (left) and near (right) Time of Flight assemblies.  The far assembly was positioned in the beamline air gap, and the near assembly directly in front of the detector's beam window.  The separation between the assemblies was 18.05~m.}
\label{fig:ToFAssemblies}
\end{figure}

For non-electron (muon, pion and proton) PID, a ToF system was designed and built at the University of Liverpool. The system is made of two identical assemblies, each consisting of a 0.15 $\times$ 150 $\times$ 150~mm EJ-212 plastic scintillator film (manufactured by Eljen Technology) covered by 3M ESR Vikuiti reflector material (the same that surrounds the ARIADNE TPC) to optimise light collection. Each scintillator film is coupled via a fishtail light-guide (also from Eljen Technology) to a 2-inch Hamamatsu H6533 PMT. The EJ-212 scintillator was chosen due to it being specifically formulated for thin films - a desirable attribute for use in-beam, as the particles will lose less energy and scatter less when passing through a thin film compared to a thick one. Each scintillator and PMT assembly is housed within a light-tight 3D-printed frame, which is held at beam height using a frame built from extruded aluminium profile. The ToF assemblies are shown in Figure~\ref{fig:ToFAssemblies}.

As indicated by its name, the key measurement for PID using the ToF system is the time taken for a particle to travel between the assemblies, which is dependent on both the particle mass and momentum. Since the momenta of the particles at T9 is known (to within 1\%), the measured flight time at any single momentum is then solely dependent on the particle mass, and therefore its type.
\\
\\
The two ToF assemblies were read out using a 2-channel, 0.25~ns per sample CAEN V1761 (10~bit, 4~GS/s) digitiser. This is sufficient resolution to distinguish between protons and other particle types at all momenta, and between muons and pions below 0.7~GeV/c.  Separating muons and pions above 0.7~GeV/c is not possible with this digitiser, due to the difference between their ToF values being less than the 0.25~ns resolution.  The same issue prevents the separation of electrons and muons at momenta above 0.8~GeV/c, hence the need for the previously described Cherenkov counter as a separate electron-tagging system.
\\
\\
In summary, a reasonable level of independent PID capability can be achieved by combining measurements from both the Cherenkov counter and the ToF system.  If a particle is detected by the Cherenkov counter and both of the ToF assemblies, it must be an electron.  If a particle is detected by both ToF assemblies, but not by the Cherenkov counter (i.e. there is no signal from the counter at the time the particle was expected to pass through it), it must be either a muon, pion or proton.  The exact time difference between the assembly signals indicates which of these particles is present, but muons and pions cannot be distinguished above 0.7~GeV/c due to the difference in their flight times being smaller than the digitiser's resolution.

\subsection{Operational Configuration}
\label{subsec:CERNOpConfig}
As noted above, operation of ARIADNE at T9 consisted of 18 days of beam running, during which time approximately 800,000 events were recorded over a range of momenta.

To reduce the possibility of pileup between beam particles - that is, two or more beam particles entering the detector within the same event window - the collimators were set so as to keep the number of particles per spill at approximately 20.  The exact collimator positions were determined empirically, with the general trend being that the collimators were more open - and the beam wider - when operating at lower momenta.  In all cases, the collimator positions varied between 0.4 and 9.0~mm (where 0mm indicates that the collimator is 'fully closed').
\\
\\
The ARIADNE detector itself was set up and operated as described above, with the segmented THGEM (previously discussed in Section~\ref{subsec:THGEM}) being used for S2 light production and charge readout. However, the additional beamline hardware described above introduced some extra requirements into the DAQ and trigger system that was previously described in Section~\ref{subsec:Trigger} and shown in Figure~\ref{fig:daq-trigger}:

\begin{itemize}
    \item The Cherenkov counter was read out using the CAEN V1720 digitiser, in addition to the existing PMT readouts on the same device.
    \item Along with the existing requirement of at least one above-threshold PMT signal and no EMCCD vetos, two additional signals were required for a global trigger to be issued. These were: 1) an above-threshold signal from the near ToF assembly (via the CAEN V1761 digitiser), and 2) the presence of the beam-spill signal (identifying that a spill has occurred, and provided by T9's own monitoring hardware). In other words, a particle would only trigger the detector if it was part of a spill and had passed through both the near ToF assembly and the main detector.
    \item A successfully issued global trigger signal was split between the now 3 digitisers as well as the EMCCDs.
\end{itemize}

\noindent Table~\ref{tab:CERNOpConfig} shows the various parameters that were used during beamline operation, arranged by device.

\begin{table}[ht]
    \centering
    \begin{tabularx}{\textwidth}{|p{3.0cm}|p{1.75cm}|p{9.1cm}|}
        \cline{1-3}
        \textbf{Parameter} & \textbf{Value} & \textbf{Comments} \\
        \cline{1-3}
        \cline{1-3}
        \textbf{EMCCDs} & & \\
        \cline{1-3}
        Gain & 1000 & Maximum available value \\
        \cline{1-3}
        Cooling & Water & Allows for operating temperature of -80\textdegree{}C, giving a mean background pixel intensity of $\approx 500$~ADU \\
        \cline{1-3}
        Binning & 4 x 4 & Gives a raw image size of 256 x 256 pixels per camera \\
        \cline{1-3}
        Exposure Time & 400 $\mu $s & \\
        \cline{1-3}
        \cline{1-3}
        \textbf{PMTs} & & \\
        \cline{1-3}
        Biases & 975V \newline 1000V \newline 1175V \newline 850V & Different voltages used so as to equalise the individual PMT responses \\
        \cline{1-3}
        Recording Length & 100,000 datapoints & Gives an event window of 400 $\mu $s, consistent with EMCCDs \\
        \cline{1-3}
        \cline{1-3}
        \textbf{ToFs} & & \\
        \cline{1-3}
        Biases & -2250V \newline -2250V & \\
        \cline{1-3}
        Recording Length & 20,000 \newline datapoints & Gives an event window of 5 $\mu $s, beginning at the same time as the global trigger is issued \\
        \cline{1-3}
        \cline{1-3}
        \textbf{Field Cage Biases} & & \\
        \cline{1-3}
        Cathode \newline Extraction Grid & -80.2 kV \newline -3.5 kV & Nominal drift field of 960~V/cm \\
        \cline{1-3}
        THGEM Top \newline THGEM Bottom & +1.4 kV \newline -1.3 kV & THGEM field of 27~kV/cm \\
        \cline{1-3}
        \cline{1-3}
        \textbf{Other} & & \\
        \cline{1-3}
        Cryostat pressure & 1.2 bar & \\
        \cline{1-3}
    \end{tabularx}
    \caption{Operational parameters for the ARIADNE detector at the T9 beamline.}
    \label{tab:CERNOpConfig}
\end{table}

\subsection{Results and Discussion}
\label{subsec:CERNResults}
\begin{figure}[ht]
\begin{center}
\begin{subfigure}{\textwidth}
  \centering
  \includegraphics[width=0.79\textwidth]{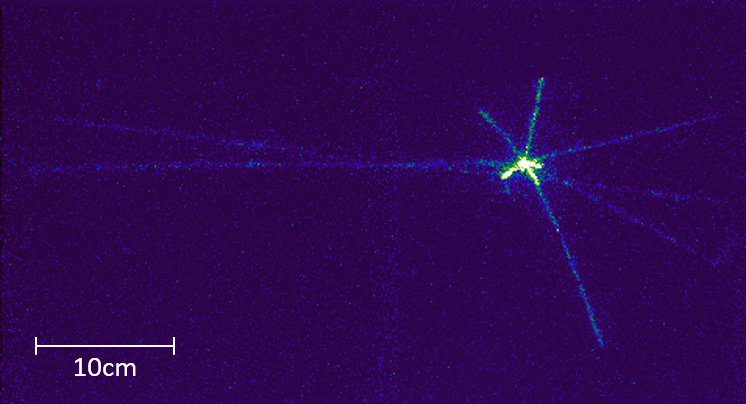}
  \caption{3.0~GeV/c anti-proton candidate, maximum intensity = 21894~ADU}
\end{subfigure}
\begin{subfigure}{\textwidth}
  \vspace{5mm}
  \centering
  \includegraphics[width=0.79\textwidth]{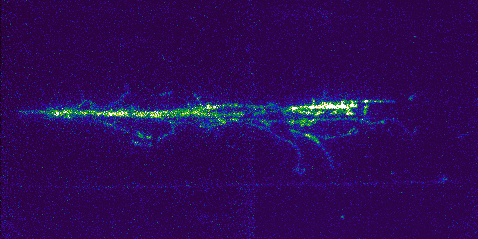}
  \caption{5.0~GeV/c electromagnetic shower, maximum intensity = 12916~ADU}
\end{subfigure}
\begin{subfigure}{\textwidth}
  \vspace{5mm}
  \centering
  \includegraphics[width=0.79\textwidth]{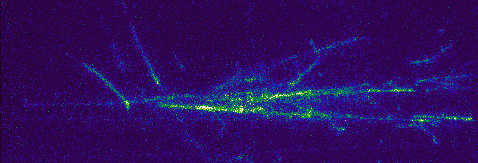}
  \caption{7.0~GeV/c electromagnetic shower, maximum intensity = 7367~ADU}
\end{subfigure}
\caption{A selection of negative polarity beam-induced events as captured by the ARIADNE detector. The indicated scale is common to all images.}
\label{fig:trackGallery1}
\end{center}
\end{figure}

\begin{figure}[ht]
\begin{subfigure}{0.49\textwidth}
  \centering
  \includegraphics[width=\textwidth]{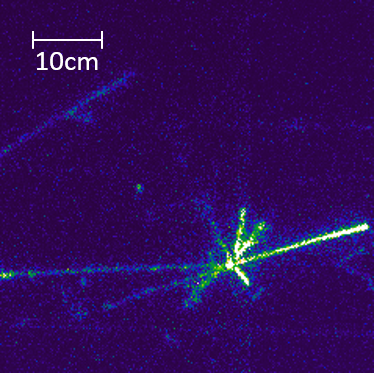}
  \caption{\centering 3.0~GeV/c anti-proton candidate, maximum intensity = 19372~ADU}
\end{subfigure}
\hspace{1mm}
\begin{subfigure}{0.49\textwidth}
  \centering
  \includegraphics[width=\textwidth]{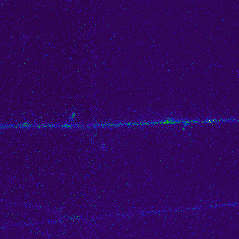}
  \caption{\centering 3.0~GeV/c muon candidate, maximum intensity = 4181~ADU}
\end{subfigure}
\begin{subfigure}{0.49\textwidth}
  \centering
  \vspace{5mm}
  \includegraphics[width=\textwidth]{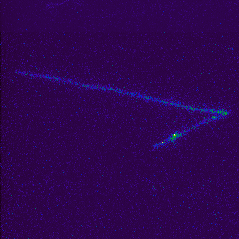}
  \caption{\centering 0.7~GeV/c nuclear recoil, maximum intensity = 5632~ADU}
\end{subfigure}
\hspace{1mm}
\begin{subfigure}{0.49\textwidth}
  \centering
  \vspace{5mm}
  \includegraphics[width=\textwidth]{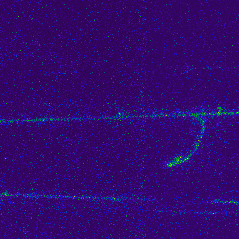}
  \caption{\centering 0.5~GeV/c delta ray candidate, maximum intensity = 3695~ADU}
\end{subfigure}
\caption{A selection of negative polarity beam-induced events as captured by the ARIADNE detector. The indicated scale is common to all images.}
\label{fig:trackGallery2}
\end{figure}

Figures~\ref{fig:trackGallery1} and \ref{fig:trackGallery2} present a selection of negative polarity beam-induced events as captured by the ARIADNE detector. These images represent the first time that argon interactions from a particle beam have been optically imaged in a dual-phase LArTPC.\\

\noindent The ARIADNE detector was operated at a THGEM field of 27~kV/cm, which is below nominal, due to a manufacturing defect on one of the channels on the 16-channel THGEM feedthrough. Figure~\ref{fig:THGEMGainPlot} (which uses the monolithic THGEM at a lower pressure of 1080~mbar, but is still representative of the segmented THGEM due to similar design and construction) indicates that the electron-multiplication electroluminescence regime becomes prominent at a THGEM field above $\approx$ 28.5~kV/cm. This means that CERN beamline operation was conducted in the proportional electroluminescence regime rather than the electron-multiplication regime, resulting in considerably reduced production of S2 light. However, even with these limitations, the EMCCD cameras were still able to reliably capture S2 light from argon interactions. The mean background pixel intensity at the EMCCD operating settings given in Table~\ref{tab:CERNOpConfig} was found to be $\approx$ 500~ADU, whereas typical pixel intensities along the particle tracks are seen to be well above this level.
\\
\\
Operation of the ARIADNE detector at the T9 beamline has demonstrated the optical readout method's capability for high spatial resolution: as previously noted in Section~\ref{subsec:Stitching}, the resolution of the images at 4 $\times$ 4 binning is 1.1~mm/pixel, which results in extremely well-defined and clear particle tracks.

This EMCCD image resolution is a contribution to the overall detector spatial resolution. Other contributions include uncertainties originating in the detector hardware, uncertainties from the analysis methods, and resolution limitations stemming from electron diffusion within the LAr. This particular contribution - which depends on the exact TPC operating parameters - can be split into transverse (in the $x-y$ plane) and longitudinal (in the drift - and therefore $z$ - direction) components, the former of which is directly comparable to the EMCCD image resolution. Studies \cite{electronDiffusion, electronDiffusion2} indicate that over a drift length of 80~cm and at drift fields between 0.5 and 1.0~kV/cm, the transverse electron diffusion is $\approx$ 1~mm - comparable in size to the EMCCD image resolution. This diffusion increases to $\approx$ 4.2~mm for a 12~m drift length at the same field (such as the TPC proposed for use in DUNE \cite{DuneCDRVol4}). This suggests that in future large LArTPCs, the overall spatial resolution would not be dominated by the resolution of EMCCD images at 4 $\times$ 4 binning, but instead by other effects such as the electron diffusion in LAr.
\\
\\
Although the 2D optical readout alone has shown excellent results, full 3D reconstruction of particle tracks requires the combination of EMCCD data with a faster readout channel, such as the PMT, for $z$ information. In the case of the PMT signal, the absolute $z$ position of a track in the TPC is determined by measuring the time difference between the S1 and S2 pulses. Relative track information, such as the angle of the track, may be determined from analysis of the S2 pulse shape alone.

\begin{figure}[ht]
\begin{center}
\vspace{5mm}
\begin{subfigure}{0.49\textwidth}
  \centering
  \includegraphics[width=\textwidth]{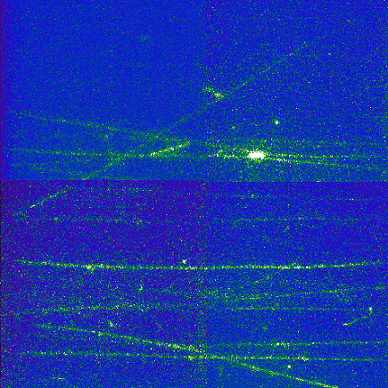}
\end{subfigure}
\hspace{1mm}
\begin{subfigure}{0.49\textwidth}
  \centering
  \includegraphics[width=\textwidth]{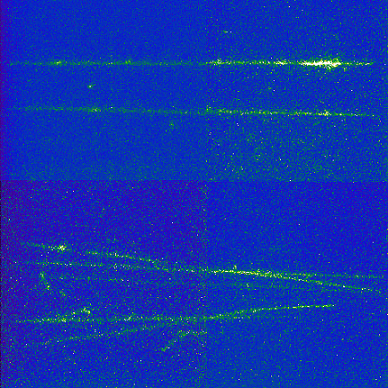}
\end{subfigure}
\caption{Examples of the ``halo'' muon background over an event window of 400~$\mu $s, generated by the PS / target interaction in the East Hall.}
\label{fig:haloMuons}
\end{center}
\end{figure}

Such a method of 3D track reconstruction has already been successfully demonstrated \cite{emccdargon}, but it requires events with low pileup - that is, an event containing a single track on the EMCCD image and a single (S1, S2) pulse pair on the PMT signal. If an event contains multiple tracks and/or pulse pairs, it becomes challenging to determine which pulse pair is associated with which track. In the case of the T9 beamline, the majority of events contained a large population of ``halo'' muons in the 400~$\mu $s event window. These are generated during the interaction of the PS proton beam and the East Hall interaction target, and travel in an increasingly wide conical spread towards and through the various experimental areas and detectors therein. These muons - some typical examples of which are shown in Figure~\ref{fig:haloMuons} - therefore act as a pileup-creating background in many events recorded by ARIADNE at the CERN T9 beamline.

\clearpage
\newpage
\FloatBarrier
\section{Cosmic Muons at the Liverpool LAr Facility}
\label{chp:Liverpool}

\subsection{Cosmic Muons}
A number of upgrades were performed on the ARIADNE detector following CERN T9 beamline operation - most notably, the change to using the monolithic THGEM and the installation of the laser calibration system. The detector as a whole has then been re-calibrated and characterised using cosmic muons.

The expected rate of cosmic muons incident on the 53 $\times$ 53~cm THGEM area is $\approx$ 50~Hz (based on the known rate of muons at the Earth's surface \cite{PDGMuons}), and this was indeed observed. This rate is significantly lower than the total rate of incident particles encountered at the CERN T9 beamline. Pileup is therefore much less of an issue when operating with cosmic muons, allowing for easier correlation between single tracks on the EMCCD images and single (S1, S2) pulse pairs in the PMT signal, and thus full 3D particle track reconstruction, as discussed previously in Section~\ref{subsec:CERNResults}. A lower pileup rate also means that a longer event window can be used.

Cosmic muons exhibit a much wider range of geometries and track lengths as they pass through the detector than beam particles. For example, a vertically incident cosmic muon can have a track length of up to 80~cm, compared to no more than 53~cm for a horizontal (beam-induced) track. This then allows for a broader characterisation of the detector response.

\subsection{Operational Configuration}
The ARIADNE detector was operational 24 hours a day for the duration of cosmic muon data-taking, in order to maximise the live time and record as many events as possible. As noted above, a longer event window - of 1~ms length - was possible due to relatively low rate of cosmic muon pileup.

The monolithic THGEM, discussed in Section~\ref{subsec:THGEM}, was used for cosmic muon data-taking, and the DAQ and trigger system was set up and operated as described in Section~\ref{subsec:Trigger} and Figure~\ref{fig:daq-trigger}, i.e. with no additional off-detector triggers. Due to the lower event readout rate, the EMCCD sensors naturally operate at a lower temperature during cosmic muon data-taking that during beamline operation, and there was no appreciable benefit to using the external liquid-cooling system. Therefore, to simplify operation of the detector, the EMCCDs were cooled only using their internal TE air-cooling system during cosmic muon data-taking.

Table~\ref{tab:LivOpConfig} shows the various parameters that were used during cosmic muon operation, arranged by device.

\begin{table}[ht]
    \vspace{5mm}
    \centering
    \begin{tabularx}{\textwidth}{|p{3.0cm}|p{1.75cm}|p{9.1cm}|}
        \cline{1-3}
        \textbf{Parameter} & \textbf{Value} & \textbf{Comments} \\
        \cline{1-3}
        \cline{1-3}
        \textbf{EMCCDs} & & \\
        \cline{1-3}
        Gain & Variable & Up to the maximum available value of 1000 \\
        \cline{1-3}
        Cooling & Air & Allows for operating temperature of -70\textdegree{}C, giving a mean background pixel intensity of $\approx 500$~ADU \\
        \cline{1-3}
        Binning & Variable & \\
        \cline{1-3}
        Exposure Time & 1 ms & \\
        \cline{1-3}
        \cline{1-3}
        \textbf{PMTs} & & \\
        \cline{1-3}
        Biases & 970V \newline 1100V \newline 1300V \newline 950V & Different voltages used so as to equalise the individual PMT responses \\
        \cline{1-3}
        Recording Length & 250,000 datapoints & Gives an event window of 1~ms, consistent with EMCCDs \\
        \cline{1-3}
        \cline{1-3}
        \textbf{Field Cage Biases} & & \\
        \cline{1-3}
        Cathode \newline Extraction Grid & -46.0 kV \newline -3.0 kV & Nominal drift field of 540~V/cm \\
        \cline{1-3}
        THGEM Top \newline THGEM Bottom & Variable \newline -1.0 kV & Between +1.0 to +2.2~kV\\
        \cline{1-3}
        \cline{1-3}
        \textbf{Other} & & \\
        \cline{1-3}
        Cryostat pressure & 1.08 bar & \\
        \cline{1-3}
    \end{tabularx}
    \caption{Operational parameters for the ARIADNE detector taking cosmic muon data at the University of Liverpool.}
    \label{tab:LivOpConfig}
\end{table}

\subsection{Results}
\label{subsec:LivResults}
\begin{figure}[ht]
\begin{center}
\begin{subfigure}{\textwidth}
  \centering
  \includegraphics[width=0.78\textwidth]{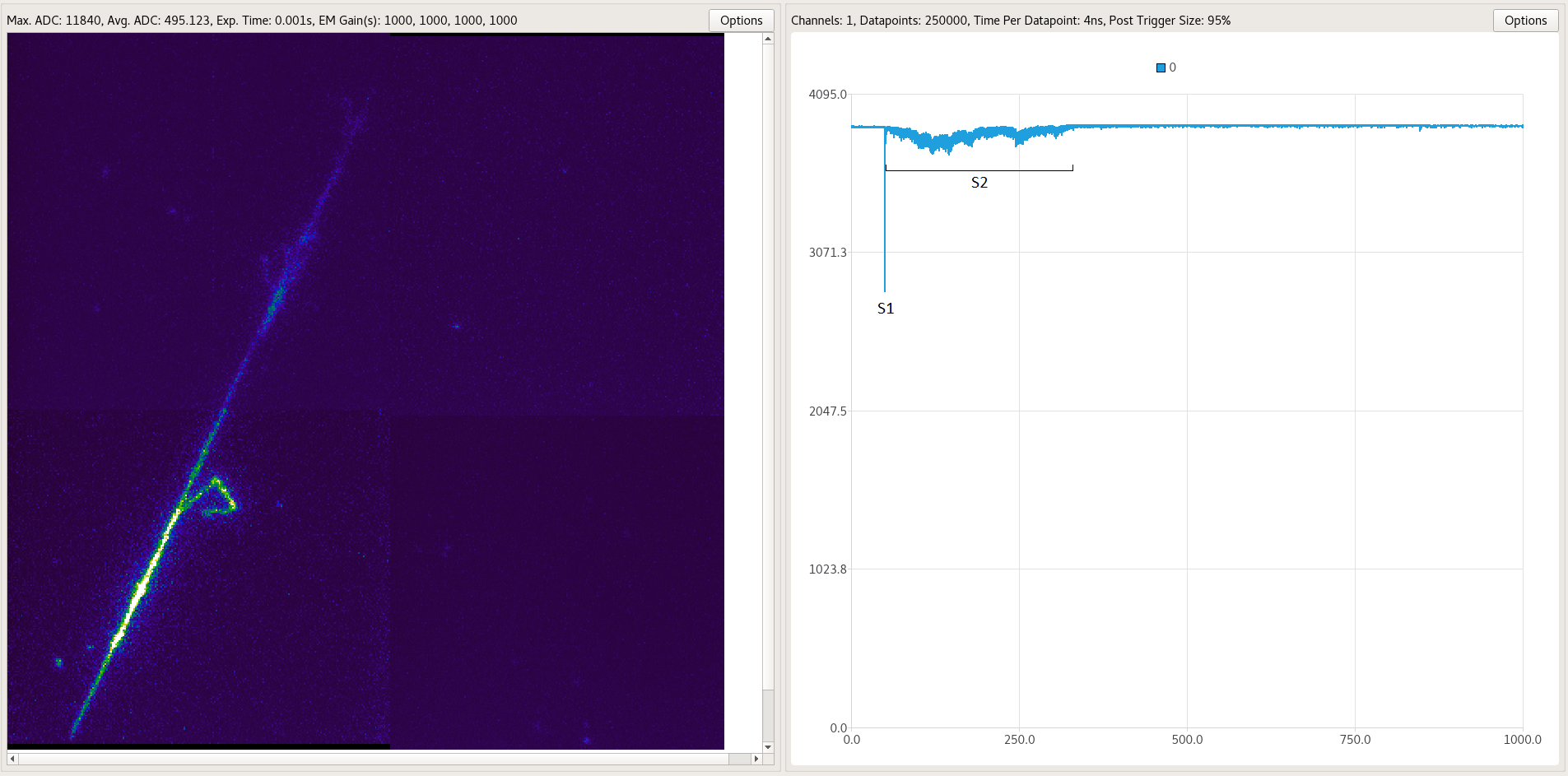}
\end{subfigure}
\begin{subfigure}{\textwidth}
  \vspace{5mm}
  \centering
  \includegraphics[width=0.78\textwidth]{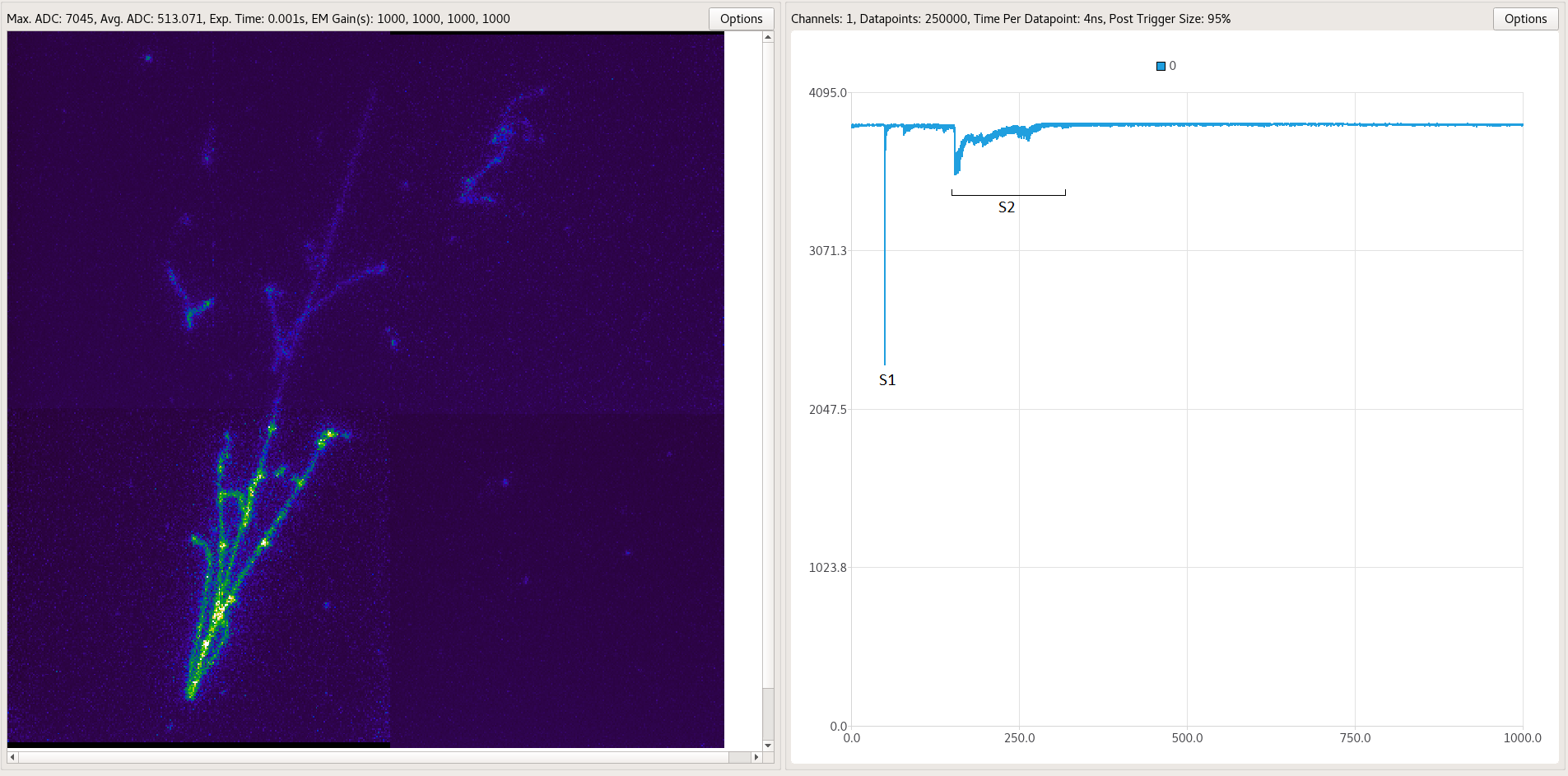}
\end{subfigure}
\begin{subfigure}{\textwidth}
  \vspace{5mm}
  \centering
  \includegraphics[width=0.78\textwidth]{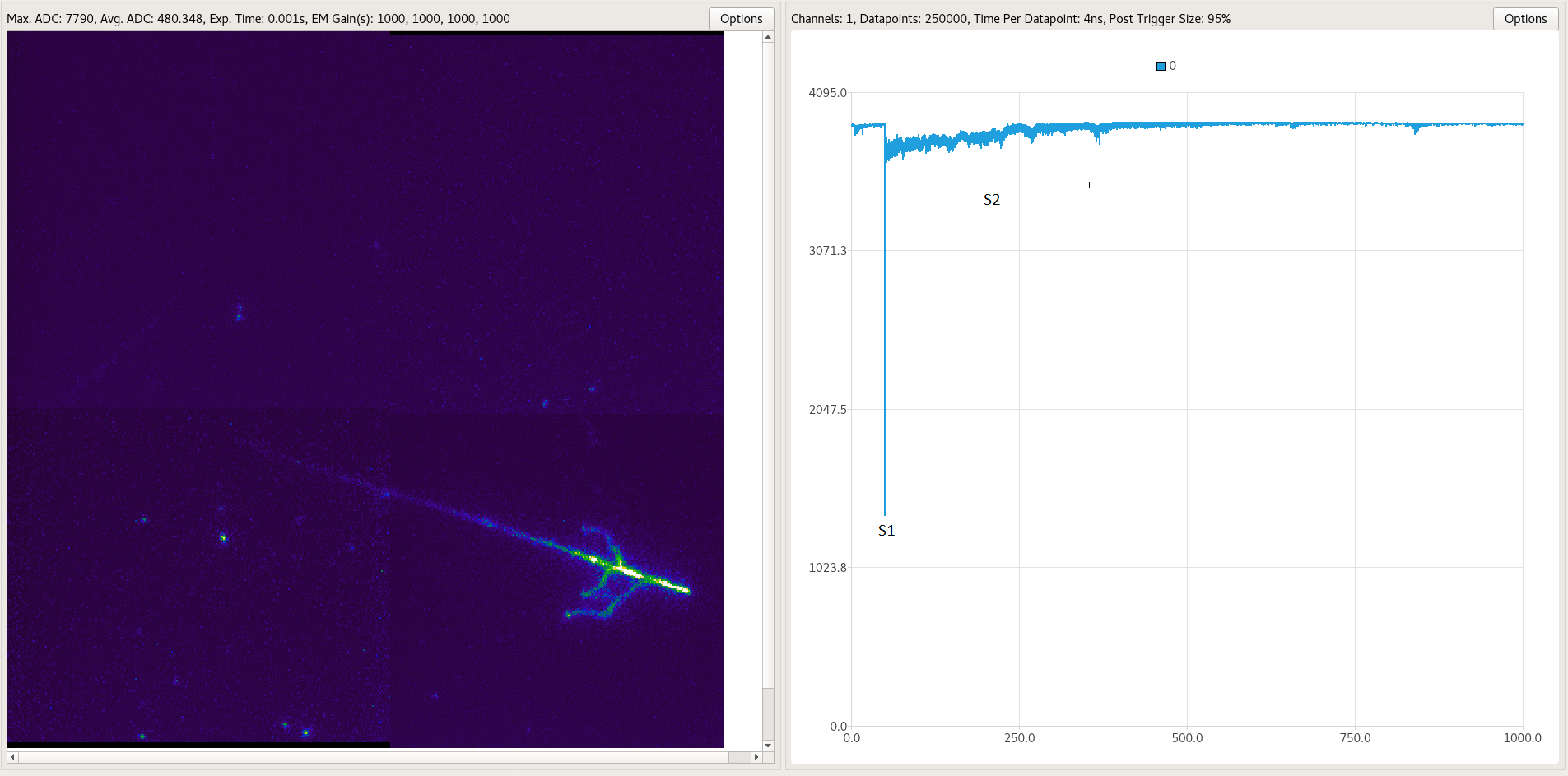}
  \label{fig:trackGallery3c}
\end{subfigure}
\caption{A selection of cosmic muon events as captured by the ARIADNE detector.  EMCCD images are shown on the left, with maximum intensities of 11840, 7045 and 7790~ADU (from top to bottom). The correlated PMT signals are shown on the right, with the S1 and S2 pulses labelled, and the signal time ($x$-axis) given in $\mu$s. All PMT signals have a full time window of 1000~$\mu$s - the same as the EMCCD exposure time.}
\label{fig:trackGallery3}
\end{center}
\end{figure}

Figure~\ref{fig:trackGallery3} presents a selection of cosmic muon events as captured by the ARIADNE detector using an EMCCD image binning of 4 $\times$ 4 pixels, an EMCCD gain of 1000, and a THGEM field of 30~kV/cm. The PMT signals and EMCCD images are correlated, and the S1 and S2 pulses can be clearly identified on the former.

Figure~\ref{fig:trackGallery3}~(top) represents a cosmic muon that entered the TPC from the top, passing through the THGEM itself. This can be inferred from the fact that the S1 and S2 pulses on the PMT signal have no separation in time, meaning that the electrons at the start of the ionisation track travelled effectively zero distance in the $z$ direction between their initial position and the THGEM. However, it is evident that this particle did not pass vertically through the TPC, but rather at an angle. If it had been completely vertical, the corresponding track would have a very short $x-y$ projection, but the EMCCD image shows an $\approx$ 55~cm long track.

Figure~\ref{fig:trackGallery3}~(middle) shows an electromagnetic shower, possibly originating from a $\gamma$, as is indicated by the lack of an ionisation track leading to the primary vertex. A number of daughter products, both charged and uncharged, can be seen and/or inferred from the shape and position of the tracks as the shower develops. Assuming a high purity and therefore a drift velocity of 1.5~mm/$\mu$s, the gap of $\approx$ 103~$\mu$s between the PMT S1 and S2 pulses indicates that the originating $\gamma$ entered the active volume $\approx$ 15.5~cm below the THGEM. The S2 pulse itself exhibits more internal structure than that in the top image, due to the numerous tracks within the shower.

Figure~\ref{fig:trackGallery3} (bottom) also depicts a cosmic muon that entered the active TPC volume from the top, with the S1 and S2 PMT pulses being separated by zero time. However, in comparison to the top image, the EMCCD track is somewhat shorter at $\approx$ 30~cm, but the S2 pulse is longer - lasting $\approx$ 450~$\mu$s. Together, these indicate that the muon in the bottom image travelled at a steeper angle through the TPC than the muon in the top image.\\

\clearpage

\begin{figure}[ht]
\begin{center}
\begin{subfigure}{\textwidth}
  \centering
  \includegraphics[width=0.32\textwidth]{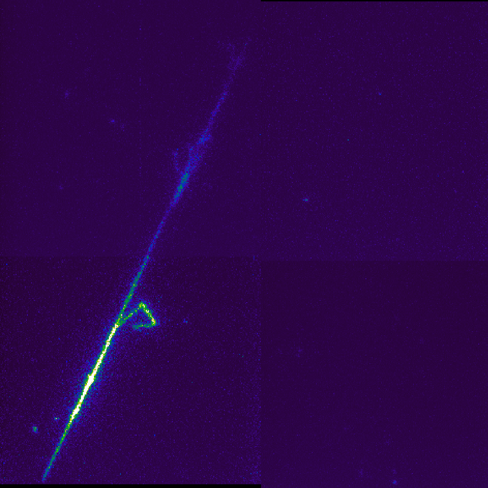}
  \hspace{1mm}
  \includegraphics[width=0.32\textwidth]{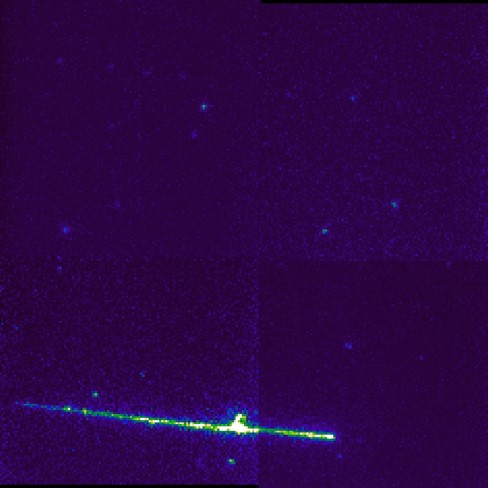}
  \hspace{1mm}
  \includegraphics[width=0.32\textwidth]{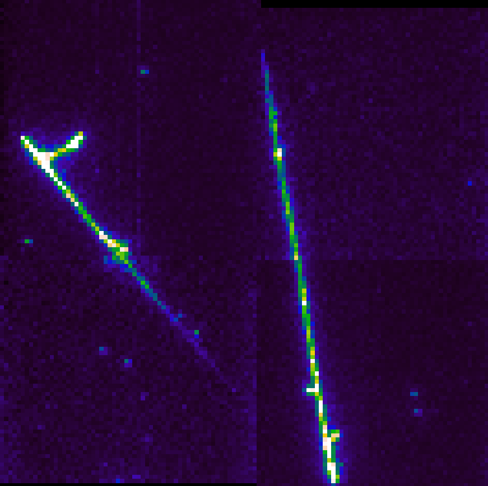}
\end{subfigure}
\caption{A selection of cosmic muon events as captured by the ARIADNE detector at different EMCCD binnings: (left) 4 $\times$ 4, (middle) 8 $\times$ 8, and (right) 16 $\times$ 16.}
\label{fig:binningComparison}
\end{center}
\end{figure}

\noindent Figure~\ref{fig:binningComparison} shows cosmic muons recorded using three different EMCCD binning scenarios: 4 $\times$ 4, 8 $\times$ 8 and 16 $\times$ 16. The left image is the same as that shown in Figure~\ref{fig:trackGallery3} (top), and has been reproduced here for comparison to the others. As a reminder of the calculation performed in Section~\ref{subsec:Stitching}, the spatial resolution of the images are 1.1, 2.2 and 4.4~mm/pixel respectively.\\

\noindent It can be seen that even 16 $\times$ 16 binning produces very clearly defined particle tracks, with an image resolution of 4.4~mm/pixel as noted in Section~\ref{subsec:Stitching} above. As previously discussed in Section~\ref{subsec:CERNResults}, this image resolution is only one contribution to the overall detector spatial resolution, and a large LArTPC of 12~m drift length operating at a drift field of 0.5~kV/cm would experience a transverse electron diffusion of up to $\approx$ 4.2~mm. This is effectively an equal contribution to the detector spatial resolution to that from the image resolution at 16 $\times$ 16 binning.

Although operating the EMCCDs at finer binning reduces their contribution to the overall detector resolution, using a coarser binning has advantages for data rates and sizes. The readout rate of the EMCCDs increases approximately linearly with binning ``coarseness'' \cite{Andor}, and file sizes are directly related to the number of post-binning pixels to be read out. Coarser binning also increases the overall light collection per pixel, reducing statistical effects that might otherwise have an impact on intensity and energy analysis, particularly for incident particles at low momenta.
\\
\\
\noindent The monolithic THGEM was able to operate stably at a higher field than the segmented one, reaching up to 32~kV/cm before discharges (previously discussed in Section~\ref{subsec:THGEM}) made further increase untenable. This increased stability is due to the simpler single-channel design: with only a single bias required for the top plane instead of 16, there is less possibility for defects to occur - such as that which affected the operation of the segmented THGEM.

\begin{figure}
\begin{subfigure}{\textwidth}
  \centering
  \includegraphics[width=\textwidth]{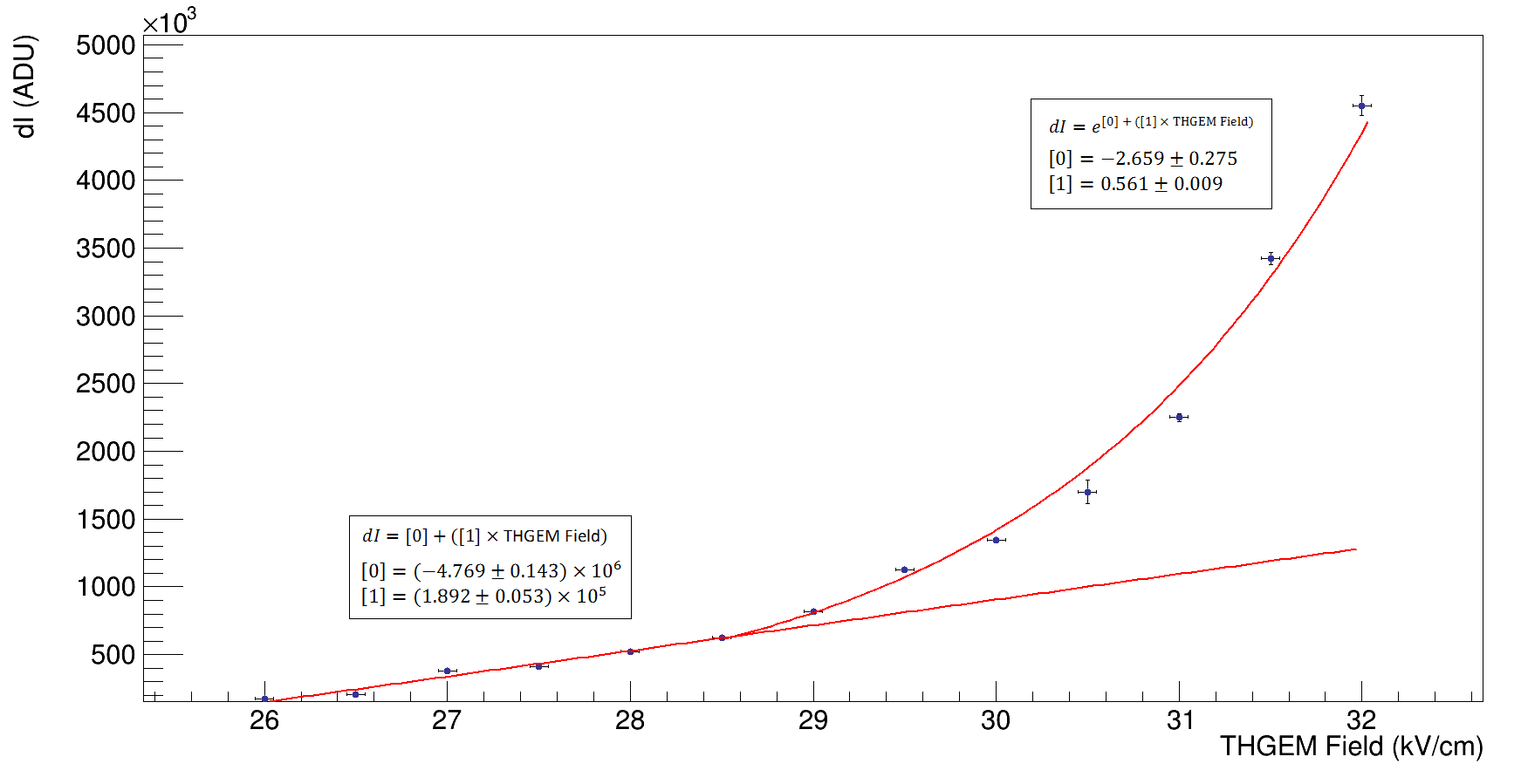}
  \caption{EMCCD image binning of 4 $\times$ 4 and EMCCD gain of 1000}
  \label{fig:THGEMGainPlot_4x4}
\end{subfigure}
\begin{subfigure}{\textwidth}
  \centering
  \vspace{5mm}
  \includegraphics[width=\textwidth]{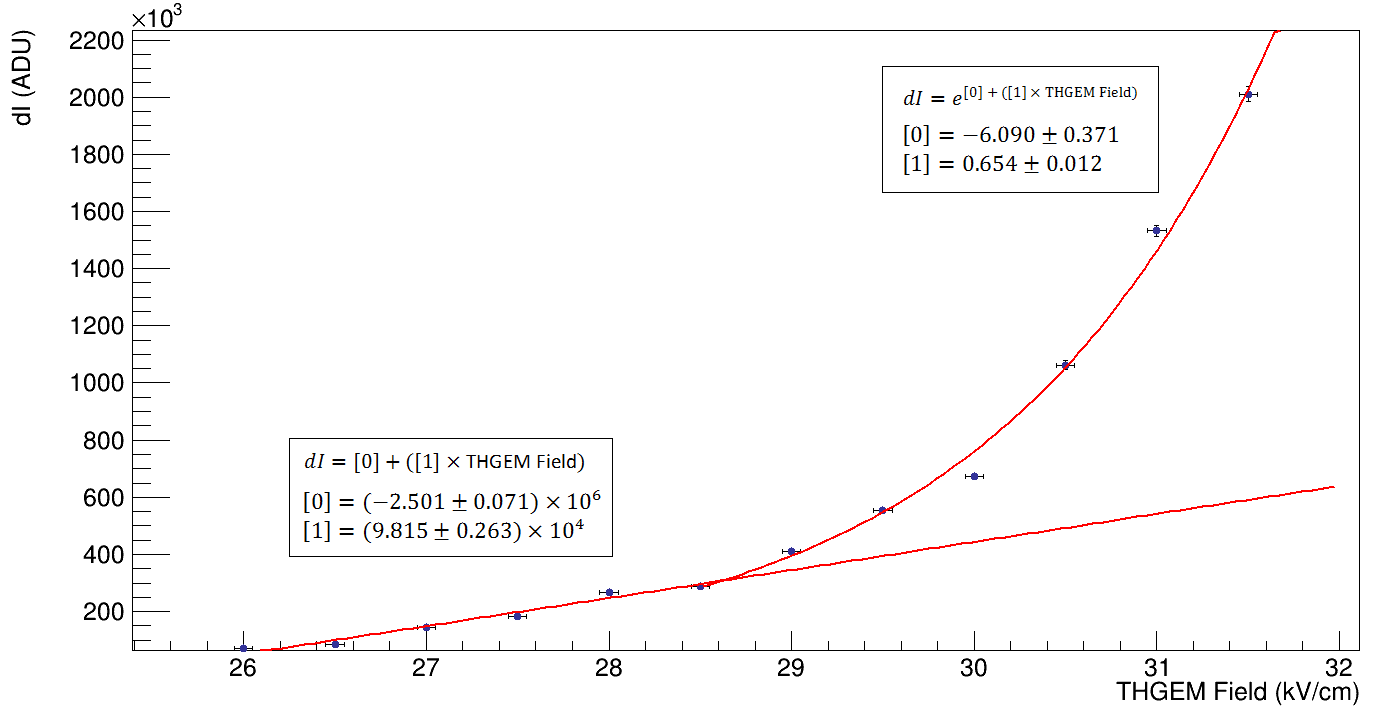}
  \caption{EMCCD image binning of 8 $\times$ 8 and EMCCD gain of 500}
  \label{fig:THGEMGainPlot_8x8}
\end{subfigure}
\caption{The relation between the total charge $dI$ - the summed intensity of all EMCCD pixels along a track - and the THGEM field. Each datapoint is the fitted Landau MPV of all single-track $dI$ values at that THGEM field. The displayed errors bars are the error on each MPV ($y$-axis) and the uncertainty in the total THGEM field ($x$-axis) due to the precision of the power supply. In both scenarios, only those cosmic muons which passed through both the THGEM and cathode grids were used, and linear and exponential functions have been fitted to the six lowest and highest datapoints respectively (with fit parameters as shown). Both functions are drawn extended to highlight the cross-over point between the proportional and electron-multiplication electroluminescence regimes.}
\label{fig:THGEMGainPlot}
\end{figure}

Stable operation allowed the characterisation of the relative light yield of the THGEM as a function of the field across it. This has been quantified by $dI$ - the summed intensity of all pixels along a given track on the EMCCD image. This value is not exactly the same as the THGEM gain, but is related to it (both being functions of the number of electrons leaving the THGEM holes) and so can still be used an reasonable indicator.
\\
\\
Figure~\ref{fig:THGEMGainPlot_4x4} shows the relation between the THGEM field and $dI$, for cosmic muons which pass through both the THGEM and the cathode grid, recorded with an EMCCD image binning of 4 $\times$ 4 and EMCCD gain of 1000. As they are minimum ionising particles (MIPs), cosmic muons have a specific energy loss of $\approx$ 2.1~MeV/cm in LAr \cite{cosmicCalibration}, and so this relation also leads to the energy calibration of the detector at these EMCCD settings. Figure~\ref{fig:THGEMGainPlot_8x8} shows the same relation for an EMCCD image binning of 8 $\times$ 8 and EMCCD gain of 500. A lower gain was required for the coarser binning due to the previously noted increased light collection per pixel, which was seen to result in saturation if an EMCCD gain of 1000 was used at 8 $\times$ 8 binning.

In the 4 $\times$ 4 binning scenario, it can be seen that there is a linear relationship below a field of $\approx$ 28.5~kV/cm, corresponding to the proportional electroluminescence regime in which the drift electrons do not produce Townsend discharges as they pass through the THGEM holes. However, as the field increases beyond 28.5~kV/cm, the value of $dI$ now increases in an exponential fashion, indicating the electron-multiplication regime.

The turn-over field value between the two regimes is effectively identical for the 8 $\times$ 8 binning scenario - as expected, since it is a function of the THGEM, and should therefore be unaffected by any changes in the EMCCD operating parameters.

\newpage
\FloatBarrier
\section{Summary and Outlook} 
The main aim of the ARIADNE experiment is to demonstrate the concept of an optical readout method for dual-phase LArTPC particle detectors. This makes use of the secondary scintillation light that is naturally produced in the THGEM of a dual-phase TPC, and could have considerable advantages over traditionally-used charge readouts. These include better energy and position sensitivity, and massive simplification of construction, operation and upgrading, together with corresponding reductions in costs. All of these factors are important to consider as kiloton-scale experiments, such as the four planned DUNE modules, are being realised.
\\
\\
ARIADNE utilises 4 single-photon sensitive EMCCD cameras to view the $x-y$ projection of secondary scintillation light that is produced by a 53 $\times$ 53~cm THGEM operating in concert with a TPC of drift length 80~cm. PMTs capture both the prompt and secondary scintillation light, allowing for measurements of the $z$ axis projection. As long as there is correlation between tracks on the EMCCD images and light pulses on the PMT signals, full 3D reconstruction of particle tracks can be achieved.

The ARIADNE detector was operated at the CERN T9 beamline between March and April 2018, in order to characterise the detector hardware with a controlled, well-described source of mixed particles over a range of momenta. The results constitute the first ever optical images of particle beam interactions in a dual-phase LArTPC. The high signal-to-background ratio that naturally comes from the use of the secondary scintillation light has been demonstrated, as has the fact that optical readout is sensitive enough to identify particle tracks even with little or no THGEM gain. Full 3D track reconstruction and analysis is ongoing, but has proved challenging due to particle pileup within the acquisition window (primarily caused by ``halo'' muons produced at the T9 source).

Following operation at the CERN T9 beamline, several upgrades were made to the detector hardware. Further data-taking was then conducted at the University of Liverpool using cosmic muons. A simpler THGEM design allowed for more stable running at higher fields, and the proportional and electron-multiplication electroluminescence regimes have been successfully demonstrated. A study on varying the EMCCD image binning indicates that even at the lowest possible spatial resolution of 4.4~mm/pixel, particle tracks are still clearly identifiable, while this coarser binning has additional operational advantages such as a faster acquisition rate, reduced data size and increased light collection per pixel.
\\
\\
Work is ongoing to further develop optical TPC data analysis software. In particular, the pixel-based nature of optical readout is well-suited to the use of neural networks and other forms of machine learning, which could result in extremely high-efficiency PID and event classification.
\\
\\
The feasibility of optical readout has been successfully demonstrated using EMCCD cameras, and future R$\&$D will be performed to enhance the performance of optical readout devices. Of particular interest is the development of camera technologies which offer extremely high-rate data acquisition and high-precision time-stamping, therefore allowing for 3D particle track reconstruction without the need for correlation to PMTs. Optical readout with a Timepix3 camera in a low pressure CF$_4$ TPC has recently been demonstrated by the authors \cite{TPX3Paper}, and implementation of this technology is planned for a future ARIADNE upgrade. Optical readout shows great promise in terms of performance and cost-efficiency for kiloton scale dual-phase LAr detectors, and could conceivably be used for the fourth DUNE module - the technology for which has not yet been decided. With this in mind, the next stage of optical readout R$\&$D would ideally include a larger-scale demonstration and optimisations in synergy with the DUNE collaboration.

\newpage
\FloatBarrier
\acknowledgments
The ARIADNE program is proudly supported by the European Research Council Grant No. 677927 and the University of Liverpool.
\\
\\
The authors are grateful to the members of the Mechanical Workshop of the University of Liverpool's Physics Department, for their expertise and dedicated contributions. We would also like to thank a number of the CERN staff for their support during operation of the ARIADNE detector at the T9 beamline - in particular: Lau Gatignon and Johannes Bernhard (beamline), Michael Jeckel (logistics), Johan Bremer and Laetitia Dufay-Chanat (cryogenics), Alexandre Desmarest and Olga Beltramello (safety and operational logistics) and Shaun Nightingale (transport). Finally, we would like to thank Philip Hamilton for his insightful comments and discussion.

\end{document}